\documentclass{aa}  

\usepackage{xcolor}
\usepackage{amssymb}
\usepackage{siunitx}

\usepackage{graphicx}
\usepackage{txfonts}
\defcitealias{Weaver2023}{W23}
\usepackage{hyperref}
\begin{document}

\title{\texttt{SEMPER} I: a novel semi-empirical model for the radio emission of star-forming galaxies at $0<$z$<5$}
\titlerunning{\texttt{SEMPER} I. Radio Predictions for Star-Forming Galaxies at $0<z<5$}

   \author{M. Giulietti
          \inst{1}\thanks{\email{m.giulietti@ira.inaf.it}}, I. Prandoni\inst{1}, M. Bonato\inst{4}, L. Bisigello\inst{1,3}, M. Bondi\inst{1}, G. Gandolfi\inst{7,8}, M. Massardi\inst{2,4}, L. Boco\inst{9}, H. J. A. Rottgering\inst{10}, 
          \and A. Lapi\inst{2,1,5,6}
          }

   \institute{INAF - Istituto di Radioastronomia,
Via Gobetti 101, 40129 Bologna, Italy             
         \and
             Scuola Internazionale Superiore di Studi Avanzati, Via Bonomea 265, 34136 Trieste, Italy 
        \and
            Dipartimento di Fisica e Astronomia "G. Galilei", Università di Padova, Via Marzolo 8, 35131 Padova, Italy
            \and
            INAF - Istituto di Radioastronomia - Italian ALMA Regional Centre, Via Gobetti 101, 40129 Bologna, Italy
        \and
            IFPU - Institute for Fundamental Physics of the Universe, Via Beirut 2, 34014 Trieste, Italy
        \and
            INFN-Sezione di Trieste, via Valerio 2, 34127 Trieste,  Italy
        \and
            Dipartimento di Fisica e Astronomia "G. Galilei", Universit\`a di Padova, Via Marzolo 8, 35131 Padova, Italy
        \and
            INAF, Osservatorio Astronomico di Padova, Vicolo dell’Osservatorio $5$, 35122, Padova, Italy
        \and
            Universit{\"a}t Heidelberg, Zentrum f{\"u}r Astronomie, Institut f{\"u}r theoretische Astrophysik, Albert-Ueberle-Str. 3, 69120 Heidelberg, Germany
        \and
             Leiden Observatory, Leiden University, PO Box 9513, NL-2300 RA Leiden, The Netherlands. 
        }            
   \date{}
\authorrunning{Giulietti et al.}
 
  \abstract
   {Star-forming galaxies (SFGs) are the dominant population in the faint radio sky, corresponding to flux densities at 1.4 GHz $< 0.1$ mJy A panchromatic approach is essential for selecting SFGs in the radio band and understanding star formation processes over cosmic time. 
   Semi-empirical models are valuable tools to effectively study galaxy formation and evolution, relying on minimal assumptions and exploiting empirical relations between galaxy properties and enabling us to take full advantage of the recent progress in radio and optical/near-infrared (NIR) observations.}   
   {In this paper, we develop the Semi-EMPirical model for Extragalactic Radio emission (\texttt{SEMPER}) to predict radio luminosity functions and number counts at 1.4 GHz and 150 MHz for SFGs. \texttt{SEMPER} is based on state-of-the-art empirical relations, with the goal of better understanding the radio properties of high-z, massive galaxy populations.}
   {We combine the redshift-dependent galaxy stellar mass functions obtained from the recent COSMOS2020 catalogue, which exploits deep near-infrared observations, with up-to-date observed scaling relations such as the galaxy main sequence and the mass-dependent far-infrared/radio correlation across cosmic time. Our luminosity functions are compared with recent observational determinations from the Very Large Array (JVLA), the Low-Frequency Array (LOFAR), the Westerbork Synthesis Radio Telescope (WSRT), the Giant Metrewave Radio Telescope (GMRT) and the Australian Telescope Compact Array (ATCA), along with previous semi-empirical models and simulations.}
   {Our semi-empirical model successfully reproduces the observed luminosity functions at 1.4 GHz and 150 MHz up to $z\sim 5$ and the most recent number count statistics from radio observations in the LOFAR Two-metre Sky Survey (LoTSS) deep fields. Our model, based on galaxies selected in the NIR, naturally predicts the presence of radio-selected massive and/or dust-obscured galaxies already in place at high redshift ($z\gtrsim3.5$), as suggested by recent results from the James Webb Space Telescope (JWST). Our predictions offer an excellent benchmark for upcoming updates from JWST and future ultra-deep radio surveys planned with the Square Kilometre Array (SKA) and its precursors.}
   {}
   \keywords{Radio continuum: galaxies --
                Galaxies: luminosity function, mass function
               }

   \maketitle
%

\section{Introduction}\label{sec:intro}

Star-forming galaxies (SFGs) emit in the radio band due to synchrotron radiation originating from electrons accelerated in supernova remnants and free-free continuum emission from hot and ionised HII regions. 
The radio luminosity, therefore, constitutes an effective tracer of the star-formation rate (SFR) of galaxies (\citealt{Condon2002AJ....124..675C}; \citealt{Kennicutt2012AR}) provided that the contribution of the Active Galactic Nuclei (AGN) is negligible. 
The link between radio emission and SFR can be derived from the tight correlation observed between the radio and Far-Infrared (FIR) luminosities (\citealt{Helou1985}; \citealt{Condon1992}; \citealt{Lacki2009}; \citealt{Murphy2011ApJ...737...67M}), with the latter tracing the star formation activity in dust-enshrouded environments. With respect to other tracers, such as the ultraviolet (UV) and H$\alpha$ luminosities, the radio emission is unaffected by dust extinction and can be efficiently exploited to study the dust-obscured star-formation (e.g. \citealt{Chapman2004}).

Radio astronomy has been transformed by the increasing capability to detect SFGs which dominate the faint radio sky, gradually emerging from radio-loud (RL) AGNs at sub-mJy flux level and becoming increasingly prominent at flux densities below $\sim 100\, \mu$Jy at 1.4 GHz (\citealt{Smolcic2008, Smolcic2017}; \citealt{Padovani2015, Padovani2016}; \citealt{Prandoni2018}; \citealt{Algera2020ApJ...903..139A}; \citealt{vanderVlugt2021ApJ...907....5V}). The faint radio populations also include a significant fraction of Radio Quiet (RQ) AGNs, which exhibit evidence of nuclear activity in the X-ray, mid-IR, or optical bands but show weak radio emission (i.e., no large-scale jets). The origin of the radio emission in these objects remains a topic of debate, primarily due to the difficulty in disentangling the contribution of the central nucleus from that of the host galaxy (\citealt{Bonato2017}; \citealt{Mancuso2017}; \citealt{Ceraj2018}; \citealt{Prandoni2018}).
Understanding these faint radio populations has been possible thanks to the significantly improved sensitivity of current deep and wide radio surveys (e.g. \citealt{Helfand15}; \citealt{Lacy2020PASP..132c5001L}; \citealt{Norris2021PASA...38...46N}; \citealt{Heywood2022};\citealt{Shimwell2022}), which have opened up new opportunities for studying the radio sky. These surveys are playing an increasingly important role in the field of galaxy evolution, providing a new avenue for investigating the star formation history of galaxies up to high redshifts (see reviews by \citealt{Dezotti2010} and \citealt{Padovani2016}; \citealt{Mancuso2015}; \citealt{Duchesne24}; \citealt{Bonato2021b}; Giulietti et al. in prep.).
This has become possible thanks to the upgraded sensitivity of radio telescopes such as the \textit{Karl G. Jansky} Very Large Array (JVLA) and the advent of new-generation radio telescopes such as MeerKAT (\citealt{Jarvis2016}), the Low-Frequency Array (LOFAR, \citealt{vanHaarlem2013}) and the Australian Square Kilometre Array Pathfinder (ASKAP, \citealt{Johnston2007PASA...24..174J}) and with the forthcoming Square Kilometre Array (SKA).

A panchromatic approach is indispensable in selecting SFGs in the radio band, for this reason, deep radio surveys have been conducted in sky fields covered by a wealth of multi-wavelength (X-ray to radio) data (e.g. \citealt{Jarvis2016}; \citealt{Novak2017}; \citealt{Smolcic2017}; \citealt{Prandoni2018}; \citealt{Bonato2021b, Bonato2021a}; \citealt{Duncan2021}; \citealt{Kondapally2021}; \citealt{Sabater2021}; \citealt{Tasse2021};\citealt{Heywood2022}; \citealt{Whittam2022};   \citealt{Best2023}). 

Recent advancements in radio observations of these sky regions have mirrored those in the Optical/Near-IR bands. For instance, in the Cosmic Evolution Survey (COSMOS) field, the latest COSMOS2020 catalogue (\citealt{Weaver2022, Weaver2023}) has benefited from the fourth data release of the UltraVISTA survey (\citealt{McCracken2012}; \citealt{Moneti2023}, Ks=25.7 at $5\sigma$), along with new observations from the James Webb Space Telescope (JWST) as part of the COSMOS-Web project (\citealt{Casey2023}; \citealt{Shuntov2025}). In the North Ecliptic Pole, the Hawaii eROSITA Ecliptic Pole Survey Catalog (HEROES, \citealt{Taylor2023}) has recently been published, and upcoming data from the \textit{Euclid} mission (\citealt{EuclidColl2022A&A...662A.112E}) will provide sub-arcsec NIR imaging down to H = 26 mag.

We now require a coherent, multi-wavelength approach that weaves together these advancements, providing a comprehensive view of the processes driving star formation as a function of redshift. Motivated by this, in this work we have developed a semi-empirical model which leverages recent estimates of the evolution of galaxies Stellar Mass Function (SMF), based on the NIR data from the COSMOS2020 catalogue (\citealt{Weaver2022,Weaver2023}), with the main goal of predicting the statistics of SFGs in the radio band.

Semi-empirical models have recently been proven to be an effective approach to galaxy formation and evolution (e.g. \citealt{Behroozi2013ApJ...770...57B, Behroozi2019MNRAS.488.3143B}; \citealt{Moster2013MNRAS.428.3121M, Moster2018MNRAS.477.1822M}; \citealt{Mancuso2016a}; \citealt{Grylls2019}; \citealt{Bisigello2022A&A...666A.193B}; \citealt{Fu2022}; \citealt{Boco2023}; see also the review by \citealt{Lapi2025}). At variance with cosmological hydrodynamic simulations (see \citealt{Vogelsberger2020NatRP...2...42V} for a review) and semi-analytic models (e.g. \citealt{Lacey2016MNRAS.462.3854L}; \citealt{Lagos2018MNRAS.481.3573L}; \citealt{Henriques2020MNRAS.491.5795H}; \citealt{Parente2023MNRAS.521.6105P}), semi-empirical models do not aim to describe the small-scale physics governing the baryon cycle from first principles. 
Instead, they rely on empirical relations between spatially averaged galaxy properties, keeping the number of parameters and assumptions to a minimum. The main limitation of this approach is that, being inherently data-driven, it is less suitable for uncovering the details behind the physical processes involved.
However, given their small number of parameters, semi-empirical models are easily expandable and computationally efficient. At the same time, they can also be exploited to identify inconsistencies between datasets by connecting different observables and efficiently performing predictions for future missions.

In this paper, we leverage state-of-the-art scaling relations to develop the up-to-date Semi-EMPirical model for Extragalactic Radio emission (\texttt{SEMPER}), used to predict luminosity functions (LFs) and number count statistics of SFGs in the radio band. In particular, we combine the redshift-dependent galaxy SMF of \cite{Weaver2023} (W23 hereafter) and various observed relations, such as the galaxy main sequence (MS; \citealt{Brinchmann2004MNRAS.351.1151B}; \citealt{Noeske2007ApJ...660L..43N}) and the recently derived mass- and redshift- dependent Far-Infrared/Radio Correlation (FIRRC, \citealt{Delvecchio2021}; \citealt{McCheyne2022}).
Our predictions are compared with state-of-the-art continuum observations at 1.4 GHz and 150 MHz from deep radio surveys conducted with the JVLA, LOFAR, the Westerbork Synthesis Radio Telescope (WSRT), the Giant Metrewave Radio Telescope (GMRT) and the Australian Telescope Compact Array (ATCA), along with the recent Tiered Radio Extragalactic Continuum Simulation (T-RECS, \citealt{Bonaldi2019MNRAS.482....2B, Bonaldi2023MNRAS.524..993B}) and previous semi-empirical models (\citealt{Mancuso2017}) for SFGs emitting in the radio-band. To extend our comparison analysis we also derive the 150 MHz radio number counts derived from the recent LOFAR Two-metre Sky Survey (LoTSS; \citealt{Shimwell2017, Shimwell2019, Shimwell2022}) three deep fields (\citealt{Duncan2021}; \citealt{Kondapally2021}; \citealt{Sabater2021}; \citealt{Tasse2021}; \citealt{Best2023}; \citealt{Cochrane2023}).

This is the first of a series of papers, in a forthcoming work, we will expand \texttt{SEMPER} to include radio emission from AGNs.

The paper is structured as follows. Sect. \ref{sec: method}, we describe our model and its main ingredients, Sect. \ref{sec: comparison samples}, we describe the datasets exploited in this work as a comparison to our model. In Sect. \ref{sec: results and discussion}, we present and discuss our results and finally, we draw our conclusions in Sect. \ref{sec:conclusion}.
\noindent
In this work, we assume a \cite{Chabrier2003} initial mass function (IMF) and a standard $\Lambda$CDM cosmology with parameters: $H_0=70 \rm \, km\,s^{-1} Mpc^{-1}$, $\Omega_{\Lambda,0}=0.7$ and $\Omega_{m,0}=0.3$, such that $h_{70} \equiv  H_0 /(70\, \rm km \, \rm s^{-1} \rm Mpc^{-1}) = 1$. All magnitudes are expressed in the AB system (\citealt{Oke1974ApJS...27...21O}). The radio source spectra are assumed to be described by a simple power law $S_{\nu} \propto \nu^{\alpha}$, where $S_{\nu}$ is the monochromatic flux density at a certain frequency $\nu$ and $\alpha$ is the radio spectral index.

\section{Method}\label{sec: method}

In this section, we describe the key ingredients of our semi-empirical model. 

\subsection{Stellar mass functions}

We exploit the recent SMFs for SFGs reported by \citetalias{Weaver2023}, expressed as $N(z, \log{M_{\star})}$ and based on the most recent data release of the COSMOS2020 catalogue (\citealt{Weaver2022}). This choice is motivated by two main reasons. First, COSMOS is one of the deepest and most studied fields, benefitting from a broad UV-to-radio coverage. The \textit{izYJHKs} coadded image of COSMOS2020 reaches magnitudes down to 26 AB and ensures completeness down to $10^9\,$M$_{\odot}$ at $z\approx 3$ for a mass-selected sample of $\sim \num{1000000}$ galaxies. 
Moreover, the objects' classifications, photometric redshifts, and physical properties are robustly supported by extensive photometry, spanning from UV to 8 $\mu$m across a total area of 2 deg$^2$ (see \citealt{Weaver2022} and \citetalias{Weaver2023} for details).

From the COSMOS2020 photo-z catalogue, \citetalias{Weaver2023} retrieved the galaxies SMF's shape and evolution from $z\approx0.2$ up to $z=7.5$ for the total mass-complete sample and up to $z=5.5$ for the quiescent and star-forming mass-complete sub-samples.
We exploit the observed measurements and the maximum likelihood parameters found by \citetalias{Weaver2023} (see their Tab. C.2.) from their Monte Carlo Markov Chain (MCMC) analysis to retrieve the best-fit curves for SFGs, described by a double (single) Schechter function for galaxies at $z<3$ ($z>3$):

\begin{equation}\label{eq:double_schechter}
\begin{aligned}
&\begin{aligned}
\Phi d \log M= & \ln (10) e^{-10^{\log M-\log M^*}} \\
& \times\left[\Phi_1^*\left(10^{\log M-\log M^*}\right)^{\alpha_1+1}\right. \\
& \left.+\Phi_2^*\left(10^{\log M-\log M^*}\right)^{\alpha_2+1}\right] d \log M.
\end{aligned}\\
\end{aligned}
\end{equation}

\noindent
In this equation, $\Phi^*_1$ and $\Phi^*_2$ are the individual normalisation of the two functions,  $\alpha_1$ and $\alpha_2$ the relative low-mass slopes and $M^*$ the characteristic stellar mass.

One of the main findings of \citetalias{Weaver2023} is the high number density of massive galaxies ($M_{\star} > 10^{11}\,$M$_{\odot}$) between $ 3 < z \leq 5.5$ compared to previous analyses using an identical selection (\citealt{Davidzon2017}). 
These massive SFGs have extreme red colours and are $\gtrsim 1$ order of magnitude fainter than the median Ks AB magnitude of the sample.
Most of these galaxies ($> 60\%$) were classified as star-forming systems at $z>1$, a fraction of which is likely to be dust-obscured and extremely faint in the NIR, being therefore missed by previous studies (see Sect. \ref{sec:dark_objects}).
Because the Schechter formalism employed in \citetalias{Weaver2023} to fit the observed SMFs fails to reproduce the excess of massive and extremely red galaxies, in this study, we re-fit the observed SMFs from \citetalias{Weaver2023}.
To do so, we employed a simple double power-law function of the form:

\begin{equation}\label{eq:double_powerlaw}
\begin{aligned}
\log \Phi d \log M=- & \log \left(10^{\left(\log M-\log M_0\right)(\alpha+1)+\log \Phi_1}\right. \\
+ & \left.10^{\left(\log M-\log M_0\right)(\beta+1)+\log \Phi_2}\right) d \log M,
\end{aligned}
\end{equation}

\noindent
where $\log \Phi_1$ and $\log \Phi_2$ are the normalisations of the two power laws, $\alpha$ and $\beta$ are the two slopes and $\log M_0$ represents the mass corresponding to the slope-change.

Since the SMFs presented in \citetalias{Weaver2023} are limited to $z\gtrsim0.2$, we included SFGs at lower redshifts by exploiting the recent SMFs reported in the Galaxy And Mass Assembly Survey Data Release 4 (GAMA DR4; \citealt{Driver2022}). 
The GAMA survey \citep{Driver2011} covers five sky regions for a total of $250\,$deg$^2$, providing images and spectroscopic redshifts for $\sim \num{230000}$ sources along with UV-to-FIR 20-band photometry. \citet{Driver2022} divided the SMF for morphological type, extending the estimates to a lower mass limit of $10^{6.75}\,$M$_{\odot}h^{-2}_{70}$ at $z < 0.1$. In this work, we select the SMFs of the morphological type "D", corresponding to single-component late-type systems, dominating the stellar mass density for local galaxies for stellar masses $M_{\star}<10^{9.25}\,$M$_{\odot}\,h^{-2}_{70}$, under the assumption that these galaxies are the primary star-forming systems in the local Universe.
To model the observed measurements of galaxies at $z<0.08$, \cite{Driver2022} elected a single Schechter function. In this case, we adopt Equation \ref{eq:double_powerlaw} to maintain consistency with our approach in fitting \citetalias{Weaver2023}'s observations. 

We reconstruct the SMF's shape via a Bayesian MCMC framework. We exploit the Python package \texttt{emcee} (\citealt{Foreman2013}).
We discuss the details of our fitting procedure in Appendix \ref{app:a_1}. The results are shown in Fig.  \ref{fig:fit_SMF_Driver_vs_dpl}  and Fig. \ref{fig:fit_SMF_Weaver_vs_dpl} and are compared with the best fits obtained respectively by \cite{Driver2022} and \citetalias{Weaver2023}. Our Double Power Law fit successfully reproduces the observed data for all the observed redshift bins and also successfully traces the high-mass points for $z>3$, where the Double Schechter profile does fail.
As a second step of our analysis, we build a continuity model to obtain the shape and evolution of SMFs by fitting the double power law's parameters as a function of redshift. This approach enables a full redshift interpolation of our model, particularly in the redshift bin $0.08<z<0.2$, where no observations are available.

\subsection{Main sequence}

The next step involves computing the star-formation rate function (SFRF) to determine the distribution in SFR for SFGs at a given redshift. We therefore exploit the well-known MS relation (\citealt{Brinchmann2004MNRAS.351.1151B}; \citealt{Noeske2007ApJ...660L..43N}).

The MS is a tight relation linking the stellar mass and the SFR ($\psi$) of a galaxy over a wide range of redshifts ($0<z<6$) through the specific star-formation rate (sSFR$\equiv \frac{\psi}{M_{\star}}$).
This relation has been extensively studied over the past decade both observationally and theoretically (see e.g. \citealt{Daddi2007,Daddi2022}; \citealt{Rodighiero2011,Rodighiero2015}; \citealt{Speagle2014}; \citealt{Whitaker2014}; \citealt{Schreiber2015}; \citealt{Mancuso2016b}; \citealt{Dunlop2017}; \citealt{Bisigello2018}; \citealt{Pantoni2019}; \citealt{Lapi2020}; \citealt{Leslie2020}; \citealt{Thorne2021}; \citealt{Leja2022}; \citealt{Popesso2023}), even though its redshift evolution, scatter and exact shape are still debated (\citealt{Peng2010}; \citealt{Rodighiero2014}; \citealt{Speagle2014}; \citealt{Whitaker2014}; \citealt{Renzini2015}; \citealt{Schreiber2015}; \citealt{Pearson2018}; \citealt{Popesso2019a, Popesso2019b}; \citealt{Leslie2020}; \citealt{Thorne2021}; \citealt{Leja2022}), in particular in the high-mass regime.

In our model, we use the recent results by \cite{Popesso2023}, which compiles numerous literature studies converted to a common calibration, covering a wide range of redshifts and stellar masses.
However, it must be noted that the MS is an average relation between the stellar mass and the SFR, and it displays some dispersion and significant outliers. Several studies suggest that SFGs, at a fixed redshift and stellar mass, are distributed in SFR following a double Gaussian shape (\citealt{Bethermin2012}; \citealt{Sargent2012}; \citealt{Ilbert2015}; \citealt{Schreiber2015}). The dominant population consists of MS galaxies and their Gaussian distribution in SFR is centred around the MS value. In contrast, the sub-complementary population of starburst galaxies have a SFR distribution centred around a value $3-4\sigma$ above the MS value. 
Several authors (e.g. \citealt{Caputi2017}; \citealt{Bisigello2018}; \citealt{Rinaldi2025}) have found an increase in starburst fraction, compared to the MS, for low mass ($M_{\star} \leq 10^9$) or higher redshift ($z\geq 2- 3$) galaxies. 

In this work, we are assuming the double-Gaussian decomposition proposed by \citet{Sargent2012} and also adopted in recent works \citep{Boco2021a}, describing the SFR distribution of a galaxy at fixed mass and redshift as:

\begin{equation}\label{eq:double_gaussian_SFR}
\begin{aligned}
\frac{d p}{d \log \psi}\left(\psi \mid z, M_{\star}\right)= & \left(\frac{A_{\mathrm{MS}}}{\sqrt{2\pi \sigma^2_{\rm MS}}}\right) \exp \left[-\frac{\left(\log \psi-\langle\log \psi\rangle_{\mathrm{MS}}\right)^2}{2 \sigma_{\mathrm{MS}}^2}\right] \\
& +\left(\frac{A_{\mathrm{SB}}}{\sqrt{2\pi \sigma^2_{\rm SB}}}\right) \exp \left[-\frac{\left(\log \psi-\langle\log \psi\rangle_{\mathrm{SB}}\right)^2}{2 \sigma_{\mathrm{SB}}^2}\right].
\end{aligned}
\end{equation}

In the above equation, the starburst fractions are fixed and do not change with redshift and mass, and the parameters A$_{\rm MS}=0.97$ and A$_{\rm SB}=0.03$ represent the fraction of MS and starburst galaxies, respectively. 
$\langle  \log\psi \rangle_{\rm MS}$ represents the first Gaussian's central value and refers to the MS and $\langle  \log\psi \rangle_{\rm SB} = \langle  \log\psi \rangle_{\rm MS} + 0.59$ is the central value of the second Gaussian referring to SB galaxies. Values are taken from \cite{Sargent2012}.
The one-sigma dispersion of the first and second Gaussian are also fixed to the values found by \cite{Sargent2012}, which are $\sigma_{\rm MS}=0.188$ and $\sigma_{\rm SB}=0.243$ respectively. By convolving Equation \ref{eq:double_gaussian_SFR} with the SMFs one obtains galaxies' SFR-functions as:

\begin{equation}\label{eq:sfrf}
\begin{aligned}
\frac{d^2 N_{\mathrm{SMF}+\mathrm{MS}}}{d \log \psi d V}(z, \log \psi)= & \int d \log M_{\star} \frac{d^2 N}{d \log M_{\star} d V}\left(z, \log M_{\star}\right) \\
& \times \frac{d p}{d \log \psi}\left(\log \psi \mid z, M_{\star}\right).
\end{aligned}
\end{equation}

\noindent

\subsection{Far-IR/radio correlation}\label{sec:FIRRC}

We now need to express the SFR as a function of galaxies' radio luminosity.
For this purpose, we adopt the FIRRC, a tight relation linking the monochromatic non-thermal radio emission and the FIR emission of SFGs (see e.g. \citealt{Helou1985}; \citealt{Condon1992}; \citealt{Yun2001}).

The FIRRC is described via the parameter $q_{\rm FIR}$ (\citealt{Bell2003}; \citealt{Ivison2010a,Ivison2010b}; \citealt{Sargent2010}), defined as:

\begin{equation}\label{eq:firrc}
q_{\mathrm{FIR}}=\log \left(\frac{L_{\mathrm{FIR}}[\mathrm{W}] / 3.75 \times 10^{12}}{L_{1.4 \mathrm{GHz}}\left[\mathrm{W} \mathrm{Hz}^{-1}\right]}\right),
\end{equation}

where $L_{\rm 1.4 GHz}$ is the rest-frame radio luminosity and $L_{\rm FIR}$ is the rest-frame FIR luminosity defined in the range 8-1000 $\mu$m.

Because of the low scatter of this relation (1$\sigma\approx 0.26$ dex), radio emission can be used as an unbiased dust-independent tracer of star formation in galaxies and as an unbiased probe of the cosmic star formation history. This is particularly promising in the context of forthcoming large-area and deep radio surveys conducted by SKA and its precursors, reaching unprecedented sensitivity and hence able to trace SFGs to very high redshift (\citealt{Jarvis2015aska.confE..68J}; \citealt{Mancuso2015}; \citealt{Schleicher2016}; \citealt{An2021}; \citealt{Cochrane2023}; \citealt{Ocran2023MNRAS.524.5229O}).

Although the FIRRC has been well established at low redshift (\citealt{Condon1992}; \citealt{Yun2001}; \citealt{Bell2003}; \citealt{Jarvis2010}; \citealt{Wang2019}; \citealt{Molnar2021}), its redshift evolution, if present, is still debated. 
On the one hand, some studies point towards a redshift evolution, expressed as $\overline{q}_{\rm FIR}\propto (1+z)^{-\beta}$ ($0.12\lesssim\beta\lesssim0.2$, \citealt{Basu2015}; \citealt{Magnelli2015}; \citealt{Tabatabaei2016}; \citealt{CalistroRivera2017}; \citealt{Delhaize2017}; \citealt{Ocran2020b}; \citealt{Sinha2022}). On the other hand, others report no significant redshift evolution (see e.g. \citealt{Sargent2010}), suggesting that the observed decreasing trend may be originated by selection effects (\citealt{Bourne2011}; \citealt{Molnar2021}). These effects have been extensively studied and can be attributed to either the different relative sensitivities of radio and FIR surveys (\citealt{Sargent2010}; \citealt{Bourne2011}; \citealt{Molnar2021}), leading to flux-limited samples that are biased towards more massive galaxies, or to different galaxy populations dominating at different redshifts (\citealt{Dezotti2024}). Recently, several studies investigated the mass-dependency of the FIRRC, reporting a decrease in $\overline{q}_{\rm FIR}$ with increasing stellar mass (\citealt{Gurkan2018}; \citealt{Delvecchio2021}; \citealt{Smith2021}; \citealt{McCheyne2022}).

In this work, we use the recent relation, including both the redshift and stellar mass dependence, presented in \cite{Delvecchio2021} and \cite{McCheyne2022} to retrieve the LF at 1.4 GHz and 150 MHz.

\cite{Delvecchio2021} retrieved the redshift and mass-dependent FIRRC at 1.4 GHz for a sample of $>\num{400000}$ SFGs in the COSMOS field. SFGs were identified in the redshift and mass ranges $0.1<z<4.0$ and $10^{8}<M_{\star}/\rm M_{\odot}<10^{12}$ through color-selection $[(NUV-r)/(r-J)]$ criteria. 
The IR information comes from the de-blended data from \cite{Jin2018}, comprising Spitzer/MIPS 24 $\mu$m data (PI: D. Sanders, \citealt{LeFloch2009}) Herschel/PACS 100 and 160 $\mu$m data from the PEP (PI: D. Lutz; \citealt{Lutz2011}) and SPIRE 250, 350 and 500 $\mu$m of the Herschel Multi-tiered Extragalactic Survey (HerMES, PI: S. Oliver, \citealt{Oliver2012}). Radio images come from the VLA COSMOS 3 GHz survey (\citealt{Smolcic2017}). Furthermore, a median stacking procedure was performed for non-detections in different $M_{\star}-z$ bins to infer the average flux densities in each band.
The relation obtained by \cite{Delvecchio2021} derived at 1.4 GHz is the following: 

\begin{equation}\label{eq:firrc_delvecchio_nouv}
\begin{aligned}
   q_{\rm FIR}(M_{\star},z)=(2.646\pm0.024)(1+z)^{(-0.023\pm0.012)} \\  -(\log({M_{\star}/M_{\odot})}-10)(0.148\pm0.013). 
   \end{aligned}
\end{equation}

However, it should be pointed out that, due to the relatively small area of the COSMOS survey ($2\, \rm deg^2$), the $q_{\rm IR}$ estimates at high stellar masses suffer from significant uncertainty at low-redshift ($z<0.4$), resulting in a higher normalisation compared to higher-redshifts ($z > 0.4$).
It should also be noticed that the L$_{\rm FIR}-\rm SFR$ conversion may not be fully applicable for low-mass/metallicity and less obscured systems (\citealt{Mannucci2010MNRAS.408.2115M}; \citealt{Pannella2015ApJ...807..141P}; \citealt{Whitaker2017ApJ...850..208W}), where the contribution of UV emission can be comparable to that of FIR emission (\citealt{Buat2012A&A...545A.141B}; \citealt{Cucciati2012A&A...539A..31C}; \citealt{Burgarella2013A&A...554A..70B}).
For this reason, \cite{Delvecchio2021} introduced a different formalism, which takes into account the contribution of dust-uncorrected UV emission through the parameter  $q_{\rm SFR_{\rm UV+FIR}}$. The redshift- and mass-dependent relation is given as:

\begin{equation}\label{eq:firrc_delvecchio}
\begin{aligned}
q_{\rm UV+ FIR}(M_{\star},z)=(2.743\pm0.034)(1+z)^{(-0.025\pm0.012)} \\ -(\log({M_{\star}/M_{\odot})}-10)(0.234\pm0.017).
   \end{aligned}
\end{equation}

In this paper, we adopt this latter relation as it better accounts for galaxies' total SFR by also including UV emission. When referring to the FIRRC of \cite{Delvecchio2021}, we specifically mean the one defined in Eq. \ref{eq:firrc_delvecchio}, even though it includes the contribution of the UV luminosity.

The FIRRC at low frequencies (150 MHz) has been less explored. Nevertheless, the mass-dependence and redshift evolution was investigated by \cite{CalistroRivera2017}, \cite{Read2018MNRAS.480.5625R} and \cite{Gurkan2018} and recently extended up to $z\sim 1$ by \cite{Smith2021} and \cite{McCheyne2022}.
\cite{McCheyne2022} derived the FIRRC in the context of the LoTSS (\citealt{Sabater2021}, see also Sect. \ref{sec:LF_150MHz}) utilising LOFAR 150 MHz observations of a mass-complete sample in the ELAIS-N1 field (\citealt{Kondapally2021}; \citealt{Duncan2021}), combined with deblended \textit{Herschel} data (\citealt{Vaccari2015salt.confE..17V}; \citealt{Hurley2017MNRAS.464..885H}). Their FIRRC is derived at 150 MHz and it is expressed as:

\begin{equation}\label{eq:firrc_mccheyne}
    \begin{aligned}
q_{\rm FIR}(z,M_{\star}) = (1.98 \pm 0.02)(1+z)^{0.02\pm0.04}\\ + (-0.22\pm 0.03)\log{(M_{\star})}-10.45.
   \end{aligned}
\end{equation}

The LOFAR observations exploited by \cite{McCheyne2022} have similar depth to the VLA COSMOS 3 GHz survey, but cover an area three times larger. This allowed the authors to probe the rare high FIR and radio luminosity populations, likely absent in smaller fields. Two samples were considered, one comprising sources with $z<1.0$ and $M_{\star}>10^{10.45}\, \rm M_{\odot}$, the other including objects at $z<0.4$ and $M_{\star}>10^{10.05}\, \rm M_{\odot}$. In both cases, the probed stellar masses reach up to a value of $M_{\star}\sim 10^{11.4}\,$M$_{\odot}$. With respect to \cite{Delvecchio2021}, \cite{McCheyne2022} FIRRC spans a lower redshift range ($z<1$) and is found to have a different normalisation, as well as a steeper dependence on $M_{\star}$. The origin of this discrepancy is still unclear.  The authors argue that it may result from either different ratios of thermal to synchrotron emission at the respective frequencies or the choice of the spectral index used to convert fluxes from 1.4 GHz and 150 MHz and vice-versa. Indeed, \cite{McCheyne2022} have shown that the difference between the two relations becomes negligible when adopting the value $\alpha = - 0.59$, which is slightly shallower than the typical value assumed for SFGs ($\alpha = -0.7$; see e.g. \citealt{Novak2017}).

\subsection{Inferring the radio luminosity function}
We express Equation \ref{eq:sfrf} in terms of $L_{\nu}$ by exploiting the mass- and redshift-dependent FIRRC. 
For this purpose, we convolve the resulting expression with a Gaussian distribution representing the probability of a given $L_{\nu}$ at fixed $\psi$, $M_{\star}$ and $z$:

\begin{equation}\label{eq:gaussianprob}
    \begin{aligned}
    \frac{d \rm p}{d \log{L_{\nu}}} \left(\log L_{\nu} \mid \psi, M_{\star},z \right) =  
       \\ \left(\frac{1}{\sqrt{2\pi\sigma^2_{\rm FIRRC}}} \right) \exp \left[ - \frac{\left( \log L_{\nu} - \langle \log L_{\nu} \rangle \right)^2}{2 \sigma^2_{\rm FIRRC}} \right].
    \end{aligned}
\end{equation}

\noindent
The term $\sigma_{\rm FIRRC}$ accounts for the scatter of the FIRRC. $\langle \log L_{\nu} \rangle$ is the radio luminosity at a frequency $\nu$ corresponding to a given $\psi$, obtained from the mean FIRRC adopting $L_{\rm FIR} = k_{\rm FIR} \psi$ \citep{Kennicutt2012AR} with $k_{\rm FIR}$ being a calibration constant rescaled for a Chabrier IMF.

The final expression for the LF of SFGs is:

\begin{equation}\label{eq:final}
    \begin{aligned}
     \frac{d^2\rm N}{d\log L_{\nu}dV}\left(\log L_{\nu}, z\right) = \\  \int  d\log M_{\star} \frac{d^2N}{d\log M_{\star}dV} \left(\log M_{\star} \mid z \right) \\ \times \int d\log \psi \frac{dp}{d\log \psi} \left(\psi \mid z, M_{\star} \right)  \\ \times \frac{d \rm p}{d \log{L_{\nu}}} \left(\log L_{\nu} \mid \psi, M_{\star},z \right).
    \end{aligned}
\end{equation}

It is important to note that, for local galaxies, the correlation between radio and FIR luminosities deviates from linearity at low radio luminosities (\citealt{Yun2001}; \citealt{Best2023}).
This nonlinearity, which translates into a systematically lower ratio between radio and FIR luminosities (see e.g. \citealt{Yun2001}), has been attributed to cosmic-ray losses suppressing the synchrotron radiation in low-mass galaxies
(\citealt{Klein1984A&A...141..241K}; \citealt{Chi1990MNRAS.245..101C}; \citealt{Price1992ApJ...401...81P}), even though other authors (e.g. \citealt{Helou1986ApJ...311L..33H}; \citealt{Lonsdale1987ApJ...314..513L}; \citealt{Fitt1988MNRAS.233..907F}) argued that this effect may also be originated by the cirrus emission produced by low-mass stars, which provide an extra contribution to the FIR emission (e.g. \citealt{Helou1986ApJ...311L..33H}; \citealt{Lonsdale1987ApJ...314..513L}; \citealt{Fitt1988MNRAS.233..907F}). 
To reproduce the local LF at low radio powers ($L_{\rm 1.4 GHz} \lesssim 10^{28.5}$ erg s$^{-1}$Hz$^{-1}$), we follow the approach adopted by \cite{Massardi2010} and \cite{Mancuso2015, Mancuso2017} and correct the radio luminosity accounting for low mass, low SFR galaxies ($\psi \lesssim$ a few $M_{\odot}\,$yr$^{-1}$) that are less efficient in producing synchrotron emission (\citealt{Bell2003}):

\begin{equation}\label{eq:correction_radio}
L_{\mathrm{synch}, \mathrm{corr}}=\frac{L_{\mathrm{synch}}}{1+\left(L_{0, \mathrm{synch}} / L_{\mathrm{synch}}\right)^\zeta},
\end{equation}

\noindent
where $\zeta = 2$ and $L_{0, \rm synch} = 3 \times 10^{28}\,$erg$\, \rm s^{-1} \, \rm Hz^{-1}$ at 1.4 GHz. We refer to \cite{Mancuso2017} for a detailed discussion of these assumptions. We anticipate that the above equation is applied both at $\nu=1.4\,$GHz and $\nu=150\,$MHz and only for the lowest redshift bins of the LF ($z\lesssim0.4$). Our choice is motivated by the fact that this effect has been primarily investigated in local galaxies and observations at higher redshifts do not cover the luminosity range where the correction is necessary. As a result, we lack clear evidence of suppression at low luminosities.

\subsection{Number counts}

The differential radio-band number counts are computed as the integral over the redshift of Equation \ref{eq:final}:

\begin{equation}\label{eq:number_counts}
\frac{d N}{d \log S_\nu d \Omega}\left(S_\nu\right)=\int d z \frac{d V}{d z d \Omega} \frac{d N}{d \log L_\nu d V}\left(L_{\nu(1+z)}, z\right),
\end{equation}

\noindent
where $dV/dz d\Omega$ is the cosmological volume per unit solid angle and the observed flux is:

\begin{equation}
S_\nu=\frac{L_{\nu(1+z)}(1+z)}{4 \pi D_L^2(z)},
\end{equation}

\noindent
with $D_L(z)$ being the luminosity distance.

\section{Comparison samples}\label{sec: comparison samples}

This section describes the comparison samples we exploit for validating our model. In the following, we focus on  LFs and number counts derived at two reference frequencies: 150 MHz and 1.4 GHz. Hence we mostly consider samples observed at these frequencies, with only a few exceptions. Samples with different observing frequencies are rescaled to 1.4 GHz or 150 MHz, assuming a spectral index $\alpha = -0.7$, unless stated otherwise. Our model is also compared with other semi-empirical predictions, which are also presented here.

For the local 1.4 GHz LF we consider  \cite{Mauch2007} sample composed by 4625 local SFGs with S$_{\rm 1.4\, GHz}>2.8$ mJy selected from the NRAO VLA Sky Survey (NVSS, \citealt{Condon1998}) and the 6 degree Field Galaxy Survey (6dFGS, \citealt{Jones2004}), covering a total area of 7076 deg$^2$.  We also make use of the work of \cite{Condon2019}, who exploited the same survey in combination with the Two Micron All Sky Survey (2MASS; \citealt{Skrutskie2006}) Extended Source Catalog (2MASSX; \citealt{Jarrett2000}), deriving a complete spectroscopic sample of 6699 SFGs, reaching a sensitivity of $\approx 0.45$ mJy beam$^{-1}$.
Additionally, we consider the sample of \cite{Padovani2015, Padovani2016}, who presented observations of the local LFs for SFGs selected in the Extended Chandra Deep Field-South (E-CDFS) through JVLA observations reaching a sensitivity of $\sim 32.5 \, \mu$Jy beam$^{-1}$.
Finally, we include the sample of 558 SFGs of \cite{Butler2019A&A...625A.111B}, including also RQ AGNs, derived from the XMM extragalactic survey south (XXL-S) covered by ATCA 2.1 GHz data, reaching a median 1$\sigma$ rms of $\approx 41\, \mu$Jy beam$^{-1}$.

For redshift-dependent ($z\gtrsim 0.1$) 1.4 GHz LFs, we exploit data from \cite{Novak2017}, which were derived from the deep VLA-COSMOS 3 GHz Large Project survey (\citealt{Smolcic2017}) conducted in the COSMOS field. The observations refer to a sample of $\sim 6000$ SFGs with reliable optical counterparts up to $z\sim5.7$. 
Deeper ($\sim \times 5$ with respect to the VLA-COSMOS 3 GHz Large Project) coverage was reached in the COSMOS-XS survey (\citealt{Algera2020ApJ...903..139A}; \citealt{vanderVlugt2021ApJ...907....5V}). From these observations, \cite{vanderVlugt2022ApJ...941...10V} identified $\approx 1300$ SFGs and derived the radio LF at 1.4 GHz up to $z\sim 4.6$. 
While the determinations from \cite{Novak2017} trace mostly the bright end of the LFs, the increased sensitivity reached in the COSMOS-XS survey allowed \cite{vanderVlugt2022ApJ...941...10V} to constrain the low-luminosity regime of the LF. From the sample of \cite{vanderVlugt2022ApJ...941...10V}, \cite{vanderVlugt2023ApJ...951..131V} added  $\approx 20$ optically dark sources (see Sec \ref{sec:dark_objects} for more details on dark galaxies), selected to be undetected up to a magnitude of $K_s = 25.9$.
We finally exploit the sample of \cite{Enia2022ApJ...927..204E}, consisting of $479$ galaxies up to z$\approx 4.5$, selected from deep VLA 1.4 GHz observations in the GOODS-North field. Fifteen galaxies in this sample are H-dark, i.e., they lack a counterpart in the HST/WFC3 \textit{H}-map down to a 5$\sigma$ detection limit up to 28.7. Only eight had enough photometric coverage to derive redshifts via spectral energy distribution fitting and were included in the LFs.

\noindent
For the 1.4 GHz number counts, we compare with those derived by \cite{Mauch2007} for local SFGs (see above), which provide good constraints at high flux densities. At the same time, for the sub-mJy regime we exploit the counts derived for both SFGs and RQ AGN by \cite{Bonato2021a}, and based on 1173 sources with $S_{\rm 1.4 GHz}> 120 \, \mu$Jy studied in the context of the Lockman Hole (LH, \citealt{Lockman1986}) project. The LH Project (\citealt{Prandoni2018}) consists of deep 1.4 GHz conducted with the WSRT, over an area of 6.6 deg$^2$, reaching a uniform rms noise of $11\, \mu$Jy beam$^{-1}$. The derived radio source statistics robustly span the flux range $0.1<\rm S<1$ mJy. \cite{Bonato2021a} re-analysed the radio sources in the central 1.4 deg$^2$ of the LH field by exploiting optical-to-FIR multi-band observations.

At 150 MHz, we mostly rely on the recent results from the LoTSS Survey (\citealt{Shimwell2017, Shimwell2019, Shimwell2022}).
In particular, we focus on the first data release (DR1) of the wide and deep fields of the LoTSS survey. The LoTSS wide DR1 covers a region of 424 deg$^2$ in the Northern sky with a median rms sensitivity of 71 $\mu$Jy beam$^{-1}$ at an angular resolution of 6 arcsec at 145 MHz (\citealt{Shimwell2019}). By cross-matching the DR1 radio source catalogue with the Sloan Digital Sky Survey (SDSS) DR7 main galaxy spectroscopic sample, \cite{Sabater2019A&A...622A..17S} derived local LFs for both AGN and SFGs.
The LoTSS deep DR1 covers three well-known regions of the sky: the ELAIS-N1 (\citealt{Oliver2000}), Bo\"{o}tes (\citealt{Jannuzi1999}) and the LH.
The LoTSS deep fields reach an rms sensitivity of $\sim 20\, \mu$Jy beam$^{-1}$ at 150 MHz, which is comparable to the depth of the VLA-COSMOS 3 GHz survey (assuming a radio spectral index $\alpha = -0.7$).
Among the $\approx 80000$ radio sources detected in the central regions of the fields (covering a total of $25\,$deg$^2$), $\sim 97\%$ were identified through a cross-matching procedure with optical and NIR data (\citealt{Kondapally2021}). The extensive multi-band coverage allowed for the determination of high-quality photometric redshifts and stellar masses (\citealt{Duncan2021}) along with the host galaxies' classification (\citealt{Best2023}). \cite{Bonato2021b} presented 150 MHz SFR and luminosity functions for the Lockman Hole field, following the classification later presented in \cite{Best2023}. More recently, also relying on the classification on \cite{Best2023}, \cite{Cochrane2023} presented 150 MHz LFs for sources with no radio excess (SFGs and RQ AGNs), extending the analysis to all three LOFAR deep fields. Luminosity functions were derived for the local Universe and in the redshift range $0.1<z<5.7$.

LoTSS LFs are complemented by the LFs presented by \cite{Ocran2020a,Ocran2020b} and based on low-frequency observations performed with the GMRT at 610 MHz, reaching a 1$\sigma$ rms of $\sim 7.1\, \mu$Jy beam$^{-1}$. Their sample consists of 4290 sources, 1685 of which are SFGs, selected in the ELAIS N1 field and spanning the redshift range $0\lesssim z<1.5$.

Finally, we exploited the DR1 catalogues of the LoTSS deep fields to derive the differential 150 MHz number counts for SFGs and RQ-AGN populations (not available in the literature).  We adopted the \citet{Best2023} radio source classification and the \cite{Cochrane2023} completeness corrections. The results are presented in Tab. \ref{tab:logcounts}.

As mentioned above, we include recent radio simulations and models in our comparison. Specifically, we use the radio number counts predicted by \cite{Mancuso2017}, based on the semi-empirical model described in \cite{Mancuso2016a, Mancuso2016b}. This model reconstructs the SFR functions and their evolution over time using observed UV/FIR data. Also, we derive simulated LFs and number counts from the T-RECS radio continuum simulation (\citealt{Bonaldi2019MNRAS.482....2B, Bonaldi2023MNRAS.524..993B}), which covers frequencies from 150 MHz to 20 GHz. T-RECS is designed to model both AGNs and SFGs, with the latter also including RQ AGNs.

\begin{table}
\caption{Estimates of the 150 MHz Euclidean normalised differential counts, $S^{2.5}_{\rm 150 MHz} dN/dS\,[\rm Jy^{1.5} sr^{-1}]$, for the SFG and RQ AGN populations obtained from the DR1 catalogues of the LoTSS deep fields.}
\centering
\renewcommand{\arraystretch}{1.3}
\begin{tabular}{ccc}
\hline
$\log S$ & $\log \rm Counts_{SFGs}$  & $\log \rm Counts_{RQ}$  \\
\(\left[\text{mJy}\right]\) & \([\text{Jy}^{1.5}\, \text{sr}^{-1}]\) & \([\text{Jy}^{1.5}\, \text{sr}^{-1}]\) \\
\hline
$-0.90$ & $1.33^{+0.003}_{-0.003}$ & $0.26^{+0.01}_{-0.01}$ \\
$-0.60$ & $1.47^{+0.003}_{-0.003}$ & $0.49^{+0.01}_{-0.01}$ \\
$-0.30$ & $1.54^{+0.003}_{-0.003}$ & $0.70^{+0.01}_{-0.01}$ \\
$0.004$  & $1.39^{+0.007}_{-0.007}$ & $0.66^{+0.02}_{-0.02}$ \\
$0.30$  & $1.16^{+0.02}_{-0.01}$   & $0.43^{+0.04}_{-0.03}$ \\
$0.60$  & $0.87^{+0.04}_{-0.03}$   & $0.34^{+0.07}_{-0.06}$ \\
$0.90$  & $0.84^{+0.07}_{-0.06}$   & $0.27^{+0.15}_{-0.11}$ \\
$1.20$  & $0.42^{+0.23}_{-0.16}$   & $0.27^{+0.29}_{-0.19}$ \\
$1.50$  & $0.50^{+0.42}_{-0.23}$   & $0.02^{+0.98}_{-0.36}$ \\
\hline
\end{tabular}
\label{tab:logcounts}
\end{table}

\section{Results and discussion} \label{sec: results and discussion}

In this section, we present the results of the comparison between our model and the comparison samples/predictions described in the previous section.

\subsection{Luminosity functions at 1.4 GHz}\label{sec:LF_1.4GHz}

\begin{figure}
    \centering
    \includegraphics[width=0.45\textwidth]{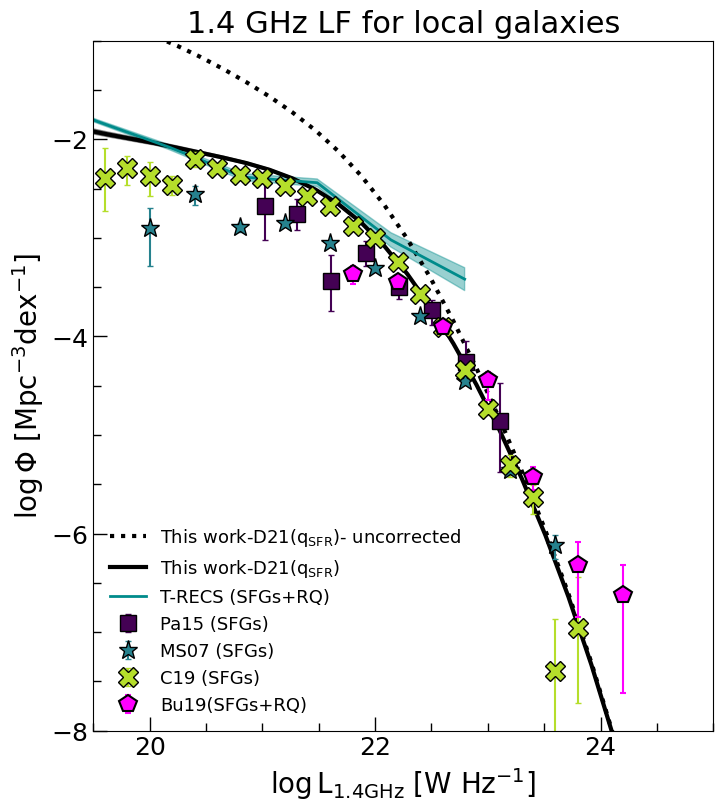}
    \caption{1.4 GHz LF for SFGs at $z\sim 0$ from \texttt{SEMPER} obtained by using \cite{Delvecchio2021} q$_{\rm SFR_{FIR+UV}}$ relation with (black solid line) and without (dotted line) the correction defined in Equation \ref{eq:correction_radio}. Data for SFGs are from \cite{Mauch2007}, \cite{Padovani2015}, \cite{Condon2019} and \cite{Butler2019A&A...625A.111B}. The cyan line indicates the LF from the T-RECS simulation with the shaded area representing its 1$\sigma$ uncertainty.}
    \label{fig:local_LF_1.4}
\end{figure}

Fig. \ref{fig:local_LF_1.4} shows the modelled local (z$<0.05$) LF for SFGs at $\nu = 1.4 \, \rm GHz$ (solid black line) obtained using Eq. \ref{eq:firrc_delvecchio} by \cite{Delvecchio2021}. We compare our results with the T-RECS simulation and with local LF estimates by \citet[][light blue stars]{Mauch2007}, \citet[][purple squares]{Padovani2015}, \citet[][magenta pentagons]{Butler2019A&A...625A.111B} and \citet[][green crosses]{Condon2019}, finding a very good match, especially with the latter. For comparison, we also show the LF obtained without applying the correction defined in Equation \ref{eq:correction_radio} (black dotted line). 
According to \cite{Delvecchio2021}, using q$_{\rm SFR_{UV+FIR}}$ instead of q$_{\rm FIR}$ should account for the lower efficiency of synchrotron emission in low-SFR, low-mass galaxies. However, Fig. \ref{fig:local_LF_1.4} shows that Eq. \ref{eq:correction_radio} is still necessary to reproduce the observed flattening of the faint end of the LF, even when adopting q$_{\rm SFR_{UV+FIR}}$ (Eq. \ref{eq:firrc_delvecchio}) instead of q$_{\rm FIR}$ (Eq. \ref{eq:firrc_delvecchio_nouv}).
In other words, the mass dependence introduced by \cite{Delvecchio2021} alone does not fully explain the observed flattening of the local LF. This discrepancy may arise because the relation from \cite{Delvecchio2021} was derived for a higher redshift range ($0.1<z<4$), whereas the local radio LF corresponds to $z<0.05$. As a result, the mass-dependency derived from the extrapolation to $z<0.1$ may not be accurate at such low redshifts.

\begin{figure*}
    \centering
    \includegraphics[width=\textwidth]{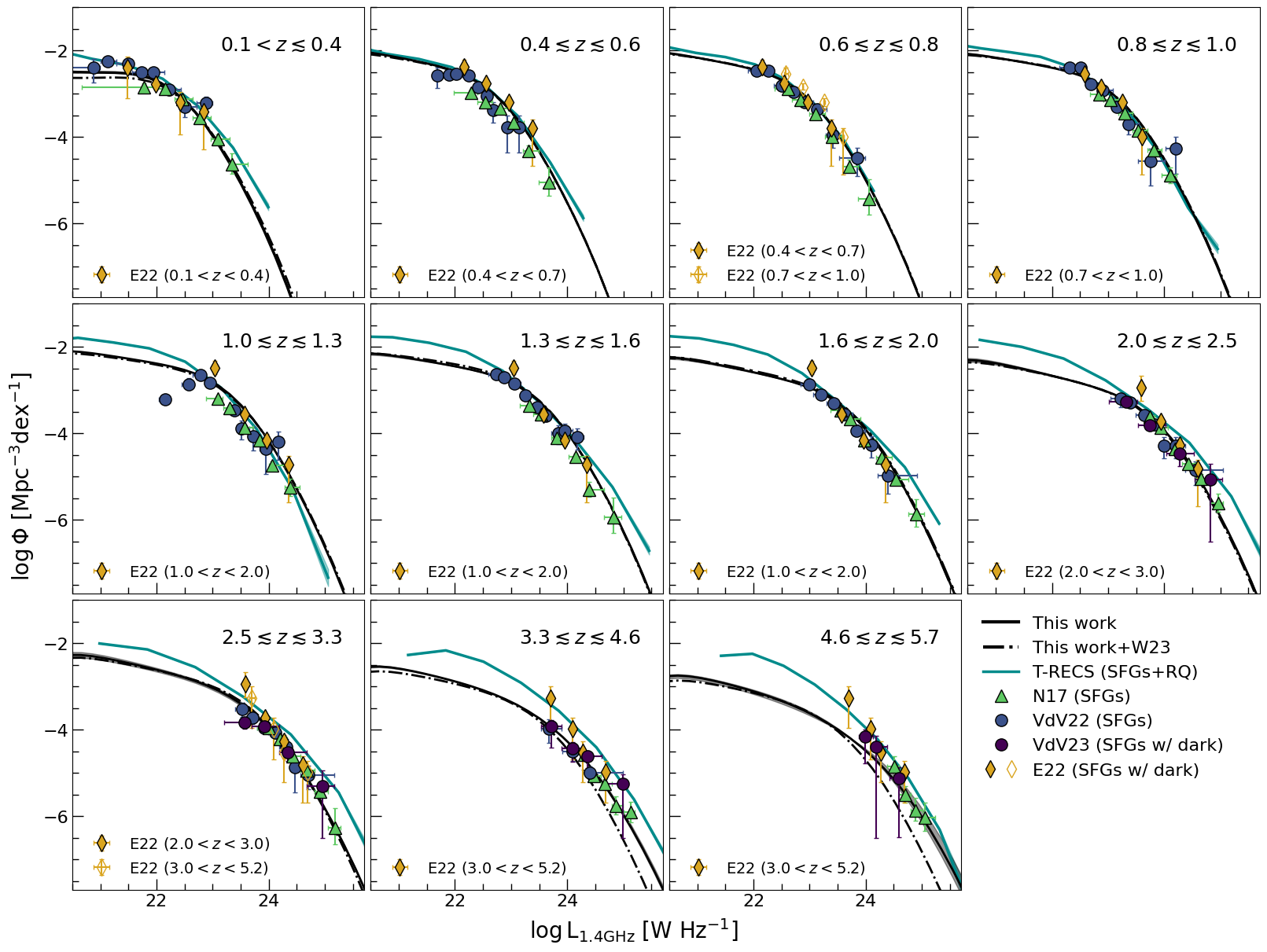}
    \caption{1.4 GHz LF for SFGs for different redshift bins (black solid lines) predicted by \texttt{SEMPER} adopting the \cite{Delvecchio2021} relation. Dot-dashed lines show how our model would change when considering the SMFs derived by \citetalias{Weaver2023} with a Double Schechter form. Eq. \ref{eq:correction_radio} is applied only for the first redshift bin. Data are from \cite{Novak2017}, \cite{Enia2022ApJ...927..204E}, \cite{vanderVlugt2022ApJ...941...10V} and \cite{vanderVlugt2023ApJ...951..131V}. \cite{Enia2022ApJ...927..204E} and \cite{vanderVlugt2023ApJ...951..131V} data include the contribution of "dark" sources. The cyan lines and shaded areas are the predictions from the T-RECS simulation by \cite{Bonaldi2019MNRAS.482....2B, Bonaldi2023MNRAS.524..993B} and the respective 1$\sigma$ uncertainties. Note that data from \cite{Enia2022ApJ...927..204E} spans redshift bins not matching the ones adopted by our model and shown in each panel's legend.}
    \label{fig:LFs_1.4}
\end{figure*}

Fig. \ref{fig:LFs_1.4} shows the modelled 1.4 GHz radio LF for different redshift bins over the range $0.1<z\lesssim5.7$ (black solid lines). 
We compare our model with the determinations from \citet[][green triangles]{Novak2017}, \citet[][gold diamonds]{Enia2022ApJ...927..204E}, \citet[][blue circles]{vanderVlugt2022ApJ...941...10V}, and \citet[][purple circles]{vanderVlugt2023ApJ...951..131V}.
Our model successfully reproduces the observations across the full redshift range ($0.1<z<5.7$). We also compare our predictions with T-RECS ones. We note that the T-RECS LFs show significant offsets with respect to our LFs (up to $\sim 0.5$ dex). Such offsets become increasingly prominent going to higher redshifts. This discrepancy may be due, at least in part, to the fact that the SFG population includes RQ AGNs in the T-RECS simulation (see also Sect. \ref{sec:LF_150MHz}). Generally, our model better reproduces the observations, except for the highest redshift bin ($4.6<z<5.7$). 
Finally, we show the LFs we would obtain by implementing the SMFs from \citetalias{Weaver2023}, which are based on a Double Schechter function fit (black dot-dashed lines). We notice a discrepancy with observations at the highest redshift bins that will be discussed in more detail in Sect. \ref{sec:dark_objects}.

\subsection{Massive galaxies}\label{sec:dark_objects}

As previously discussed, \citetalias{Weaver2023} found a large number of massive, red SFGs,  which are typically  located at
high-redshift (z>3) and are heavily dust-obscured (AV > 3). 
Such sources challenge galaxy formation and evolution simulations, which predict 1.8$\times$ less numerous sources at $2.5<z\leq 5.5$ over the same mass range (\citetalias{Weaver2023}).
Because of the deeper NIR UltraVISTA images, COSMOS2020 is more sensitive to faint red sources with respect to previous surveys conducted at optical/NIR wavelengths, which likely missed most of these objects (e.g. \citealt{Simpson2014}; \citealt{Franco2018}; \citealt{Wang2019}; \citealt{Barrufet2023}; \citealt{Williams2024}). 
A fraction of them were, however, observed at longer wavelengths. Such galaxies are usually referred to as "optically dark" (\citealt{Simpson2014}; \citealt{Wang2019}; \citealt{Gruppioni2020}; \citealt{Sun2021}; \citealt{Fudamoto2021}; \citealt{Smail2021}; \citealt{Talia2021}; \citealt{Enia2022ApJ...927..204E}; \citealt{Shu2022}; \citealt{Behiri2023}; \citealt{vanderVlugt2023ApJ...951..131V}; \citealt{Gentile2024a}; \citealt{Williams2024}), although various terminologies exist depending on their selection method (e.g. HST-dark, Rs-NIR dark, H-dropout, NIR-dark/faint, OIR-dark). In recent years, several observational studies (e.g. \citealt{Wang2019}; \citealt{Gruppioni2020}) have highlighted that this population of "dark" galaxies can significantly impact the cosmic SFR density, with a contribution up to $\sim 40 \%$ the contribution of high-redshift Lyman-Break Galaxies (\citealt{Talia2021}; \citealt{Enia2022ApJ...927..204E}; \citealt{Behiri2023}; \citealt{vanderVlugt2023ApJ...951..131V}; \citealt{Gentile2024a}).

\begin{figure*}
    \centering
    \includegraphics[width=0.7\linewidth]{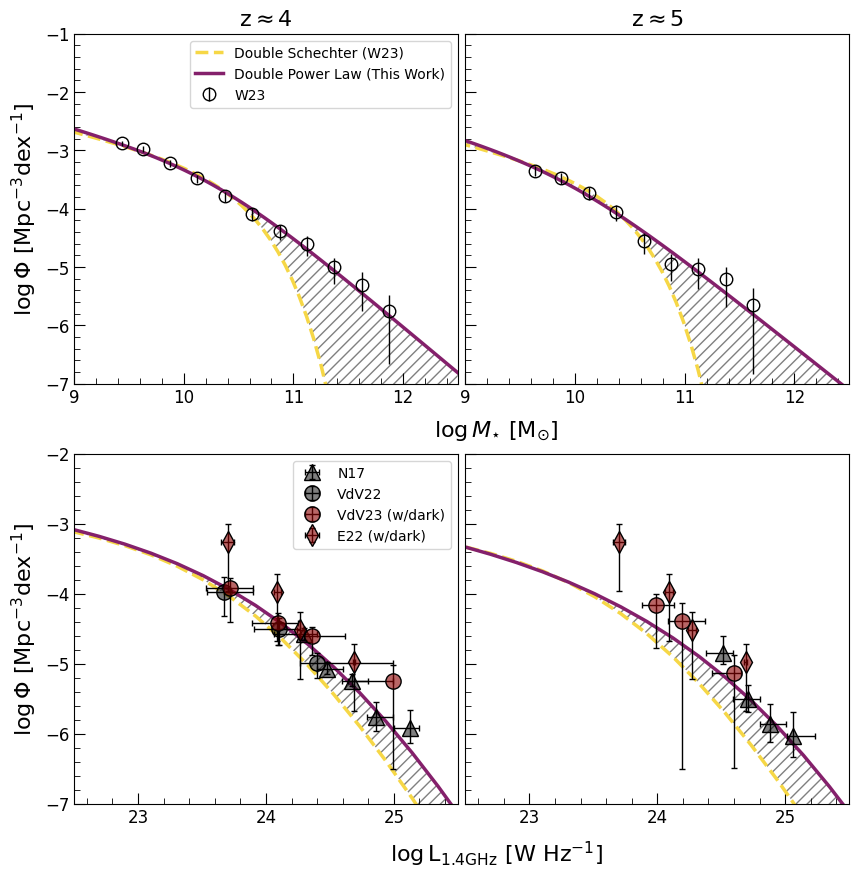}
    \caption{This figure summarises the results presented in Sect. \ref{sec:dark_objects}. The two top panels show the SMFs for the highest redshift bins discussed in this paper. The bottom panels show the LF at 1.4 GHz derived from our model in the same redshift bins. The yellow dashed and purple solid lines respectively indicate the predictions obtained when adopting a Double Schechter Function (\citetalias{Weaver2023}) and a Double Power Law (this paper) for the SMFs fitting. The hatched grey area highlights the difference between the two curves. Data points are the same as in Fig. \ref{fig:fit_SMF_Weaver_vs_dpl} for the SMFs and Figs \ref{fig:LFs_1.4} for the LFs, highlighting in red samples which include "dark" galaxies.}
    \label{fig:smf_vs_lf}
\end{figure*}

Our model, derived from SMFs based on ultra-deep NIR observations, is sensitive to this red high-mass galaxy population and likely includes at least a fraction of "dark" (and/or extremely NIR faint) objects. 
Therefore, it is interesting to explore the link between massive and NIR-faint galaxies, accounted for in our model, and radio-selected SFG samples, which should be unbiased against optical colour. 

In Fig. \ref{fig:smf_vs_lf}, we directly compare the SMFs presented in Appendix \ref{app:a_1} and our 1.4 GHz LF predictions for the two latter redshift bins of Fig. \ref{fig:LFs_1.4}. We note that when the double power law fit is used in the derivation of the LF, the model can reproduce the observations of high redshift (z$\gtrsim 3.3$) radio-selected samples  (\citealt{Novak2017}; \citealt{Enia2022ApJ...927..204E}; \citealt{vanderVlugt2022ApJ...941...10V,vanderVlugt2023ApJ...951..131V}). 
In contrast, the observed 1.4 GHz LFs are not reproduced when using a Double Schechter fit to the SMFs, which does not account for galaxies with $M_{\star} > 10^{11}\,$M$_{\odot}$. The observed discrepancy is more prominent when considering radio samples, including dark galaxies (\citealt{Enia2022ApJ...927..204E}; \citealt{vanderVlugt2023ApJ...951..131V}; red points). Remarkably, this comparison suggests that the high-z massive and red objects detected in excess in the NIR (\citetalias{Weaver2023}) likely correspond to a population of bright radio sources at similar redshifts. This result is in line with observational studies (e.g. \citealt{Talia2021}; \citealt{Enia2022ApJ...927..204E}; \citealt{Behiri2023}; \citealt{Gentile2024a}) showing that radio band observations are sensitive to massive and obscured galaxies. This evidence further validates our choice to adopt a Double Power Law profile for the SMFs' reconstruction.

Thanks to its unprecedented sensitivity at near and mid-infrared bands, JWST is finding significant numbers of massive, dust-obscured SFGs already in place at high redshifts (\citealt{Barrufet2023}; \citealt{Labbe2023Natur.616..266L}; \citealt{Nelson2023ApJ...948L..18N}; \citealt{Perez2023ApJ...946L..16P}).
Future deeper JWST observations (e.g. in the context of the COSMOS-Web survey, see \citealt{Casey2023}) will better characterise this population by reducing the existing uncertainties on their photometric redshifts and stellar masses (e.g. \citealt{Barrufet2023}). These observations will provide more stringent constraints that we will use to refine our model, possibly improving further the matching with the observed radio LFs at $z>4$.

\subsection{Number counts at 1.4 GHz}\label{sec:numbercounts_1.4GHz}

\begin{figure}
    \centering
    \includegraphics[width=0.49\textwidth]{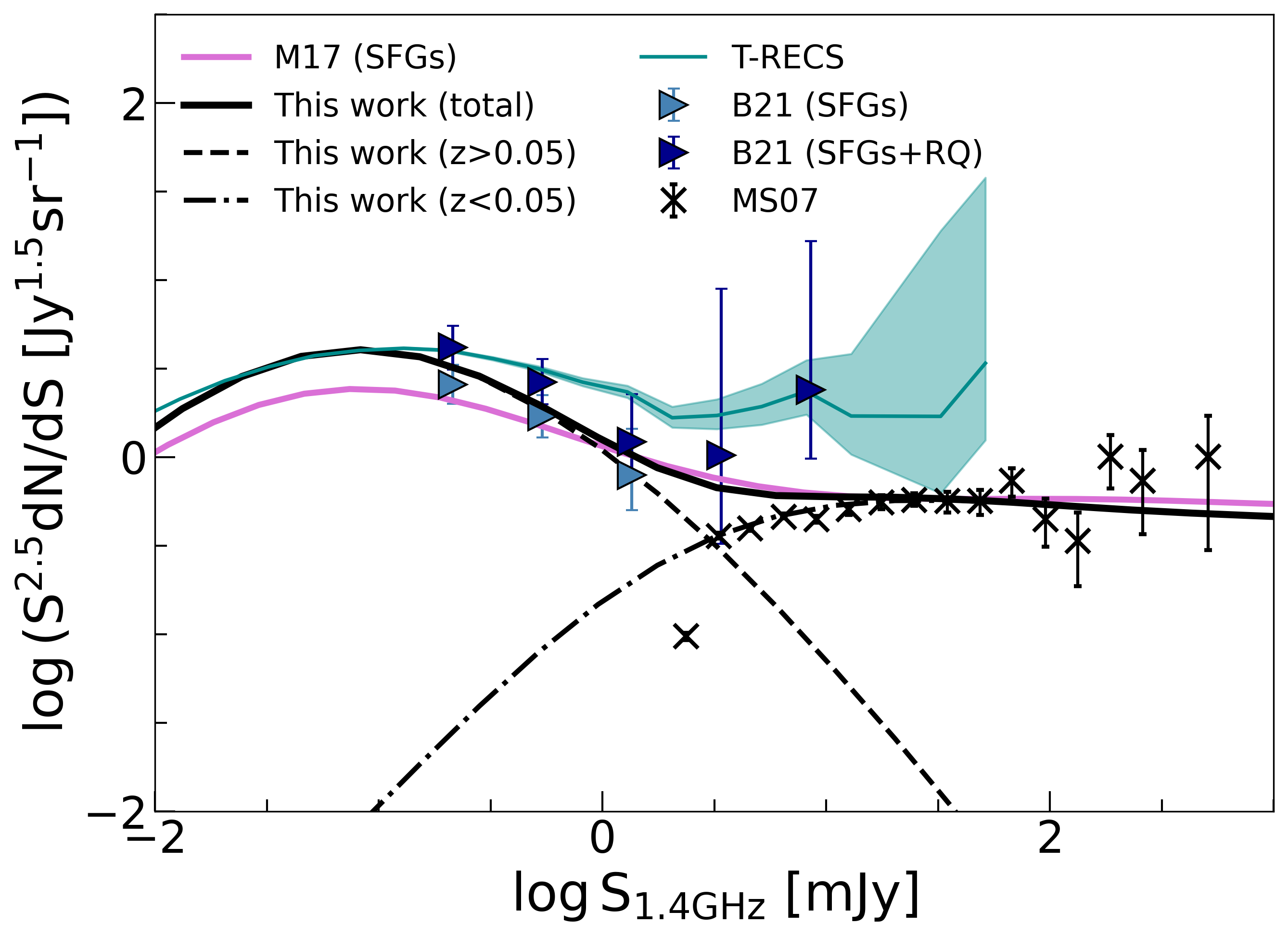}
    \caption{1.4 GHz Euclidean normalised differential number counts from \texttt{SEMPER} for SFGs (black solid line) obtained adopting the relation of \cite{Delvecchio2021}. The dashed and dot-dashed lines represent the contribution for galaxies located respectively above and below $z\sim 0.05$. The solid pink line is the prediction from \cite{Mancuso2017}. Triangles are data from \cite{Bonato2021a} for SFGs (light blue) and SFGs and RQ sources combined (dark blue). In cyan, we display the predictions from the T-RECS simulation by \cite{Bonaldi2019MNRAS.482....2B, Bonaldi2023MNRAS.524..993B}. Black crosses show the number counts for local galaxies from \cite{Mauch2007}.}
    \label{fig:numbercounts_1.4}
\end{figure}

Fig. \ref{fig:numbercounts_1.4} shows the 1.4 GHz number counts predicted by our model by adopting Eq. \ref{eq:firrc_delvecchio}. 
Our predictions agree with the number counts derived for SFGs by \citet[][light blue triangles]{Bonato2021a}, and also reproduce the observed number counts for local galaxies obtained by \citet[][black crosses]{Mauch2007} at $\log S_{\rm 1.4 GHz}\gtrsim 1$ mJy.  
The overall trend is similar to the one of the \citet[][pink solid line]{Mancuso2017} model, even if we predict a higher ($\sim 0.2$ dex) number of SFGs at fluxes $S_{\rm 1.4 GHz}\lesssim 0.5$ mJy.
The 1.4 GHz number counts from the T-RECS simulation are up to $\sim 0.7$ dex higher than our predictions for $S_{1.4\, \rm GHz} \gtrsim 0.1$ mJy, but are consistent with observations when including the contribution of RQ AGNs (\citealt{Bonato2021a}, dark blue triangles).

\subsection{Luminosity functions at 150 MHz}\label{sec:LF_150MHz}

\begin{figure}
    \centering
    \includegraphics[width=0.45\textwidth]{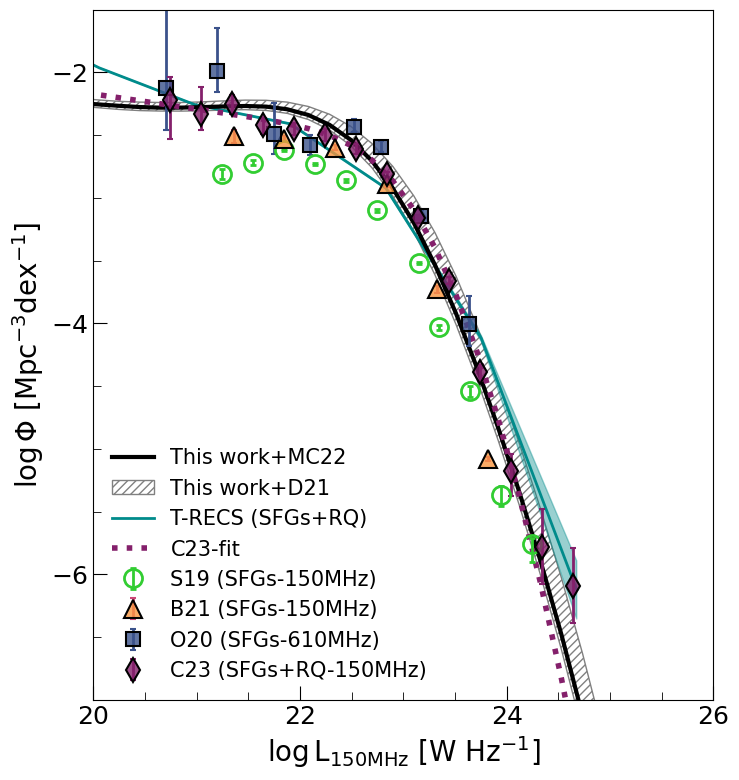}
    \caption{150 MHz LF for SFGs from \texttt{SEMPER} at $z\sim 0$. The model is obtained from Eq. \ref{eq:firrc_mccheyne} and Eq. \ref{eq:firrc_delvecchio} (hatched grey area) using a spectral index range of $\alpha=[-0.8, -0.6]$ and by applying Eq. \ref{eq:correction_radio}. Data points are from \cite{Sabater2019A&A...622A..17S}, \cite{Bonato2021b} and \cite{Cochrane2023} at 150 MHz and \cite{Ocran2020b} at 640 MHz (rescaled assuming $\alpha=-0.7)$. The dotted line refers to the best-fit modified Schechter function presented in \cite{Cochrane2023}. The shaded area is from the T-RECS simulation (\citealt{Bonaldi2019MNRAS.482....2B, Bonaldi2023MNRAS.524..993B}).}
    \label{fig:local_LF_150}
\end{figure}

\begin{figure*}
    \centering
    \includegraphics[width=\textwidth]{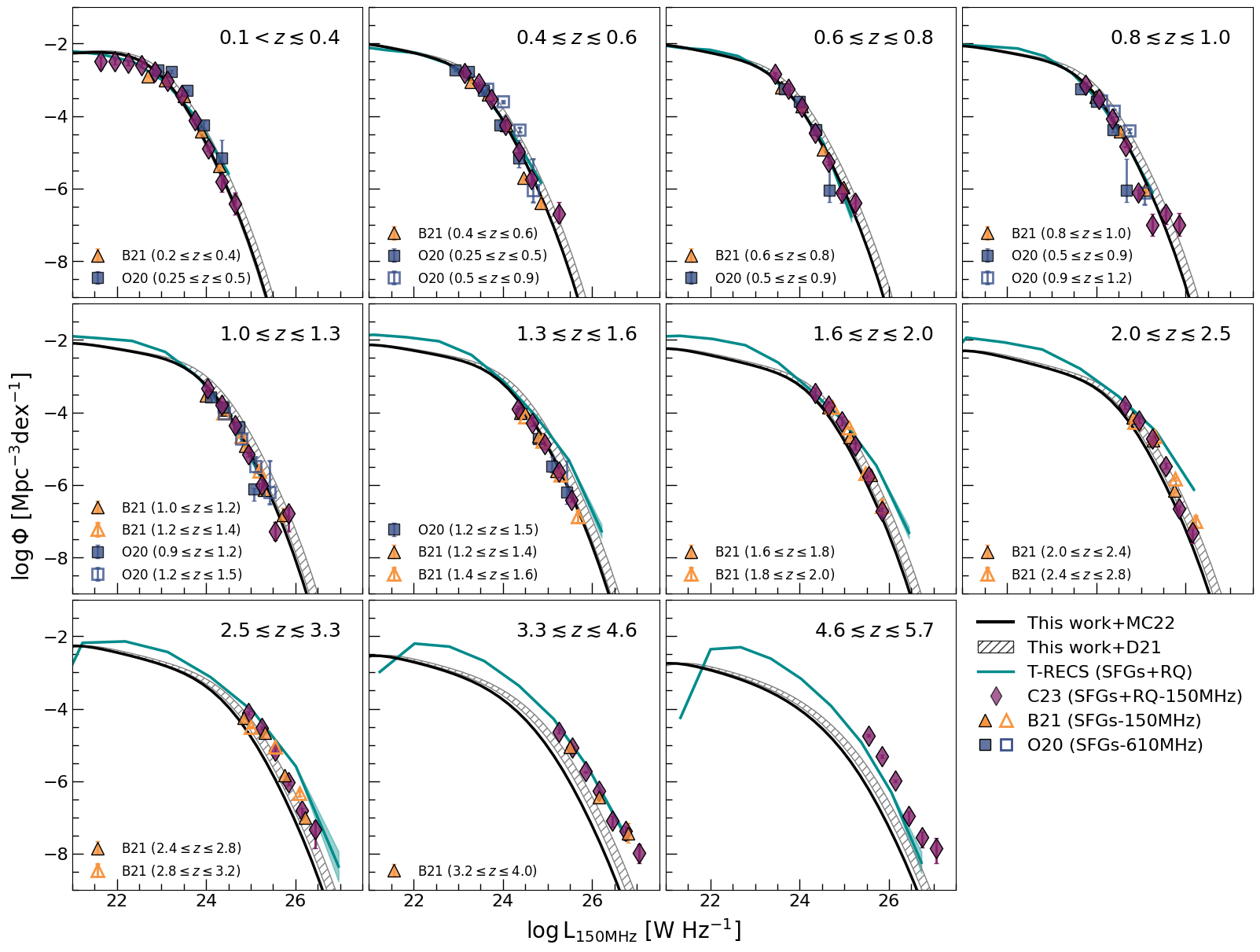}
    \caption{150 MHz LFs for SFGs at different redshift bins predicted by \texttt{SEMPER}. The model is derived by assuming the mass-dependent FIRRC from \cite{Delvecchio2021} (solid black line) and \cite{McCheyne2022} (hatched grey area). As in Fig. \ref{fig:LFs_1.4}, Eq. \ref{eq:correction_radio} is applied only to the first redshift bin. Data points (symbols as in legend) are from  \cite{Cochrane2023}, \cite{Bonato2021b} and \cite{Ocran2020b}. The latter are scaled from 610 MHz to 150 MHz assuming $\alpha=-0.7$. Cyan lines and shaded areas are the predictions from the T-RECS simulation by \cite{Bonaldi2019MNRAS.482....2B, Bonaldi2023MNRAS.524..993B} and their $1\sigma$ uncertainties, respectively.}
    \label{fig:LFs_150}
\end{figure*}

Fig. \ref{fig:local_LF_150} shows our predictions for the local ($0.03<z<0.3$) 150 MHz LF, compared with recent determinations at 150 and 610 MHz (\citealt[][purple diamonds;]{Cochrane2023} \citealt[][empty green circles;]{Sabater2019A&A...622A..17S} \citealt[][gold triangles;]{Bonato2021b} \citealt{Ocran2020b}, blue squares).
 Moreover, we include the parametrised expression (dotted purple line) derived by \cite{Cochrane2023} adopting a modified Schechter function.
The 150 MHz LF predictions presented in this section are derived using both the \citet[][Eq. \ref{eq:firrc_mccheyne}; black solid line]{McCheyne2022} and the \citet{Delvecchio2021}[Eq. \ref{eq:firrc_delvecchio}; hatched grey area] FIRRC relations. The latter has been rescaled to 150 MHz using a range of spectral indexes ($\alpha=[-0.8, -0.6]$), to better account for a realistic spectral index distribution around the typical SFG value of -0.7, and/or any spectral index variations with redshift (in the case of redshift-dependent LFs, see Fig.~\ref{fig:LFs_150}).  

Our results agree very well with the determinations of \cite{Cochrane2023} and \cite{Ocran2020b}. 
The data points from \cite{Sabater2019A&A...622A..17S}, \cite{Butler2019A&A...625A.111B} and \cite{Bonato2021b} lie below our model and \cite{Cochrane2023} determinations (up to 1 dex offsets). As discussed in \cite{Cochrane2023}, this difference likely arises from the lack of corrections for radio completeness in previous data sets and the different choices of redshift bins. This and the strong redshift evolution of the LFs may indeed explain the offset. The predictions from T-RECS (cyan shaded area) are roughly consistent with our model and with data from \cite{Cochrane2023}.  
Fig. \ref{fig:local_LF_150} also shows that the predictions derived from Eq. \ref{eq:firrc_delvecchio} are in better agreement with the predictions obtained from Eq. \ref{eq:firrc_mccheyne} when rescaled to 150 MHz using a flatter spectral index ($\alpha \approx -0.6$; see lower edge of the grey hatched area). This is consistent with what was found by \cite{McCheyne2022}, i.e. that their FIRRC normalisation closely matches that of \cite{Delvecchio2021} when rescaled using a spectral index value $\alpha = - 0.59$.

Fig. \ref{fig:LFs_150} displays the 150 MHz LFs obtained for different redshift bins in the range $0.1<z\lesssim 5.7$. Our predictions are compared with data from \cite{Cochrane2023}, \cite{Bonato2021b} and \cite{Ocran2020b}, as well as with 
predictions from the T-RECS simulation. 
The model based on Eq. \ref{eq:firrc_mccheyne} (black solid line) reproduces the observational data accurately up to redshift $z\approx 1.5$. This suggests that the FIRRC from \cite{McCheyne2022}, originally derived up to $z\approx 1$, is able to represent the data up to slightly higher redshifts. At redshifts $z\gtrsim 1.6$ the observations better agree with the model predictions derived using Eq. \ref{eq:firrc_delvecchio} (hatched grey area). This may be due to the fact that \cite{Delvecchio2021} relation was derived from samples spanning a wider redshift range ($1<z<4$).
As for $\nu=1.4$ GHz, we find good agreement between our model and the T-RECS simulation up to $z\approx 1.5$. At higher redshifts, our model shifts towards lower luminosities compared to T-RECS. 
At z$\gtrsim 3.3$ the data better agree with T-RECS than with our model, which is located $\sim 0.5$ dex below the observations from \cite{Cochrane2023} and \cite{Bonato2021b}. This may be ascribed, at least partially, to the fact that both the T-RECS simulation and \cite{Cochrane2023} sample include the contribution from RQ AGNs. 
At redshift $z>4.6$, both our model and T-RECS fall short. However, it is important to notice that the photometric redshifts of the LoTSS deep fields can be considered highly reliable only up to z$\sim$1.5 for SFGs (as extensively discussed in \cite{Duncan2021}), implying that the highest-z bins of the LFs can be affected by large uncertainties.  

\subsection{Number counts at 150 MHz}\label{sec:numbercounts_150MHz}

\begin{figure}
    \centering
    
    \includegraphics[width=0.48\textwidth]{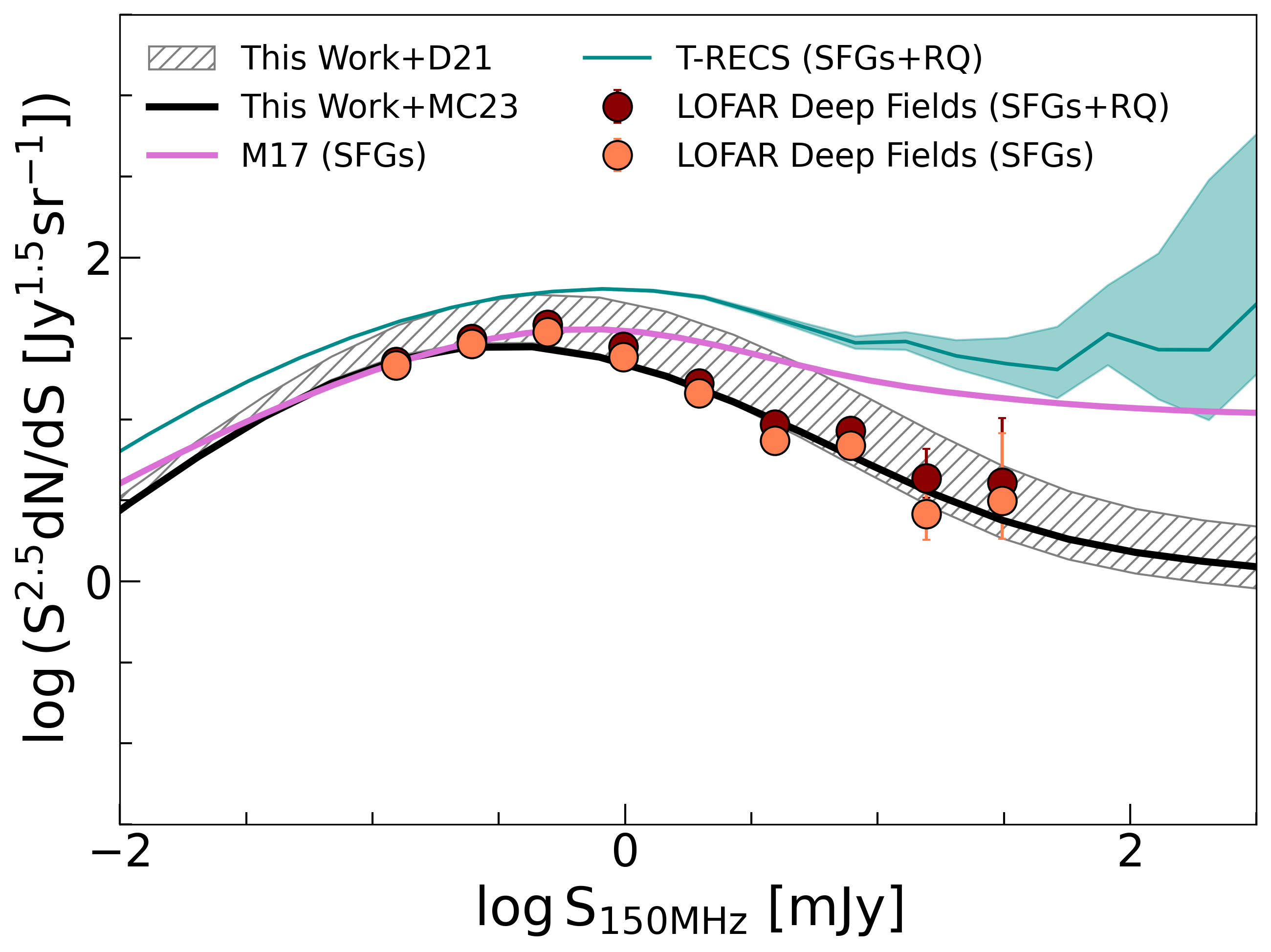}
    \caption{150 MHz Euclidean normalized differential number counts for SFGs from \texttt{SEMPER}. The solid black line (hatched area) represents our model obtained adopting Eq. \ref{eq:firrc_mccheyne} (Eq. \ref{eq:firrc_delvecchio}, scaled to 150 MHz using a spectral index range $-0.8<\alpha<-0.6$). The cyan and pink solid lines represent predictions from the T-RECS simulation (\citealt{Bonaldi2019MNRAS.482....2B, Bonaldi2023MNRAS.524..993B}) and the model from \cite{Mancuso2017}, respectively. Circles are the differential source counts we computed from the LoTSS deep fields DR1 catalogues.}
    \label{fig:numbercounts_150}
\end{figure}

In Fig. \ref{fig:numbercounts_150} we show the number counts at $\nu=150\,$MHz as derived from our model, compared with the predictions of the T-RECS simulation (cyan shaded area) and of the model by \cite{Mancuso2017} (pink solid line). We also present the number counts for SFGs and RQ-AGNs that we derived from the DR1 catalogues of the LoTSS deep fields 
(see Tab. \ref{tab:logcounts}).
As for the 150 MHz LFs, the modelled counts are obtained adopting either the FIRRC described in Eq.~\ref{eq:firrc_mccheyne} \cite[][black solid line]{McCheyne2022} or the one described by Eq.~\ref{eq:firrc_delvecchio} \cite[][hatched grey area]{Delvecchio2021}.
Whichever FIRRC relation we adopt, our model agrees very well with the observed source counts. 
In particular, our model can better reproduce the decrease of the number counts observed at $\gtrsim 1.5$ mJy than the \cite{Mancuso2017} model. The T-RECS simulation lies above our predictions (and the observed source counts) by at least  $\sim 0.3$ dex overall. A better agreement with T-RECS is however observed in the range $0.1\lesssim S_{\rm 150\, MHz} \lesssim1\,$mJy, when we adopt the \cite{Delvecchio2021} FIRRC.

\section{Conclusions}\label{sec:conclusion}

In this paper, we have developed \texttt{SEMPER}, a semi-empirical model to predict radio LFs and number counts for SFGs.  
Our model combines the evolving SMFs for SFGs, based on the recent deep NIR data from the COSMOS2020 catalogue (\citetalias{Weaver2023}), with up-to-date observed scaling relations, such as the galaxy main sequence from \cite{Popesso2023} and the mass- and redshift-dependent FIRRC from \cite{Delvecchio2021} and \cite{McCheyne2022}. 
We refitted the redshift-dependent \citetalias{Weaver2023} SMFs with a Double Power Law function instead of with the Double Schechter function originally adopted by the authors. We modelled the fit parameters in the range $0<z<5$ and obtained the time and shape evolution of the SMFs.

We tested our model against radio data at two reference frequencies: 1.4 GHz and 150 MHz. In general, we find remarkably good agreement with the latest observational constraints on LFs and number counts from the JVLA, LOFAR, GMRT, WSRT and ATCA. Our main results can be summarised as follows:

\begin{itemize}

    \item Our SMFs, reconstructed via a Double Power Law function, nicely reproduce observations for all the redshift bins, including the observed massive galaxies ($M_{\star}\gtrsim 10^{11}\,$M$_{\odot}$) at high redshifts ($z\gtrsim 2.5$), which were attributed by \citetalias{Weaver2023} to a novel population of dust-obscured SFGs, only revealed thanks to ultra-deep NIR imaging in the COSMOS field. This population could not be reproduced by a Double Schechter function fitting.
    
    \item Our predictions align with observational determinations of the 1.4 GHz LFs up to redshift $z\sim 5$, while deviating from the T-RECS sky simulation at $z\gtrsim 1$, particularly at bright and faint luminosities.

    \item 
    Our model can reproduce the observed 1.4 GHz LFs obtained for samples of radio-selected massive and dust-obscured SFGs at $z>3.5$, some of which contain a fraction of optically dark galaxies. 
    This indirectly suggests a natural link between this population and the radio-bright SFGs population observed at high redshift in deep radio fields. 
    
    \item The 1.4 GHz normalised number counts are well in agreement with determinations for local and high redshift SFGs obtained from the NVSS survey (\citealt{Mauch2007}) and WSRT deep observations of the Lockman Hole (\citealt{Prandoni2018}; \citealt{Bonato2021a}), respectively. Our model shows a similar trend to the one of \cite{Mancuso2017}, but predicts a slightly higher ($\sim 0.2$ dex) number of SFGs for $S_{1.4 \rm GHz}\lesssim0.5\,$mJy, while it deviates from the T-RECS predictions at $S_{1.4 \rm GHz}\gtrsim0.1\,$mJy.

    \item We find good agreement between our predicted and observed LFs also at low frequency ($\nu=$150 MHz), particularly when compared with the recent determinations from the LoTSS deep fields (\citealt{Cochrane2023}). Using the FIRRC of \cite{McCheyne2022} yields better results for $z\lesssim1.5$, while better consistency is found with the data at higher redshifts when using the \cite{Delvecchio2021} relation. The origin of this discrepancy is likely associated with the different redshift ranges sampled by the two correlations. At $z\gtrsim 3.3$, the LFs observed in the LoTSS deep fields show an excess with respect to our model predictions. We attribute this discrepancy to either larger photometric redshift uncertainties in the LoTSS-Deep DR1 radio catalogues or to the fact that both SFGs and RQ AGN are included in the observed LFs.
  
  \item We derive the 150 MHz number counts for SFGs and RQ AGN from the LoTSS deep fields DR1 catalogues. Our 150 MHz number count predictions closely match the observed LoTSS-deep number counts either when we consider SFGs alone or when we include RQ AGNs. We only find partial agreement between our results and what is predicted by \cite{Mancuso2017}, as their model produces flatter counts at $S_{\rm 150 MHz} \gtrsim 1.5\,$mJy. The T-RECS simulation, on the other hand, predicts higher counts than our model over the entire flux range explored.

\end{itemize} 

Our predictions will be further tested with upcoming deep radio continuum surveys planned for SKA precursors (\citealt{Norris2013}) and, on a longer timescale, for the SKA (\citealt{Prandoni2015}). We also plan to update our model by exploiting optical/NIR JWST observations, which will soon provide new SMFs expanding the redshift and mass ranges explored so far. 

\begin{acknowledgements}
M. G. acknowledges support from INAF under the Large Grant 2022 funding scheme (project "MeerKAT and LOFAR Team up: a Unique Radio Window on Galaxy/AGN co-Evolution”). M. B. acknowledges support from INAF under the mini-grant ``A systematic search for ultra-bright high-z strongly lensed galaxies in Planck catalogues''. L.B. acknowledges financial support from the German Excellence Strategy via the Heidelberg Cluster of Excellence (EXC 2181 - 390900948) STRUCTURES. A. L. has been supported by: European Union - NextGenerationEU under the PRIN MUR 2022 project n. 20224JR28W "Charting unexplored avenues in Dark Matter"; INAF Large Grant 2022 funding scheme with the project "MeerKAT and LOFAR Team up: a Unique Radio Window on Galaxy/AGN co-Evolution"; INAF GO-GTO Normal 2023 funding scheme with the project "Serendipitous H-ATLAS-fields Observations of Radio Extragalactic Sources (SHORES)"; project ``Data Science methods for MultiMessenger Astrophysics \& Multi-Survey Cosmology'' funded by the Italian Ministry of University and Research, Programmazione triennale 2021/2023 (DM n.2503 dd. 9 December 2019), Programma Congiunto Scuole; Italian Research Center on High Performance Computing Big Data and Quantum Computing (ICSC), project funded by European Union - NextGenerationEU - and National Recovery and Resilience Plan (NRRP) - Mission 4 Component 2 within the activities of Spoke 3 (Astrophysics and Cosmos Observations).
\end{acknowledgements}

\bibliographystyle{aa} 
\bibliography{bibliography}

\begin{thebibliography}{178}
\expandafter\ifx\csname natexlab\endcsname\relax\def\natexlab#1{#1}\fi

\bibitem[{{Algera} {et~al.}(2020){Algera}, {van der Vlugt}, {Hodge}, {Smail}, {Novak}, {Radcliffe}, {Riechers}, {R{\"o}ttgering}, {Smol{\v{c}}i{\'c}}, \& {Walter}}]{Algera2020ApJ...903..139A}
{Algera}, H.~S.~B., {van der Vlugt}, D., {Hodge}, J.~A., {et~al.} 2020, \apj, 903, 139

\bibitem[{{An} {et~al.}(2021){An}, {Vaccari}, {Smail}, {Jarvis}, {Whittam}, {Hale}, {Jin}, {Collier}, {Daddi}, {Delhaize}, {Frank}, {Murphy}, {Prescott}, {Sekhar}, {Taylor}, {Ao}, {Knowles}, {Marchetti}, {Randriamampandry}, \& {Randriamanakoto}}]{An2021}
{An}, F., {Vaccari}, M., {Smail}, I., {et~al.} 2021, \mnras, 507, 2643

\bibitem[{{Barrufet} {et~al.}(2023){Barrufet}, {Oesch}, {Weibel}, {Brammer}, {Bezanson}, {Bouwens}, {Fudamoto}, {Gonzalez}, {Gottumukkala}, {Illingworth}, {Heintz}, {Holden}, {Labbe}, {Magee}, {Naidu}, {Nelson}, {Stefanon}, {Smit}, {van Dokkum}, {Weaver}, \& {Williams}}]{Barrufet2023}
{Barrufet}, L., {Oesch}, P.~A., {Weibel}, A., {et~al.} 2023, \mnras, 522, 449

\bibitem[{{Basu} {et~al.}(2015){Basu}, {Wadadekar}, {Beelen}, {Singh}, {Archana}, {Sirothia}, \& {Ishwara-Chandra}}]{Basu2015}
{Basu}, A., {Wadadekar}, Y., {Beelen}, A., {et~al.} 2015, \apj, 803, 51

\bibitem[{{Behiri} {et~al.}(2023){Behiri}, {Talia}, {Cimatti}, {Lapi}, {Massardi}, {Enia}, {Vignali}, {Bethermin}, {Faisst}, {Gentile}, {Giulietti}, {Gruppioni}, {Pozzi}, {Smol{\c{c}}i{\'c}}, \& {Zamorani}}]{Behiri2023}
{Behiri}, M., {Talia}, M., {Cimatti}, A., {et~al.} 2023, \apj, 957, 63

\bibitem[{{Behroozi} {et~al.}(2019){Behroozi}, {Wechsler}, {Hearin}, \& {Conroy}}]{Behroozi2019MNRAS.488.3143B}
{Behroozi}, P., {Wechsler}, R.~H., {Hearin}, A.~P., \& {Conroy}, C. 2019, \mnras, 488, 3143

\bibitem[{{Behroozi} {et~al.}(2013){Behroozi}, {Wechsler}, \& {Conroy}}]{Behroozi2013ApJ...770...57B}
{Behroozi}, P.~S., {Wechsler}, R.~H., \& {Conroy}, C. 2013, \apj, 770, 57

\bibitem[{{Bell}(2003)}]{Bell2003}
{Bell}, E.~F. 2003, \apj, 586, 794

\bibitem[{{Best} {et~al.}(2023){Best}, {Kondapally}, {Williams}, {Cochrane}, {Duncan}, {Hale}, {Haskell}, {Ma{\l}ek}, {McCheyne}, {Smith}, {Wang}, {Botteon}, {Bonato}, {Bondi}, {Calistro Rivera}, {Gao}, {G{\"u}rkan}, {Hardcastle}, {Jarvis}, {Mingo}, {Miraghaei}, {Morabito}, {Nisbet}, {Prandoni}, {R{\"o}ttgering}, {Sabater}, {Shimwell}, {Tasse}, \& {van Weeren}}]{Best2023}
{Best}, P.~N., {Kondapally}, R., {Williams}, W.~L., {et~al.} 2023, \mnras, 523, 1729

\bibitem[{{B{\'e}thermin} {et~al.}(2012){B{\'e}thermin}, {Daddi}, {Magdis}, {Sargent}, {Hezaveh}, {Elbaz}, {Le Borgne}, {Mullaney}, {Pannella}, {Buat}, {Charmandaris}, {Lagache}, \& {Scott}}]{Bethermin2012}
{B{\'e}thermin}, M., {Daddi}, E., {Magdis}, G., {et~al.} 2012, \apjl, 757, L23

\bibitem[{{Bisigello} {et~al.}(2018){Bisigello}, {Caputi}, {Grogin}, \& {Koekemoer}}]{Bisigello2018}
{Bisigello}, L., {Caputi}, K.~I., {Grogin}, N., \& {Koekemoer}, A. 2018, \aap, 609, A82

\bibitem[{{Bisigello} {et~al.}(2022){Bisigello}, {Vallini}, {Gruppioni}, {Esposito}, {Calura}, {Delvecchio}, {Feltre}, {Pozzi}, \& {Rodighiero}}]{Bisigello2022A&A...666A.193B}
{Bisigello}, L., {Vallini}, L., {Gruppioni}, C., {et~al.} 2022, \aap, 666, A193

\bibitem[{{Boco} {et~al.}(2021){Boco}, {Lapi}, {Chruslinska}, {Donevski}, {Sicilia}, \& {Danese}}]{Boco2021a}
{Boco}, L., {Lapi}, A., {Chruslinska}, M., {et~al.} 2021, \apj, 907, 110

\bibitem[{{Boco} {et~al.}(2023){Boco}, {Lapi}, {Shankar}, {Fu}, {Gabrielli}, \& {Sicilia}}]{Boco2023}
{Boco}, L., {Lapi}, A., {Shankar}, F., {et~al.} 2023, \apj, 954, 97

\bibitem[{{Bonaldi} {et~al.}(2019){Bonaldi}, {Bonato}, {Galluzzi}, {Harrison}, {Massardi}, {Kay}, {De Zotti}, \& {Brown}}]{Bonaldi2019MNRAS.482....2B}
{Bonaldi}, A., {Bonato}, M., {Galluzzi}, V., {et~al.} 2019, \mnras, 482, 2

\bibitem[{{Bonaldi} {et~al.}(2023){Bonaldi}, {Hartley}, {Ronconi}, {De Zotti}, \& {Bonato}}]{Bonaldi2023MNRAS.524..993B}
{Bonaldi}, A., {Hartley}, P., {Ronconi}, T., {De Zotti}, G., \& {Bonato}, M. 2023, \mnras, 524, 993

\bibitem[{{Bonato} {et~al.}(2017){Bonato}, {Negrello}, {Mancuso}, {De Zotti}, {Ciliegi}, {Cai}, {Lapi}, {Massardi}, {Bonaldi}, {Sajina}, {Smol{\v{c}}i{\'c}}, \& {Schinnerer}}]{Bonato2017}
{Bonato}, M., {Negrello}, M., {Mancuso}, C., {et~al.} 2017, \mnras, 469, 1912

\bibitem[{{Bonato} {et~al.}(2021{\natexlab{a}}){Bonato}, {Prandoni}, {De Zotti}, {Best}, {Bondi}, {Calistro Rivera}, {Cochrane}, {G{\"u}rkan}, {Haskell}, {Kondapally}, {Magliocchetti}, {Leslie}, {Malek}, {R{\"o}ttgering}, {Smith}, {Tasse}, \& {Wang}}]{Bonato2021b}
{Bonato}, M., {Prandoni}, I., {De Zotti}, G., {et~al.} 2021{\natexlab{a}}, \aap, 656, A48

\bibitem[{{Bonato} {et~al.}(2021{\natexlab{b}}){Bonato}, {Prandoni}, {De Zotti}, {Brienza}, {Morganti}, \& {Vaccari}}]{Bonato2021a}
{Bonato}, M., {Prandoni}, I., {De Zotti}, G., {et~al.} 2021{\natexlab{b}}, \mnras, 500, 22

\bibitem[{{Bourne} {et~al.}(2011){Bourne}, {Dunne}, {Ivison}, {Maddox}, {Dickinson}, \& {Frayer}}]{Bourne2011}
{Bourne}, N., {Dunne}, L., {Ivison}, R.~J., {et~al.} 2011, \mnras, 410, 1155

\bibitem[{{Brinchmann} {et~al.}(2004){Brinchmann}, {Charlot}, {White}, {Tremonti}, {Kauffmann}, {Heckman}, \& {Brinkmann}}]{Brinchmann2004MNRAS.351.1151B}
{Brinchmann}, J., {Charlot}, S., {White}, S.~D.~M., {et~al.} 2004, \mnras, 351, 1151

\bibitem[{{Buat} {et~al.}(2012){Buat}, {Noll}, {Burgarella}, {Giovannoli}, {Charmandaris}, {Pannella}, {Hwang}, {Elbaz}, {Dickinson}, {Magdis}, {Reddy}, \& {Murphy}}]{Buat2012A&A...545A.141B}
{Buat}, V., {Noll}, S., {Burgarella}, D., {et~al.} 2012, \aap, 545, A141

\bibitem[{{Burgarella} {et~al.}(2013){Burgarella}, {Buat}, {Gruppioni}, {Cucciati}, {Heinis}, {Berta}, {B{\'e}thermin}, {Bock}, {Cooray}, {Dunlop}, {Farrah}, {Franceschini}, {Le Floc'h}, {Lutz}, {Magnelli}, {Nordon}, {Oliver}, {Page}, {Popesso}, {Pozzi}, {Riguccini}, {Vaccari}, \& {Viero}}]{Burgarella2013A&A...554A..70B}
{Burgarella}, D., {Buat}, V., {Gruppioni}, C., {et~al.} 2013, \aap, 554, A70

\bibitem[{{Butler} {et~al.}(2019){Butler}, {Huynh}, {Kapi{\'n}ska}, {Delvecchio}, {Smol{\v{c}}i{\'c}}, {Chiappetti}, {Koulouridis}, \& {Pierre}}]{Butler2019A&A...625A.111B}
{Butler}, A., {Huynh}, M., {Kapi{\'n}ska}, A., {et~al.} 2019, \aap, 625, A111

\bibitem[{{Calistro Rivera} {et~al.}(2017){Calistro Rivera}, {Williams}, {Hardcastle}, {Duncan}, {R{\"o}ttgering}, {Best}, {Br{\"u}ggen}, {Chy{\.z}y}, {Conselice}, {de Gasperin}, {Engels}, {G{\"u}rkan}, {Intema}, {Jarvis}, {Mahony}, {Miley}, {Morabito}, {Prandoni}, {Sabater}, {Smith}, {Tasse}, {van der Werf}, \& {White}}]{CalistroRivera2017}
{Calistro Rivera}, G., {Williams}, W.~L., {Hardcastle}, M.~J., {et~al.} 2017, \mnras, 469, 3468

\bibitem[{{Caputi} {et~al.}(2017){Caputi}, {Deshmukh}, {Ashby}, {Cowley}, {Bisigello}, {Fazio}, {Fynbo}, {Le F{\`e}vre}, {Milvang-Jensen}, \& {Ilbert}}]{Caputi2017}
{Caputi}, K.~I., {Deshmukh}, S., {Ashby}, M.~L.~N., {et~al.} 2017, \apj, 849, 45

\bibitem[{{Casey} {et~al.}(2023){Casey}, {Kartaltepe}, {Drakos}, {Franco}, {Harish}, {Paquereau}, {Ilbert}, {Rose}, {Cox}, {Nightingale}, {Robertson}, {Silverman}, {Koekemoer}, {Massey}, {McCracken}, {Rhodes}, {Akins}, {Allen}, {Amvrosiadis}, {Arango-Toro}, {Bagley}, {Bongiorno}, {Capak}, {Champagne}, {Chartab}, {Ch{\'a}vez Ortiz}, {Chworowsky}, {Cooke}, {Cooper}, {Darvish}, {Ding}, {Faisst}, {Finkelstein}, {Fujimoto}, {Gentile}, {Gillman}, {Gould}, {Gozaliasl}, {Hayward}, {He}, {Hemmati}, {Hirschmann}, {Jahnke}, {Jin}, {Khostovan}, {Kokorev}, {Lambrides}, {Laigle}, {Larson}, {Leung}, {Liu}, {Liaudat}, {Long}, {Magdis}, {Mahler}, {Mainieri}, {Manning}, {Maraston}, {Martin}, {McCleary}, {McKinney}, {McPartland}, {Mobasher}, {Pattnaik}, {Renzini}, {Rich}, {Sanders}, {Sattari}, {Scognamiglio}, {Scoville}, {Sheth}, {Shuntov}, {Sparre}, {Suzuki}, {Talia}, {Toft}, {Trakhtenbrot}, {Urry}, {Valentino}, {Vanderhoof}, {Vardoulaki}, {Weaver}, {Whitaker}, {Wilkins}, {Yang}, \& {Zavala}}]{Casey2023}
{Casey}, C.~M., {Kartaltepe}, J.~S., {Drakos}, N.~E., {et~al.} 2023, \apj, 954, 31

\bibitem[{{Ceraj} {et~al.}(2018){Ceraj}, {Smol{\v{c}}i{\'c}}, {Delvecchio}, {Novak}, {Zamorani}, {Delhaize}, {Schinnerer}, {Vardoulaki}, \& {Herrera Ruiz}}]{Ceraj2018}
{Ceraj}, L., {Smol{\v{c}}i{\'c}}, V., {Delvecchio}, I., {et~al.} 2018, \aap, 620, A192

\bibitem[{{Chabrier}(2003)}]{Chabrier2003}
{Chabrier}, G. 2003, \pasp, 115, 763

\bibitem[{{Chapman} {et~al.}(2004){Chapman}, {Smail}, {Blain}, \& {Ivison}}]{Chapman2004}
{Chapman}, S.~C., {Smail}, I., {Blain}, A.~W., \& {Ivison}, R.~J. 2004, \apj, 614, 671

\bibitem[{{Chi} \& {Wolfendale}(1990)}]{Chi1990MNRAS.245..101C}
{Chi}, X. \& {Wolfendale}, A.~W. 1990, \mnras, 245, 101

\bibitem[{{Cochrane} {et~al.}(2023){Cochrane}, {Kondapally}, {Best}, {Sabater}, {Duncan}, {Smith}, {Hardcastle}, {R{\"o}ttgering}, {Prandoni}, {Haskell}, {G{\"u}rkan}, \& {Miley}}]{Cochrane2023}
{Cochrane}, R.~K., {Kondapally}, R., {Best}, P.~N., {et~al.} 2023, \mnras, 523, 6082

\bibitem[{{Condon}(1992)}]{Condon1992}
{Condon}, J.~J. 1992, \araa, 30, 575

\bibitem[{{Condon} {et~al.}(2002){Condon}, {Cotton}, \& {Broderick}}]{Condon2002AJ....124..675C}
{Condon}, J.~J., {Cotton}, W.~D., \& {Broderick}, J.~J. 2002, \aj, 124, 675

\bibitem[{{Condon} {et~al.}(1998){Condon}, {Cotton}, {Greisen}, {Yin}, {Perley}, {Taylor}, \& {Broderick}}]{Condon1998}
{Condon}, J.~J., {Cotton}, W.~D., {Greisen}, E.~W., {et~al.} 1998, \aj, 115, 1693

\bibitem[{{Condon} {et~al.}(2019){Condon}, {Matthews}, \& {Broderick}}]{Condon2019}
{Condon}, J.~J., {Matthews}, A.~M., \& {Broderick}, J.~J. 2019, \apj, 872, 148

\bibitem[{{Cucciati} {et~al.}(2012){Cucciati}, {Tresse}, {Ilbert}, {Le F{\`e}vre}, {Garilli}, {Le Brun}, {Cassata}, {Franzetti}, {Maccagni}, {Scodeggio}, {Zucca}, {Zamorani}, {Bardelli}, {Bolzonella}, {Bielby}, {McCracken}, {Zanichelli}, \& {Vergani}}]{Cucciati2012A&A...539A..31C}
{Cucciati}, O., {Tresse}, L., {Ilbert}, O., {et~al.} 2012, \aap, 539, A31

\bibitem[{{Daddi} {et~al.}(2022){Daddi}, {Delvecchio}, {Dimauro}, {Magnelli}, {Gomez-Guijarro}, {Coogan}, {Elbaz}, {Kalita}, {Le Bail}, {Rich}, \& {Tan}}]{Daddi2022}
{Daddi}, E., {Delvecchio}, I., {Dimauro}, P., {et~al.} 2022, \aap, 661, L7

\bibitem[{{Daddi} {et~al.}(2007){Daddi}, {Dickinson}, {Morrison}, {Chary}, {Cimatti}, {Elbaz}, {Frayer}, {Renzini}, {Pope}, {Alexander}, {Bauer}, {Giavalisco}, {Huynh}, {Kurk}, \& {Mignoli}}]{Daddi2007}
{Daddi}, E., {Dickinson}, M., {Morrison}, G., {et~al.} 2007, \apj, 670, 156

\bibitem[{{Davidzon} {et~al.}(2017){Davidzon}, {Ilbert}, {Laigle}, {Coupon}, {McCracken}, {Delvecchio}, {Masters}, {Capak}, {Hsieh}, {Le F{\`e}vre}, {Tresse}, {Bethermin}, {Chang}, {Faisst}, {Le Floc'h}, {Steinhardt}, {Toft}, {Aussel}, {Dubois}, {Hasinger}, {Salvato}, {Sanders}, {Scoville}, \& {Silverman}}]{Davidzon2017}
{Davidzon}, I., {Ilbert}, O., {Laigle}, C., {et~al.} 2017, \aap, 605, A70

\bibitem[{{De Zotti} {et~al.}(2024){De Zotti}, {Bonato}, {Giulietti}, {Massardi}, {Negrello}, {Algera}, \& {Delhaize}}]{Dezotti2024}
{De Zotti}, G., {Bonato}, M., {Giulietti}, M., {et~al.} 2024, \aap, 689, A272

\bibitem[{{de Zotti} {et~al.}(2010){de Zotti}, {Massardi}, {Negrello}, \& {Wall}}]{Dezotti2010}
{de Zotti}, G., {Massardi}, M., {Negrello}, M., \& {Wall}, J. 2010, \aapr, 18, 1

\bibitem[{{Delhaize} {et~al.}(2017){Delhaize}, {Smol{\v{c}}i{\'c}}, {Delvecchio}, {Novak}, {Sargent}, {Baran}, {Magnelli}, {Zamorani}, {Schinnerer}, {Murphy}, {Aravena}, {Berta}, {Bondi}, {Capak}, {Carilli}, {Ciliegi}, {Civano}, {Ilbert}, {Karim}, {Laigle}, {Le F{\`e}vre}, {Marchesi}, {McCracken}, {Salvato}, {Seymour}, \& {Tasca}}]{Delhaize2017}
{Delhaize}, J., {Smol{\v{c}}i{\'c}}, V., {Delvecchio}, I., {et~al.} 2017, \aap, 602, A4

\bibitem[{{Delvecchio} {et~al.}(2021){Delvecchio}, {Daddi}, {Sargent}, {Jarvis}, {Elbaz}, {Jin}, {Liu}, {Whittam}, {Algera}, {Carraro}, {D'Eugenio}, {Delhaize}, {Kalita}, {Leslie}, {Moln{\'a}r}, {Novak}, {Prandoni}, {Smol{\v{c}}i{\'c}}, {Ao}, {Aravena}, {Bournaud}, {Collier}, {Randriamampandry}, {Randriamanakoto}, {Rodighiero}, {Schober}, {White}, \& {Zamorani}}]{Delvecchio2021}
{Delvecchio}, I., {Daddi}, E., {Sargent}, M.~T., {et~al.} 2021, \aap, 647, A123

\bibitem[{{Driver} {et~al.}(2022){Driver}, {Bellstedt}, {Robotham}, {Baldry}, {Davies}, {Liske}, {Obreschkow}, {Taylor}, {Wright}, {Alpaslan}, {Bamford}, {Bauer}, {Bland-Hawthorn}, {Bilicki}, {Bravo}, {Brough}, {Casura}, {Cluver}, {Colless}, {Conselice}, {Croom}, {de Jong}, {D'Eugenio}, {De Propris}, {Dogruel}, {Drinkwater}, {Dvornik}, {Farrow}, {Frenk}, {Giblin}, {Graham}, {Grootes}, {Gunawardhana}, {Hashemizadeh}, {H{\"a}u{\ss}ler}, {Heymans}, {Hildebrandt}, {Holwerda}, {Hopkins}, {Jarrett}, {Heath Jones}, {Kelvin}, {Koushan}, {Kuijken}, {Lara-L{\'o}pez}, {Lange}, {L{\'o}pez-S{\'a}nchez}, {Loveday}, {Mahajan}, {Meyer}, {Moffett}, {Napolitano}, {Norberg}, {Owers}, {Radovich}, {Raouf}, {Peacock}, {Phillipps}, {Pimbblet}, {Popescu}, {Said}, {Sansom}, {Seibert}, {Sutherland}, {Thorne}, {Tuffs}, {Turner}, {van der Wel}, {van Kampen}, \& {Wilkins}}]{Driver2022}
{Driver}, S.~P., {Bellstedt}, S., {Robotham}, A. S.~G., {et~al.} 2022, \mnras, 513, 439

\bibitem[{{Driver} {et~al.}(2011){Driver}, {Hill}, {Kelvin}, {Robotham}, {Liske}, {Norberg}, {Baldry}, {Bamford}, {Hopkins}, {Loveday}, {Peacock}, {Andrae}, {Bland-Hawthorn}, {Brough}, {Brown}, {Cameron}, {Ching}, {Colless}, {Conselice}, {Croom}, {Cross}, {de Propris}, {Dye}, {Drinkwater}, {Ellis}, {Graham}, {Grootes}, {Gunawardhana}, {Jones}, {van Kampen}, {Maraston}, {Nichol}, {Parkinson}, {Phillipps}, {Pimbblet}, {Popescu}, {Prescott}, {Roseboom}, {Sadler}, {Sansom}, {Sharp}, {Smith}, {Taylor}, {Thomas}, {Tuffs}, {Wijesinghe}, {Dunne}, {Frenk}, {Jarvis}, {Madore}, {Meyer}, {Seibert}, {Staveley-Smith}, {Sutherland}, \& {Warren}}]{Driver2011}
{Driver}, S.~P., {Hill}, D.~T., {Kelvin}, L.~S., {et~al.} 2011, \mnras, 413, 971

\bibitem[{{Duchesne} {et~al.}(2024){Duchesne}, {Grundy}, {Heald}, {Lenc}, {Leung}, {McConnell}, {Murphy}, {Pritchard}, {Rose}, {Thomson}, {Wang}, {Wang}, \& {Whiting}}]{Duchesne24}
{Duchesne}, S.~W., {Grundy}, J.~A., {Heald}, G.~H., {et~al.} 2024, \pasa, 41, e003

\bibitem[{{Duncan} {et~al.}(2021){Duncan}, {Kondapally}, {Brown}, {Bonato}, {Best}, {R{\"o}ttgering}, {Bondi}, {Bowler}, {Cochrane}, {G{\"u}rkan}, {Hardcastle}, {Jarvis}, {Kunert-Bajraszewska}, {Leslie}, {Ma{\l}ek}, {Morabito}, {O'Sullivan}, {Prandoni}, {Sabater}, {Shimwell}, {Smith}, {Wang}, {Wo{\l}owska}, \& {Tasse}}]{Duncan2021}
{Duncan}, K.~J., {Kondapally}, R., {Brown}, M.~J.~I., {et~al.} 2021, \aap, 648, A4

\bibitem[{{Dunlop} {et~al.}(2017){Dunlop}, {McLure}, {Biggs}, {Geach}, {Micha{\l}owski}, {Ivison}, {Rujopakarn}, {van Kampen}, {Kirkpatrick}, {Pope}, {Scott}, {Swinbank}, {Targett}, {Aretxaga}, {Austermann}, {Best}, {Bruce}, {Chapin}, {Charlot}, {Cirasuolo}, {Coppin}, {Ellis}, {Finkelstein}, {Hayward}, {Hughes}, {Ibar}, {Jagannathan}, {Khochfar}, {Koprowski}, {Narayanan}, {Nyland}, {Papovich}, {Peacock}, {Rieke}, {Robertson}, {Vernstrom}, {Werf}, {Wilson}, \& {Yun}}]{Dunlop2017}
{Dunlop}, J.~S., {McLure}, R.~J., {Biggs}, A.~D., {et~al.} 2017, \mnras, 466, 861

\bibitem[{{Enia} {et~al.}(2022){Enia}, {Talia}, {Pozzi}, {Cimatti}, {Delvecchio}, {Zamorani}, {D'Amato}, {Bisigello}, {Gruppioni}, {Rodighiero}, {Calura}, {Dallacasa}, {Giulietti}, {Barchiesi}, {Behiri}, \& {Romano}}]{Enia2022ApJ...927..204E}
{Enia}, A., {Talia}, M., {Pozzi}, F., {et~al.} 2022, \apj, 927, 204

\bibitem[{{Euclid Collaboration}:~Scaramella {et~al.}(2022){Euclid Collaboration}:~Scaramella, {Amiaux}, {Mellier}, {Burigana}, {Carvalho}, {Cuillandre}, {Da Silva}, {Derosa}, {Dinis}, {Maiorano}, {Maris}, {Tereno}, {Laureijs}, {Boenke}, {Buenadicha}, {Dupac}, {Gaspar Venancio}, {G{\'o}mez-{\'A}lvarez}, {Hoar}, {Lorenzo Alvarez}, {Racca}, {Saavedra-Criado}, {Schwartz}, {Vavrek}, {Schirmer}, {Aussel}, {Azzollini}, {Cardone}, {Cropper}, {Ealet}, {Garilli}, {Gillard}, {Granett}, {Guzzo}, {Hoekstra}, {Jahnke}, {Kitching}, {Maciaszek}, {Meneghetti}, {Miller}, {Nakajima}, {Niemi}, {Pasian}, {Percival}, {Pottinger}, {Sauvage}, {Scodeggio}, {Wachter}, {Zacchei}, {Aghanim}, {Amara}, {Auphan}, {Auricchio}, {Awan}, {Balestra}, {Bender}, {Bodendorf}, {Bonino}, {Branchini}, {Brau-Nogue}, {Brescia}, {Candini}, {Capobianco}, {Carbone}, {Carlberg}, {Carretero}, {Casas}, {Castander}, {Castellano}, {Cavuoti}, {Cimatti}, {Cledassou}, {Congedo}, {Conselice}, {Conversi}, {Copin}, {Corcione}, {Costille}, {Courbin}, {Degaudenzi},
  {Douspis}, {Dubath}, {Duncan}, {Dusini}, {Farrens}, {Ferriol}, {Fosalba}, {Fourmanoit}, {Frailis}, {Franceschi}, {Franzetti}, {Fumana}, {Gillis}, {Giocoli}, {Grazian}, {Grupp}, {Haugan}, {Holmes}, {Hormuth}, {Hudelot}, {Kermiche}, {Kiessling}, {Kilbinger}, {Kohley}, {Kubik}, {K{\"u}mmel}, {Kunz}, {Kurki-Suonio}, {Lahav}, {Ligori}, {Lilje}, {Lloro}, {Mansutti}, {Marggraf}, {Markovic}, {Marulli}, {Massey}, {Maurogordato}, {Melchior}, {Merlin}, {Meylan}, {Mohr}, {Moresco}, {Morin}, {Moscardini}, {Munari}, {Nichol}, {Padilla}, {Paltani}, {Peacock}, {Pedersen}, {Pettorino}, {Pires}, {Poncet}, {Popa}, {Pozzetti}, {Raison}, {Rebolo}, {Rhodes}, {Rix}, {Roncarelli}, {Rossetti}, {Saglia}, {Schneider}, {Schrabback}, {Secroun}, {Seidel}, {Serrano}, {Sirignano}, {Sirri}, {Skottfelt}, {Stanco}, {Starck}, {Tallada-Cresp{\'\i}}, {Tavagnacco}, {Taylor}, {Teplitz}, {Toledo-Moreo}, {Torradeflot}, {Trifoglio}, {Valentijn}, {Valenziano}, {Verdoes Kleijn}, {Wang}, {Welikala}, {Weller}, {Wetzstein}, {Zamorani}, {Zoubian},
  {Andreon}, {Baldi}, {Bardelli}, {Boucaud}, {Camera}, {Di Ferdinando}, {Fabbian}, {Farinelli}, {Galeotta}, {Graci{\'a}-Carpio}, {Maino}, {Medinaceli}, {Mei}, {Neissner}, {Polenta}, {Renzi}, {Romelli}, {Rosset}, {Sureau}, {Tenti}, {Vassallo}, {Zucca}, {Baccigalupi}, {Balaguera-Antol{\'\i}nez}, {Battaglia}, {Biviano}, {Borgani}, {Bozzo}, {Cabanac}, {Cappi}, {Casas}, {Castignani}, {Colodro-Conde}, {Coupon}, {Courtois}, {Cuby}, {de la Torre}, {Desai}, {Dole}, {Fabricius}, {Farina}, {Ferreira}, {Finelli}, {Flose-Reimberg}, {Fotopoulou}, {Ganga}, {Gozaliasl}, {Hook}, {Keihanen}, {Kirkpatrick}, {Liebing}, {Lindholm}, {Mainetti}, {Martinelli}, {Martinet}, {Maturi}, {McCracken}, {Metcalf}, {Morgante}, {Nightingale}, {Nucita}, {Patrizii}, {Potter}, {Riccio}, {S{\'a}nchez}, {Sapone}, {Schewtschenko}, {Schultheis}, {Scottez}, {Teyssier}, {Tutusaus}, {Valiviita}, {Viel}, {Vriend}, \& {Whittaker}}]{EuclidColl2022A&A...662A.112E}
{Euclid Collaboration}:~Scaramella, R., {Amiaux}, J., {Mellier}, Y., {et~al.} 2022, \aap, 662, A112

\bibitem[{{Fitt} {et~al.}(1988){Fitt}, {Alexander}, \& {Cox}}]{Fitt1988MNRAS.233..907F}
{Fitt}, A.~J., {Alexander}, P., \& {Cox}, M.~J. 1988, \mnras, 233, 907

\bibitem[{{Foreman-Mackey} {et~al.}(2013){Foreman-Mackey}, {Hogg}, {Lang}, \& {Goodman}}]{Foreman2013}
{Foreman-Mackey}, D., {Hogg}, D.~W., {Lang}, D., \& {Goodman}, J. 2013, \pasp, 125, 306

\bibitem[{{Franco} {et~al.}(2018){Franco}, {Elbaz}, {B{\'e}thermin}, {Magnelli}, {Schreiber}, {Ciesla}, {Dickinson}, {Nagar}, {Silverman}, {Daddi}, {Alexander}, {Wang}, {Pannella}, {Le Floc'h}, {Pope}, {Giavalisco}, {Maury}, {Bournaud}, {Chary}, {Demarco}, {Ferguson}, {Finkelstein}, {Inami}, {Iono}, {Juneau}, {Lagache}, {Leiton}, {Lin}, {Magdis}, {Messias}, {Motohara}, {Mullaney}, {Okumura}, {Papovich}, {Pforr}, {Rujopakarn}, {Sargent}, {Shu}, \& {Zhou}}]{Franco2018}
{Franco}, M., {Elbaz}, D., {B{\'e}thermin}, M., {et~al.} 2018, \aap, 620, A152

\bibitem[{{Fu} {et~al.}(2022){Fu}, {Shankar}, {Ayromlou}, {Dickson}, {Koutsouridou}, {Rosas-Guevara}, {Marsden}, {Brocklebank}, {Bernardi}, {Shiamtanis}, {Williams}, {Zanisi}, {Allevato}, {Boco}, {Bonoli}, {Cattaneo}, {Dimauro}, {Jiang}, {Lapi}, {Menci}, {Petropoulou}, \& {Villforth}}]{Fu2022}
{Fu}, H., {Shankar}, F., {Ayromlou}, M., {et~al.} 2022, \mnras, 516, 3206

\bibitem[{{Fudamoto} {et~al.}(2021){Fudamoto}, {Oesch}, {Schouws}, {Stefanon}, {Smit}, {Bouwens}, {Bowler}, {Endsley}, {Gonzalez}, {Inami}, {Labbe}, {Stark}, {Aravena}, {Barrufet}, {da Cunha}, {Dayal}, {Ferrara}, {Graziani}, {Hodge}, {Hutter}, {Li}, {De Looze}, {Nanayakkara}, {Pallottini}, {Riechers}, {Schneider}, {Ucci}, {van der Werf}, \& {White}}]{Fudamoto2021}
{Fudamoto}, Y., {Oesch}, P.~A., {Schouws}, S., {et~al.} 2021, \nat, 597, 489

\bibitem[{{Gentile} {et~al.}(2024){Gentile}, {Talia}, {Behiri}, {Zamorani}, {Barchiesi}, {Vignali}, {Pozzi}, {Bethermin}, {Enia}, {Faisst}, {Giulietti}, {Gruppioni}, {Lapi}, {Massardi}, {Smol{\v{c}}i{\'c}}, {Vaccari}, \& {Cimatti}}]{Gentile2024a}
{Gentile}, F., {Talia}, M., {Behiri}, M., {et~al.} 2024, \apj, 962, 26

\bibitem[{Gruppioni {et~al.}(2020)Gruppioni, B{\'e}thermin, Loiacono, Le~F{\`e}vre, Capak, Cassata, Faisst, Schaerer, Silverman, Yan, Bardelli, Boquien, Carraro, Cimatti, Dessauges-Zavadsky, Ginolfi, Fujimoto, Hathi, Jones, Khusanova, Koekemoer, Lagache, Lemaux, Oesch, Pozzi, Riechers, Rodighiero, Romano, Talia, Vallini, Vergani, Zamorani, \& Zucca}]{Gruppioni2020}
Gruppioni, C., B{\'e}thermin, M., Loiacono, F., {et~al.} 2020, \aap, 643, A8

\bibitem[{{Grylls} {et~al.}(2019){Grylls}, {Shankar}, {Zanisi}, \& {Bernardi}}]{Grylls2019}
{Grylls}, P.~J., {Shankar}, F., {Zanisi}, L., \& {Bernardi}, M. 2019, \mnras, 483, 2506

\bibitem[{{G{\"u}rkan} {et~al.}(2018){G{\"u}rkan}, {Hardcastle}, {Smith}, {Best}, {Bourne}, {Calistro-Rivera}, {Heald}, {Jarvis}, {Prandoni}, {R{\"o}ttgering}, {Sabater}, {Shimwell}, {Tasse}, \& {Williams}}]{Gurkan2018}
{G{\"u}rkan}, G., {Hardcastle}, M.~J., {Smith}, D.~J.~B., {et~al.} 2018, \mnras, 475, 3010

\bibitem[{{Helfand} {et~al.}(2015){Helfand}, {White}, \& {Becker}}]{Helfand15}
{Helfand}, D.~J., {White}, R.~L., \& {Becker}, R.~H. 2015, \apj, 801, 26

\bibitem[{{Helou}(1986)}]{Helou1986ApJ...311L..33H}
{Helou}, G. 1986, \apjl, 311, L33

\bibitem[{{Helou} {et~al.}(1985){Helou}, {Soifer}, \& {Rowan-Robinson}}]{Helou1985}
{Helou}, G., {Soifer}, B.~T., \& {Rowan-Robinson}, M. 1985, \apjl, 298, L7

\bibitem[{{Henriques} {et~al.}(2020){Henriques}, {Yates}, {Fu}, {Guo}, {Kauffmann}, {Srisawat}, {Thomas}, \& {White}}]{Henriques2020MNRAS.491.5795H}
{Henriques}, B. M.~B., {Yates}, R.~M., {Fu}, J., {et~al.} 2020, \mnras, 491, 5795

\bibitem[{{Heywood} {et~al.}(2022){Heywood}, {Jarvis}, {Hale}, {Whittam}, {Bester}, {Hugo}, {Kenyon}, {Prescott}, {Smirnov}, {Tasse}, {Afonso}, {Best}, {Collier}, {Deane}, {Frank}, {Hardcastle}, {Knowles}, {Maddox}, {Murphy}, {Prandoni}, {Randriamampandry}, {Santos}, {Sekhar}, {Tabatabaei}, {Taylor}, \& {Thorat}}]{Heywood2022}
{Heywood}, I., {Jarvis}, M.~J., {Hale}, C.~L., {et~al.} 2022, \mnras, 509, 2150

\bibitem[{{Hurley} {et~al.}(2017){Hurley}, {Oliver}, {Betancourt}, {Clarke}, {Cowley}, {Duivenvoorden}, {Farrah}, {Griffin}, {Lacey}, {Le Floc'h}, {Papadopoulos}, {Sargent}, {Scudder}, {Vaccari}, {Valtchanov}, \& {Wang}}]{Hurley2017MNRAS.464..885H}
{Hurley}, P.~D., {Oliver}, S., {Betancourt}, M., {et~al.} 2017, \mnras, 464, 885

\bibitem[{{Ilbert} {et~al.}(2015){Ilbert}, {Arnouts}, {Le Floc'h}, {Aussel}, {Bethermin}, {Capak}, {Hsieh}, {Kajisawa}, {Karim}, {Le F{\`e}vre}, {Lee}, {Lilly}, {McCracken}, {Michel-Dansac}, {Moutard}, {Renzini}, {Salvato}, {Sanders}, {Scoville}, {Sheth}, {Silverman}, {Smol{\v{c}}i{\'c}}, {Taniguchi}, \& {Tresse}}]{Ilbert2015}
{Ilbert}, O., {Arnouts}, S., {Le Floc'h}, E., {et~al.} 2015, \aap, 579, A2

\bibitem[{{Ivison} {et~al.}(2010{\natexlab{a}}){Ivison}, {Alexander}, {Biggs}, {Brandt}, {Chapin}, {Coppin}, {Devlin}, {Dickinson}, {Dunlop}, {Dye}, {Eales}, {Frayer}, {Halpern}, {Hughes}, {Ibar}, {Kov{\'a}cs}, {Marsden}, {Moncelsi}, {Netterfield}, {Pascale}, {Patanchon}, {Rafferty}, {Rex}, {Schinnerer}, {Scott}, {Semisch}, {Smail}, {Swinbank}, {Truch}, {Tucker}, {Viero}, {Walter}, {Wei{\ss}}, {Wiebe}, \& {Xue}}]{Ivison2010a}
{Ivison}, R.~J., {Alexander}, D.~M., {Biggs}, A.~D., {et~al.} 2010{\natexlab{a}}, \mnras, 402, 245

\bibitem[{{Ivison} {et~al.}(2010{\natexlab{b}}){Ivison}, {Magnelli}, {Ibar}, {Andreani}, {Elbaz}, {Altieri}, {Amblard}, {Arumugam}, {Auld}, {Aussel}, {Babbedge}, {Berta}, {Blain}, {Bock}, {Bongiovanni}, {Boselli}, {Buat}, {Burgarella}, {Castro-Rodr{\'\i}guez}, {Cava}, {Cepa}, {Chanial}, {Cimatti}, {Cirasuolo}, {Clements}, {Conley}, {Conversi}, {Cooray}, {Daddi}, {Dominguez}, {Dowell}, {Dwek}, {Eales}, {Farrah}, {F{\"o}rster Schreiber}, {Fox}, {Franceschini}, {Gear}, {Genzel}, {Glenn}, {Griffin}, {Gruppioni}, {Halpern}, {Hatziminaoglou}, {Isaak}, {Lagache}, {Levenson}, {Lu}, {Lutz}, {Madden}, {Maffei}, {Magdis}, {Mainetti}, {Maiolino}, {Marchetti}, {Morrison}, {Mortier}, {Nguyen}, {Nordon}, {O'Halloran}, {Oliver}, {Omont}, {Owen}, {Page}, {Panuzzo}, {Papageorgiou}, {Pearson}, {P{\'e}rez-Fournon}, {P{\'e}rez Garc{\'\i}a}, {Poglitsch}, {Pohlen}, {Popesso}, {Pozzi}, {Rawlings}, {Raymond}, {Rigopoulou}, {Riguccini}, {Rizzo}, {Rodighiero}, {Roseboom}, {Rowan-Robinson}, {Saintonge}, {Sanchez Portal}, {Santini},
  {Schulz}, {Scott}, {Seymour}, {Shao}, {Shupe}, {Smith}, {Stevens}, {Sturm}, {Symeonidis}, {Tacconi}, {Trichas}, {Tugwell}, {Vaccari}, {Valtchanov}, {Vieira}, {Vigroux}, {Wang}, {Ward}, {Wright}, {Xu}, \& {Zemcov}}]{Ivison2010b}
{Ivison}, R.~J., {Magnelli}, B., {Ibar}, E., {et~al.} 2010{\natexlab{b}}, \aap, 518, L31

\bibitem[{{Jannuzi} \& {Dey}(1999)}]{Jannuzi1999}
{Jannuzi}, B.~T. \& {Dey}, A. 1999, in Astronomical Society of the Pacific Conference Series, Vol. 191, Photometric Redshifts and the Detection of High Redshift Galaxies, ed. R.~{Weymann}, L.~{Storrie-Lombardi}, M.~{Sawicki}, \& R.~{Brunner}, 111

\bibitem[{{Jarrett} {et~al.}(2000){Jarrett}, {Chester}, {Cutri}, {Schneider}, {Skrutskie}, \& {Huchra}}]{Jarrett2000}
{Jarrett}, T.~H., {Chester}, T., {Cutri}, R., {et~al.} 2000, \aj, 119, 2498

\bibitem[{{Jarvis} {et~al.}(2015){Jarvis}, {Seymour}, {Afonso}, {Best}, {Beswick}, {Heywood}, {Huynh}, {Murphy}, {Prandoni}, {Schinnerer}, {Simpson}, {Vaccari}, \& {White}}]{Jarvis2015aska.confE..68J}
{Jarvis}, M., {Seymour}, N., {Afonso}, J., {et~al.} 2015, in Advancing Astrophysics with the Square Kilometre Array (AASKA14), 68

\bibitem[{{Jarvis} {et~al.}(2016){Jarvis}, {Taylor}, {Agudo}, {Allison}, {Deane}, {Frank}, {Gupta}, {Heywood}, {Maddox}, {McAlpine}, {Santos}, {Scaife}, {Vaccari}, {Zwart}, {Adams}, {Bacon}, {Baker}, {Bassett}, {Best}, {Beswick}, {Blyth}, {Brown}, {Bruggen}, {Cluver}, {Colafrancesco}, {Cotter}, {Cress}, {Dav{\'e}}, {Ferrari}, {Hardcastle}, {Hale}, {Harrison}, {Hatfield}, {Klockner}, {Kolwa}, {Malefahlo}, {Marubini}, {Mauch}, {Moodley}, {Morganti}, {Norris}, {Peters}, {Prandoni}, {Prescott}, {Oliver}, {Oozeer}, {Rottgering}, {Seymour}, {Simpson}, {Smirnov}, \& {Smith}}]{Jarvis2016}
{Jarvis}, M., {Taylor}, R., {Agudo}, I., {et~al.} 2016, in MeerKAT Science: On the Pathway to the SKA, 6

\bibitem[{{Jarvis} {et~al.}(2010){Jarvis}, {Smith}, {Bonfield}, {Hardcastle}, {Falder}, {Stevens}, {Ivison}, {Auld}, {Baes}, {Baldry}, {Bamford}, {Bourne}, {Buttiglione}, {Cava}, {Cooray}, {Dariush}, {de Zotti}, {Dunlop}, {Dunne}, {Dye}, {Eales}, {Fritz}, {Hill}, {Hopwood}, {Hughes}, {Ibar}, {Jones}, {Kelvin}, {Lawrence}, {Leeuw}, {Loveday}, {Maddox}, {Micha{\l}owski}, {Negrello}, {Norberg}, {Pohlen}, {Prescott}, {Rigby}, {Robotham}, {Rodighiero}, {Scott}, {Sharp}, {Temi}, {Thompson}, {van der Werf}, {van Kampen}, {Vlahakis}, \& {White}}]{Jarvis2010}
{Jarvis}, M.~J., {Smith}, D.~J.~B., {Bonfield}, D.~G., {et~al.} 2010, \mnras, 409, 92

\bibitem[{{Jin} {et~al.}(2018){Jin}, {Daddi}, {Liu}, {Smol{\v{c}}i{\'c}}, {Schinnerer}, {Calabr{\`o}}, {Gu}, {Delhaize}, {Delvecchio}, {Gao}, {Salvato}, {Puglisi}, {Dickinson}, {Bertoldi}, {Sargent}, {Novak}, {Magdis}, {Aretxaga}, {Wilson}, \& {Capak}}]{Jin2018}
{Jin}, S., {Daddi}, E., {Liu}, D., {et~al.} 2018, \apj, 864, 56

\bibitem[{{Johnston} {et~al.}(2007){Johnston}, {Bailes}, {Bartel}, {Baugh}, {Bietenholz}, {Blake}, {Braun}, {Brown}, {Chatterjee}, {Darling}, {Deller}, {Dodson}, {Edwards}, {Ekers}, {Ellingsen}, {Feain}, {Gaensler}, {Haverkorn}, {Hobbs}, {Hopkins}, {Jackson}, {James}, {Joncas}, {Kaspi}, {Kilborn}, {Koribalski}, {Kothes}, {Landecker}, {Lenc}, {Lovell}, {Macquart}, {Manchester}, {Matthews}, {McClure-Griffiths}, {Norris}, {Pen}, {Phillips}, {Power}, {Protheroe}, {Sadler}, {Schmidt}, {Stairs}, {Staveley-Smith}, {Stil}, {Taylor}, {Tingay}, {Tzioumis}, {Walker}, {Wall}, \& {Wolleben}}]{Johnston2007PASA...24..174J}
{Johnston}, S., {Bailes}, M., {Bartel}, N., {et~al.} 2007, \pasa, 24, 174

\bibitem[{{Jones} {et~al.}(2004){Jones}, {Saunders}, {Colless}, {Read}, {Parker}, {Watson}, {Campbell}, {Burkey}, {Mauch}, {Moore}, {Hartley}, {Cass}, {James}, {Russell}, {Fiegert}, {Dawe}, {Huchra}, {Jarrett}, {Lahav}, {Lucey}, {Mamon}, {Proust}, {Sadler}, \& {Wakamatsu}}]{Jones2004}
{Jones}, D.~H., {Saunders}, W., {Colless}, M., {et~al.} 2004, \mnras, 355, 747

\bibitem[{{Kennicutt} \& {Evans}(2012)}]{Kennicutt2012AR}
{Kennicutt}, R.~C. \& {Evans}, N.~J. 2012, \araa, 50, 531

\bibitem[{{Klein} {et~al.}(1984){Klein}, {Wielebinski}, \& {Thuan}}]{Klein1984A&A...141..241K}
{Klein}, U., {Wielebinski}, R., \& {Thuan}, T.~X. 1984, \aap, 141, 241

\bibitem[{{Kondapally} {et~al.}(2021){Kondapally}, {Best}, {Hardcastle}, {Nisbet}, {Bonato}, {Sabater}, {Duncan}, {McCheyne}, {Cochrane}, {Bowler}, {Williams}, {Shimwell}, {Tasse}, {Croston}, {Goyal}, {Jamrozy}, {Jarvis}, {Mahatma}, {R{\"o}ttgering}, {Smith}, {Wo{\l}owska}, {Bondi}, {Brienza}, {Brown}, {Br{\"u}ggen}, {Chambers}, {Garrett}, {G{\"u}rkan}, {Huber}, {Kunert-Bajraszewska}, {Magnier}, {Mingo}, {Mostert}, {Nikiel-Wroczy{\'n}ski}, {O'Sullivan}, {Paladino}, {Ploeckinger}, {Prandoni}, {Rosenthal}, {Schwarz}, {Shulevski}, {Wagenveld}, \& {Wang}}]{Kondapally2021}
{Kondapally}, R., {Best}, P.~N., {Hardcastle}, M.~J., {et~al.} 2021, \aap, 648, A3

\bibitem[{{Labb{\'e}} {et~al.}(2023){Labb{\'e}}, {van Dokkum}, {Nelson}, {Bezanson}, {Suess}, {Leja}, {Brammer}, {Whitaker}, {Mathews}, {Stefanon}, \& {Wang}}]{Labbe2023Natur.616..266L}
{Labb{\'e}}, I., {van Dokkum}, P., {Nelson}, E., {et~al.} 2023, \nat, 616, 266

\bibitem[{{Lacey} {et~al.}(2016){Lacey}, {Baugh}, {Frenk}, {Benson}, {Bower}, {Cole}, {Gonzalez-Perez}, {Helly}, {Lagos}, \& {Mitchell}}]{Lacey2016MNRAS.462.3854L}
{Lacey}, C.~G., {Baugh}, C.~M., {Frenk}, C.~S., {et~al.} 2016, \mnras, 462, 3854

\bibitem[{Lacki \& Thompson(2010)}]{Lacki2009}
Lacki, B.~C. \& Thompson, T.~A. 2010, \apj, 717, 196–208

\bibitem[{{Lacy} {et~al.}(2020){Lacy}, {Baum}, {Chandler}, {Chatterjee}, {Clarke}, {Deustua}, {English}, {Farnes}, {Gaensler}, {Gugliucci}, {Hallinan}, {Kent}, {Kimball}, {Law}, {Lazio}, {Marvil}, {Mao}, {Medlin}, {Mooley}, {Murphy}, {Myers}, {Osten}, {Richards}, {Rosolowsky}, {Rudnick}, {Schinzel}, {Sivakoff}, {Sjouwerman}, {Taylor}, {White}, {Wrobel}, {Andernach}, {Beasley}, {Berger}, {Bhatnager}, {Birkinshaw}, {Bower}, {Brandt}, {Brown}, {Burke-Spolaor}, {Butler}, {Comerford}, {Demorest}, {Fu}, {Giacintucci}, {Golap}, {G{\"u}th}, {Hales}, {Hiriart}, {Hodge}, {Horesh}, {Ivezi{\'c}}, {Jarvis}, {Kamble}, {Kassim}, {Liu}, {Loinard}, {Lyons}, {Masters}, {Mezcua}, {Moellenbrock}, {Mroczkowski}, {Nyland}, {O'Dea}, {O'Sullivan}, {Peters}, {Radford}, {Rao}, {Robnett}, {Salcido}, {Shen}, {Sobotka}, {Witz}, {Vaccari}, {van Weeren}, {Vargas}, {Williams}, \& {Yoon}}]{Lacy2020PASP..132c5001L}
{Lacy}, M., {Baum}, S.~A., {Chandler}, C.~J., {et~al.} 2020, \pasp, 132, 035001

\bibitem[{{Lagos} {et~al.}(2018){Lagos}, {Tobar}, {Robotham}, {Obreschkow}, {Mitchell}, {Power}, \& {Elahi}}]{Lagos2018MNRAS.481.3573L}
{Lagos}, C. d.~P., {Tobar}, R.~J., {Robotham}, A. S.~G., {et~al.} 2018, \mnras, 481, 3573

\bibitem[{{Lapi} {et~al.}(2025){Lapi}, {Boco}, \& {Shankar}}]{Lapi2025}
{Lapi}, A., {Boco}, L., \& {Shankar}, F. 2025, ArXiv e-prints, arXiv:2502.12764

\bibitem[{{Lapi} {et~al.}(2020){Lapi}, {Pantoni}, {Boco}, \& {Danese}}]{Lapi2020}
{Lapi}, A., {Pantoni}, L., {Boco}, L., \& {Danese}, L. 2020, \apj, 897, 81

\bibitem[{{Le Floc'h} {et~al.}(2009){Le Floc'h}, {Aussel}, {Ilbert}, {Riguccini}, {Frayer}, {Salvato}, {Arnouts}, {Surace}, {Feruglio}, {Rodighiero}, {Capak}, {Kartaltepe}, {Heinis}, {Sheth}, {Yan}, {McCracken}, {Thompson}, {Sanders}, {Scoville}, \& {Koekemoer}}]{LeFloch2009}
{Le Floc'h}, E., {Aussel}, H., {Ilbert}, O., {et~al.} 2009, \apj, 703, 222

\bibitem[{{Leja} {et~al.}(2022){Leja}, {Speagle}, {Ting}, {Johnson}, {Conroy}, {Whitaker}, {Nelson}, {van Dokkum}, \& {Franx}}]{Leja2022}
{Leja}, J., {Speagle}, J.~S., {Ting}, Y.-S., {et~al.} 2022, \apj, 936, 165

\bibitem[{{Leslie} {et~al.}(2020){Leslie}, {Schinnerer}, {Liu}, {Magnelli}, {Algera}, {Karim}, {Davidzon}, {Gozaliasl}, {Jim{\'e}nez-Andrade}, {Lang}, {Sargent}, {Novak}, {Groves}, {Smol{\v{c}}i{\'c}}, {Zamorani}, {Vaccari}, {Battisti}, {Vardoulaki}, {Peng}, \& {Kartaltepe}}]{Leslie2020}
{Leslie}, S.~K., {Schinnerer}, E., {Liu}, D., {et~al.} 2020, \apj, 899, 58

\bibitem[{{Lockman} {et~al.}(1986){Lockman}, {Jahoda}, \& {McCammon}}]{Lockman1986}
{Lockman}, F.~J., {Jahoda}, K., \& {McCammon}, D. 1986, \apj, 302, 432

\bibitem[{{Lonsdale Persson} \& {Helou}(1987)}]{Lonsdale1987ApJ...314..513L}
{Lonsdale Persson}, C.~J. \& {Helou}, G. 1987, \apj, 314, 513

\bibitem[{{Lutz} {et~al.}(2011){Lutz}, {Poglitsch}, {Altieri}, {Andreani}, {Aussel}, {Berta}, {Bongiovanni}, {Brisbin}, {Cava}, {Cepa}, {Cimatti}, {Daddi}, {Dominguez-Sanchez}, {Elbaz}, {F{\"o}rster Schreiber}, {Genzel}, {Grazian}, {Gruppioni}, {Harwit}, {Le Floc'h}, {Magdis}, {Magnelli}, {Maiolino}, {Nordon}, {P{\'e}rez Garc{\'\i}a}, {Popesso}, {Pozzi}, {Riguccini}, {Rodighiero}, {Saintonge}, {Sanchez Portal}, {Santini}, {Shao}, {Sturm}, {Tacconi}, {Valtchanov}, {Wetzstein}, \& {Wieprecht}}]{Lutz2011}
{Lutz}, D., {Poglitsch}, A., {Altieri}, B., {et~al.} 2011, \aap, 532, A90

\bibitem[{{Magnelli} {et~al.}(2015){Magnelli}, {Ivison}, {Lutz}, {Valtchanov}, {Farrah}, {Berta}, {Bertoldi}, {Bock}, {Cooray}, {Ibar}, {Karim}, {Le Floc'h}, {Nordon}, {Oliver}, {Page}, {Popesso}, {Pozzi}, {Rigopoulou}, {Riguccini}, {Rodighiero}, {Rosario}, {Roseboom}, {Wang}, \& {Wuyts}}]{Magnelli2015}
{Magnelli}, B., {Ivison}, R.~J., {Lutz}, D., {et~al.} 2015, \aap, 573, A45

\bibitem[{{Mancuso} {et~al.}(2015){Mancuso}, {Lapi}, {Cai}, {Negrello}, {De Zotti}, {Bressan}, {Bonato}, {Perrotta}, \& {Danese}}]{Mancuso2015}
{Mancuso}, C., {Lapi}, A., {Cai}, Z.-Y., {et~al.} 2015, \apj, 810, 72

\bibitem[{{Mancuso} {et~al.}(2017){Mancuso}, {Lapi}, {Prandoni}, {Obi}, {Gonzalez-Nuevo}, {Perrotta}, {Bressan}, {Celotti}, \& {Danese}}]{Mancuso2017}
{Mancuso}, C., {Lapi}, A., {Prandoni}, I., {et~al.} 2017, \apj, 842, 95

\bibitem[{{Mancuso} {et~al.}(2016{\natexlab{a}}){Mancuso}, {Lapi}, {Shi}, {Cai}, {Gonzalez-Nuevo}, {B{\'e}thermin}, \& {Danese}}]{Mancuso2016b}
{Mancuso}, C., {Lapi}, A., {Shi}, J., {et~al.} 2016{\natexlab{a}}, \apj, 833, 152

\bibitem[{{Mancuso} {et~al.}(2016{\natexlab{b}}){Mancuso}, {Lapi}, {Shi}, {Gonzalez-Nuevo}, {Aversa}, \& {Danese}}]{Mancuso2016a}
{Mancuso}, C., {Lapi}, A., {Shi}, J., {et~al.} 2016{\natexlab{b}}, \apj, 823, 128

\bibitem[{{Mannucci} {et~al.}(2010){Mannucci}, {Cresci}, {Maiolino}, {Marconi}, \& {Gnerucci}}]{Mannucci2010MNRAS.408.2115M}
{Mannucci}, F., {Cresci}, G., {Maiolino}, R., {Marconi}, A., \& {Gnerucci}, A. 2010, \mnras, 408, 2115

\bibitem[{{Massardi} {et~al.}(2010){Massardi}, {Bonaldi}, {Negrello}, {Ricciardi}, {Raccanelli}, \& {de Zotti}}]{Massardi2010}
{Massardi}, M., {Bonaldi}, A., {Negrello}, M., {et~al.} 2010, \mnras, 404, 532

\bibitem[{{Mauch} \& {Sadler}(2007)}]{Mauch2007}
{Mauch}, T. \& {Sadler}, E.~M. 2007, \mnras, 375, 931

\bibitem[{{McCheyne} {et~al.}(2022){McCheyne}, {Oliver}, {Sargent}, {Kondapally}, {Smith}, {Haskell}, {Duncan}, {Best}, {Sabater}, {Bonato}, {Calistro Rivera}, {Cochrane}, {Campos Varillas}, {Hurley}, {Leslie}, {Ma{\l}ek}, {Magliocchetti}, {Prandoni}, {Read}, {Rottgering}, {Tasse}, {Vaccari}, \& {Wang}}]{McCheyne2022}
{McCheyne}, I., {Oliver}, S., {Sargent}, M., {et~al.} 2022, \aap, 662, A100

\bibitem[{{McCracken} {et~al.}(2012){McCracken}, {Milvang-Jensen}, {Dunlop}, {Franx}, {Fynbo}, {Le F{\`e}vre}, {Holt}, {Caputi}, {Goranova}, {Buitrago}, {Emerson}, {Freudling}, {Hudelot}, {L{\'o}pez-Sanjuan}, {Magnard}, {Mellier}, {M{\o}ller}, {Nilsson}, {Sutherland}, {Tasca}, \& {Zabl}}]{McCracken2012}
{McCracken}, H.~J., {Milvang-Jensen}, B., {Dunlop}, J., {et~al.} 2012, \aap, 544, A156

\bibitem[{{Moln{\'a}r} {et~al.}(2021){Moln{\'a}r}, {Sargent}, {Leslie}, {Magnelli}, {Schinnerer}, {Zamorani}, {Delhaize}, {Smol{\v{c}}i{\'c}}, {Tisani{\'c}}, \& {Vardoulaki}}]{Molnar2021}
{Moln{\'a}r}, D.~C., {Sargent}, M.~T., {Leslie}, S., {et~al.} 2021, \mnras, 504, 118

\bibitem[{{Moneti} {et~al.}(2023){Moneti}, {McCracken}, {Hudelot}, {Rouberol}, {Herent}, {Mellier}, {Dunlop}, {Le Fevre}, {Franx}, {Fynbo}, {Bowler}, {Caputi}, {Kauffmann}, {Milvang-Jensen}, {Gonzalez-Fernandez}, {Gonzalez-Solares}, {Irwin}, {Lewis}, {Blake}, {Cross}, {Read}, \& {Sutorius}}]{Moneti2023}
{Moneti}, A., {McCracken}, H.~J., {Hudelot}, W., {et~al.} 2023, {VizieR Online Data Catalog: The fourth UltraVISTA data release (DR4) (Moneti+, 2019)}, VizieR On-line Data Catalog: II/373. Originally published in: 2012A\&A...544A.156M

\bibitem[{{Moster} {et~al.}(2013){Moster}, {Naab}, \& {White}}]{Moster2013MNRAS.428.3121M}
{Moster}, B.~P., {Naab}, T., \& {White}, S. D.~M. 2013, \mnras, 428, 3121

\bibitem[{{Moster} {et~al.}(2018){Moster}, {Naab}, \& {White}}]{Moster2018MNRAS.477.1822M}
{Moster}, B.~P., {Naab}, T., \& {White}, S. D.~M. 2018, \mnras, 477, 1822

\bibitem[{{Murphy} {et~al.}(2011){Murphy}, {Condon}, {Schinnerer}, {Kennicutt}, {Calzetti}, {Armus}, {Helou}, {Turner}, {Aniano}, {Beir{\~a}o}, {Bolatto}, {Brandl}, {Croxall}, {Dale}, {Donovan Meyer}, {Draine}, {Engelbracht}, {Hunt}, {Hao}, {Koda}, {Roussel}, {Skibba}, \& {Smith}}]{Murphy2011ApJ...737...67M}
{Murphy}, E.~J., {Condon}, J.~J., {Schinnerer}, E., {et~al.} 2011, \apj, 737, 67

\bibitem[{{Nelson} {et~al.}(2023){Nelson}, {Suess}, {Bezanson}, {Price}, {van Dokkum}, {Leja}, {Wang}, {Whitaker}, {Labb{\'e}}, {Barrufet}, {Brammer}, {Eisenstein}, {Gibson}, {Hartley}, {Johnson}, {Heintz}, {Mathews}, {Miller}, {Oesch}, {Sandles}, {Setton}, {Speagle}, {Tacchella}, {Tadaki}, {{\"U}bler}, \& {Weaver}}]{Nelson2023ApJ...948L..18N}
{Nelson}, E.~J., {Suess}, K.~A., {Bezanson}, R., {et~al.} 2023, \apjl, 948, L18

\bibitem[{{Noeske} {et~al.}(2007){Noeske}, {Weiner}, {Faber}, {Papovich}, {Koo}, {Somerville}, {Bundy}, {Conselice}, {Newman}, {Schiminovich}, {Le Floc'h}, {Coil}, {Rieke}, {Lotz}, {Primack}, {Barmby}, {Cooper}, {Davis}, {Ellis}, {Fazio}, {Guhathakurta}, {Huang}, {Kassin}, {Martin}, {Phillips}, {Rich}, {Small}, {Willmer}, \& {Wilson}}]{Noeske2007ApJ...660L..43N}
{Noeske}, K.~G., {Weiner}, B.~J., {Faber}, S.~M., {et~al.} 2007, \apjl, 660, L43

\bibitem[{{Norris} {et~al.}(2013){Norris}, {Afonso}, {Bacon}, {Beck}, {Bell}, {Beswick}, {Best}, {Bhatnagar}, {Bonafede}, {Brunetti}, {Budav{\'a}ri}, {Cassano}, {Condon}, {Cress}, {Dabbech}, {Feain}, {Fender}, {Ferrari}, {Gaensler}, {Giovannini}, {Haverkorn}, {Heald}, {Van der Heyden}, {Hopkins}, {Jarvis}, {Johnston-Hollitt}, {Kothes}, {Van Langevelde}, {Lazio}, {Mao}, {Mart{\'\i}nez-Sansigre}, {Mary}, {Mcalpine}, {Middelberg}, {Murphy}, {Padovani}, {Paragi}, {Prandoni}, {Raccanelli}, {Rigby}, {Roseboom}, {R{\"o}ttgering}, {Sabater}, {Salvato}, {Scaife}, {Schilizzi}, {Seymour}, {Smith}, {Umana}, {Zhao}, \& {Zinn}}]{Norris2013}
{Norris}, R.~P., {Afonso}, J., {Bacon}, D., {et~al.} 2013, \pasa, 30, e020

\bibitem[{{Norris} {et~al.}(2021){Norris}, {Marvil}, {Collier}, {Kapi{\'n}ska}, {O'Brien}, {Rudnick}, {Andernach}, {Asorey}, {Brown}, {Br{\"u}ggen}, {Crawford}, {English}, {Rahman}, {Filipovi{\'c}}, {Gordon}, {G{\"u}rkan}, {Hale}, {Hopkins}, {Huynh}, {HyeongHan}, {James Jee}, {Koribalski}, {Lenc}, {Luken}, {Parkinson}, {Prandoni}, {Raja}, {Reiprich}, {Riseley}, {Shabala}, {Sheil}, {Vernstrom}, {Whiting}, {Allison}, {Anderson}, {Ball}, {Bell}, {Bunton}, {Galvin}, {Gupta}, {Hotan}, {Jacka}, {Macgregor}, {Mahony}, {Maio}, {Moss}, {Pandey-Pommier}, \& {Voronkov}}]{Norris2021PASA...38...46N}
{Norris}, R.~P., {Marvil}, J., {Collier}, J.~D., {et~al.} 2021, \pasa, 38, e046

\bibitem[{{Novak} {et~al.}(2017){Novak}, {Smol{\v{c}}i{\'c}}, {Delhaize}, {Delvecchio}, {Zamorani}, {Baran}, {Bondi}, {Capak}, {Carilli}, {Ciliegi}, {Civano}, {Ilbert}, {Karim}, {Laigle}, {Le F{\`e}vre}, {Marchesi}, {McCracken}, {Miettinen}, {Salvato}, {Sargent}, {Schinnerer}, \& {Tasca}}]{Novak2017}
{Novak}, M., {Smol{\v{c}}i{\'c}}, V., {Delhaize}, J., {et~al.} 2017, \aap, 602, A5

\bibitem[{{Ocran} {et~al.}(2020{\natexlab{a}}){Ocran}, {Taylor}, {Vaccari}, {Ishwara-Chandra}, \& {Prandoni}}]{Ocran2020a}
{Ocran}, E.~F., {Taylor}, A.~R., {Vaccari}, M., {Ishwara-Chandra}, C.~H., \& {Prandoni}, I. 2020{\natexlab{a}}, \mnras, 491, 1127

\bibitem[{{Ocran} {et~al.}(2020{\natexlab{b}}){Ocran}, {Taylor}, {Vaccari}, {Ishwara-Chandra}, {Prandoni}, {Prescott}, \& {Mancuso}}]{Ocran2020b}
{Ocran}, E.~F., {Taylor}, A.~R., {Vaccari}, M., {et~al.} 2020{\natexlab{b}}, \mnras, 491, 5911

\bibitem[{{Ocran} {et~al.}(2023){Ocran}, {Vaccari}, {Stil}, {Taylor}, {Ishwara-Chandra}, \& {Kim}}]{Ocran2023MNRAS.524.5229O}
{Ocran}, E.~F., {Vaccari}, M., {Stil}, J.~M., {et~al.} 2023, \mnras, 524, 5229

\bibitem[{{Oke}(1974)}]{Oke1974ApJS...27...21O}
{Oke}, J.~B. 1974, \apjs, 27, 21

\bibitem[{{Oliver} {et~al.}(2000){Oliver}, {Rowan-Robinson}, {Alexander}, {Almaini}, {Balcells}, {Baker}, {Barcons}, {Barden}, {Bellas-Velidis}, {Cabrera-Guerra}, {Carballo}, {Cesarsky}, {Ciliegi}, {Clements}, {Crockett}, {Danese}, {Dapergolas}, {Drolias}, {Eaton}, {Efstathiou}, {Egami}, {Elbaz}, {Fadda}, {Fox}, {Franceschini}, {Genzel}, {Goldschmidt}, {Graham}, {Gonzalez-Serrano}, {Gonzalez-Solares}, {Granato}, {Gruppioni}, {Herbstmeier}, {H{\'e}raudeau}, {Joshi}, {Kontizas}, {Kontizas}, {Kotilainen}, {Kunze}, {La Franca}, {Lari}, {Lawrence}, {Lemke}, {Linden-V{\o}rnle}, {Mann}, {M{\'a}rquez}, {Masegosa}, {Mattila}, {McMahon}, {Miley}, {Missoulis}, {Mobasher}, {Morel}, {N{\o}rgaard-Nielsen}, {Omont}, {Papadopoulos}, {Perez-Fournon}, {Puget}, {Rigopoulou}, {Rocca-Volmerange}, {Serjeant}, {Silva}, {Sumner}, {Surace}, {Vaisanen}, {van der Werf}, {Verma}, {Vigroux}, {Villar-Martin}, \& {Willott}}]{Oliver2000}
{Oliver}, S., {Rowan-Robinson}, M., {Alexander}, D.~M., {et~al.} 2000, \mnras, 316, 749

\bibitem[{{Oliver} {et~al.}(2012){Oliver}, {Bock}, {Altieri}, {Amblard}, {Arumugam}, {Aussel}, {Babbedge}, {Beelen}, {B{\'e}thermin}, {Blain}, {Boselli}, {Bridge}, {Brisbin}, {Buat}, {Burgarella}, {Castro-Rodr{\'\i}guez}, {Cava}, {Chanial}, {Cirasuolo}, {Clements}, {Conley}, {Conversi}, {Cooray}, {Dowell}, {Dubois}, {Dwek}, {Dye}, {Eales}, {Elbaz}, {Farrah}, {Feltre}, {Ferrero}, {Fiolet}, {Fox}, {Franceschini}, {Gear}, {Giovannoli}, {Glenn}, {Gong}, {Gonz{\'a}lez Solares}, {Griffin}, {Halpern}, {Harwit}, {Hatziminaoglou}, {Heinis}, {Hurley}, {Hwang}, {Hyde}, {Ibar}, {Ilbert}, {Isaak}, {Ivison}, {Lagache}, {Le Floc'h}, {Levenson}, {Faro}, {Lu}, {Madden}, {Maffei}, {Magdis}, {Mainetti}, {Marchetti}, {Marsden}, {Marshall}, {Mortier}, {Nguyen}, {O'Halloran}, {Omont}, {Page}, {Panuzzo}, {Papageorgiou}, {Patel}, {Pearson}, {P{\'e}rez-Fournon}, {Pohlen}, {Rawlings}, {Raymond}, {Rigopoulou}, {Riguccini}, {Rizzo}, {Rodighiero}, {Roseboom}, {Rowan-Robinson}, {S{\'a}nchez Portal}, {Schulz}, {Scott}, {Seymour}, {Shupe},
  {Smith}, {Stevens}, {Symeonidis}, {Trichas}, {Tugwell}, {Vaccari}, {Valtchanov}, {Vieira}, {Viero}, {Vigroux}, {Wang}, {Ward}, {Wardlow}, {Wright}, {Xu}, \& {Zemcov}}]{Oliver2012}
{Oliver}, S.~J., {Bock}, J., {Altieri}, B., {et~al.} 2012, \mnras, 424, 1614

\bibitem[{{Padovani}(2016)}]{Padovani2016}
{Padovani}, P. 2016, \aapr, 24, 13

\bibitem[{{Padovani} {et~al.}(2015){Padovani}, {Bonzini}, {Kellermann}, {Miller}, {Mainieri}, \& {Tozzi}}]{Padovani2015}
{Padovani}, P., {Bonzini}, M., {Kellermann}, K.~I., {et~al.} 2015, \mnras, 452, 1263

\bibitem[{{Pannella} {et~al.}(2015){Pannella}, {Elbaz}, {Daddi}, {Dickinson}, {Hwang}, {Schreiber}, {Strazzullo}, {Aussel}, {Bethermin}, {Buat}, {Charmandaris}, {Cibinel}, {Juneau}, {Ivison}, {Le Borgne}, {Le Floc'h}, {Leiton}, {Lin}, {Magdis}, {Morrison}, {Mullaney}, {Onodera}, {Renzini}, {Salim}, {Sargent}, {Scott}, {Shu}, \& {Wang}}]{Pannella2015ApJ...807..141P}
{Pannella}, M., {Elbaz}, D., {Daddi}, E., {et~al.} 2015, \apj, 807, 141

\bibitem[{{Pantoni} {et~al.}(2019){Pantoni}, {Lapi}, {Massardi}, {Goswami}, \& {Danese}}]{Pantoni2019}
{Pantoni}, L., {Lapi}, A., {Massardi}, M., {Goswami}, S., \& {Danese}, L. 2019, \apj, 880, 129

\bibitem[{{Parente} {et~al.}(2023){Parente}, {Ragone-Figueroa}, {Granato}, \& {Lapi}}]{Parente2023MNRAS.521.6105P}
{Parente}, M., {Ragone-Figueroa}, C., {Granato}, G.~L., \& {Lapi}, A. 2023, \mnras, 521, 6105

\bibitem[{{Pearson} {et~al.}(2018){Pearson}, {Wang}, {Hurley}, {Ma{\l}ek}, {Buat}, {Burgarella}, {Farrah}, {Oliver}, {Smith}, \& {van der Tak}}]{Pearson2018}
{Pearson}, W.~J., {Wang}, L., {Hurley}, P.~D., {et~al.} 2018, \aap, 615, A146

\bibitem[{{Peng} {et~al.}(2010){Peng}, {Lilly}, {Kova{\v{c}}}, {Bolzonella}, {Pozzetti}, {Renzini}, {Zamorani}, {Ilbert}, {Knobel}, {Iovino}, {Maier}, {Cucciati}, {Tasca}, {Carollo}, {Silverman}, {Kampczyk}, {de Ravel}, {Sanders}, {Scoville}, {Contini}, {Mainieri}, {Scodeggio}, {Kneib}, {Le F{\`e}vre}, {Bardelli}, {Bongiorno}, {Caputi}, {Coppa}, {de la Torre}, {Franzetti}, {Garilli}, {Lamareille}, {Le Borgne}, {Le Brun}, {Mignoli}, {Perez Montero}, {Pello}, {Ricciardelli}, {Tanaka}, {Tresse}, {Vergani}, {Welikala}, {Zucca}, {Oesch}, {Abbas}, {Barnes}, {Bordoloi}, {Bottini}, {Cappi}, {Cassata}, {Cimatti}, {Fumana}, {Hasinger}, {Koekemoer}, {Leauthaud}, {Maccagni}, {Marinoni}, {McCracken}, {Memeo}, {Meneux}, {Nair}, {Porciani}, {Presotto}, \& {Scaramella}}]{Peng2010}
{Peng}, Y.-j., {Lilly}, S.~J., {Kova{\v{c}}}, K., {et~al.} 2010, \apj, 721, 193

\bibitem[{{P{\'e}rez-Gonz{\'a}lez} {et~al.}(2023){P{\'e}rez-Gonz{\'a}lez}, {Barro}, {Annunziatella}, {Costantin}, {Garc{\'\i}a-Argum{\'a}nez}, {McGrath}, {M{\'e}rida}, {Zavala}, {Arrabal Haro}, {Bagley}, {Backhaus}, {Behroozi}, {Bell}, {Bisigello}, {Buat}, {Calabr{\`o}}, {Casey}, {Cleri}, {Coogan}, {Cooper}, {Cooray}, {Dekel}, {Dickinson}, {Elbaz}, {Ferguson}, {Finkelstein}, {Fontana}, {Franco}, {Gardner}, {Giavalisco}, {G{\'o}mez-Guijarro}, {Grazian}, {Grogin}, {Guo}, {Huertas-Company}, {Jogee}, {Kartaltepe}, {Kewley}, {Kirkpatrick}, {Kocevski}, {Koekemoer}, {Long}, {Lotz}, {Lucas}, {Papovich}, {Pirzkal}, {Ravindranath}, {Somerville}, {Tacchella}, {Trump}, {Wang}, {Wilkins}, {Wuyts}, {Yang}, \& {Yung}}]{Perez2023ApJ...946L..16P}
{P{\'e}rez-Gonz{\'a}lez}, P.~G., {Barro}, G., {Annunziatella}, M., {et~al.} 2023, \apjl, 946, L16

\bibitem[{{Popesso} {et~al.}(2023){Popesso}, {Concas}, {Cresci}, {Belli}, {Rodighiero}, {Inami}, {Dickinson}, {Ilbert}, {Pannella}, \& {Elbaz}}]{Popesso2023}
{Popesso}, P., {Concas}, A., {Cresci}, G., {et~al.} 2023, \mnras, 519, 1526

\bibitem[{{Popesso} {et~al.}(2019{\natexlab{a}}){Popesso}, {Concas}, {Morselli}, {Schreiber}, {Rodighiero}, {Cresci}, {Belli}, {Erfanianfar}, {Mancini}, {Inami}, {Dickinson}, {Ilbert}, {Pannella}, \& {Elbaz}}]{Popesso2019a}
{Popesso}, P., {Concas}, A., {Morselli}, L., {et~al.} 2019{\natexlab{a}}, \mnras, 483, 3213

\bibitem[{{Popesso} {et~al.}(2019{\natexlab{b}}){Popesso}, {Morselli}, {Concas}, {Schreiber}, {Rodighiero}, {Cresci}, {Belli}, {Ilbert}, {Erfanianfar}, {Mancini}, {Inami}, {Dickinson}, {Pannella}, \& {Elbaz}}]{Popesso2019b}
{Popesso}, P., {Morselli}, L., {Concas}, A., {et~al.} 2019{\natexlab{b}}, \mnras, 490, 5285

\bibitem[{{Prandoni} {et~al.}(2018){Prandoni}, {Guglielmino}, {Morganti}, {Vaccari}, {Maini}, {R{\"o}ttgering}, {Jarvis}, \& {Garrett}}]{Prandoni2018}
{Prandoni}, I., {Guglielmino}, G., {Morganti}, R., {et~al.} 2018, \mnras, 481, 4548

\bibitem[{{Prandoni} \& {Seymour}(2015)}]{Prandoni2015}
{Prandoni}, I. \& {Seymour}, N. 2015, in Advancing Astrophysics with the Square Kilometre Array (AASKA14), 67

\bibitem[{{Price} \& {Duric}(1992)}]{Price1992ApJ...401...81P}
{Price}, R. \& {Duric}, N. 1992, \apj, 401, 81

\bibitem[{{Read} {et~al.}(2018){Read}, {Smith}, {G{\"u}rkan}, {Hardcastle}, {Williams}, {Best}, {Brinks}, {Calistro-Rivera}, {Chy{\.Z}y}, {Duncan}, {Dunne}, {Jarvis}, {Morabito}, {Prandoni}, {R{\"o}ttgering}, {Sabater}, \& {Viaene}}]{Read2018MNRAS.480.5625R}
{Read}, S.~C., {Smith}, D.~J.~B., {G{\"u}rkan}, G., {et~al.} 2018, \mnras, 480, 5625

\bibitem[{{Renzini} \& {Peng}(2015)}]{Renzini2015}
{Renzini}, A. \& {Peng}, Y.-j. 2015, \apjl, 801, L29

\bibitem[{{Rinaldi} {et~al.}(2025){Rinaldi}, {Navarro-Carrera}, {Caputi}, {Iani}, {{\"O}stlin}, {Colina}, {Alberts}, {{\'A}lvarez-M{\'a}rquez}, {Annunziatella}, {Boogaard}, {Costantin}, {Hjorth}, {Langeroodi}, {Melinder}, {Moutard}, \& {Walter}}]{Rinaldi2025}
{Rinaldi}, P., {Navarro-Carrera}, R., {Caputi}, K.~I., {et~al.} 2025, \apj, 981, 161

\bibitem[{{Rodighiero} {et~al.}(2015){Rodighiero}, {Brusa}, {Daddi}, {Negrello}, {Mullaney}, {Delvecchio}, {Lutz}, {Renzini}, {Franceschini}, {Baronchelli}, {Pozzi}, {Gruppioni}, {Strazzullo}, {Cimatti}, \& {Silverman}}]{Rodighiero2015}
{Rodighiero}, G., {Brusa}, M., {Daddi}, E., {et~al.} 2015, \apjl, 800, L10

\bibitem[{{Rodighiero} {et~al.}(2011){Rodighiero}, {Daddi}, {Baronchelli}, {Cimatti}, {Renzini}, {Aussel}, {Popesso}, {Lutz}, {Andreani}, {Berta}, {Cava}, {Elbaz}, {Feltre}, {Fontana}, {F{\"o}rster Schreiber}, {Franceschini}, {Genzel}, {Grazian}, {Gruppioni}, {Ilbert}, {Le Floch}, {Magdis}, {Magliocchetti}, {Magnelli}, {Maiolino}, {McCracken}, {Nordon}, {Poglitsch}, {Santini}, {Pozzi}, {Riguccini}, {Tacconi}, {Wuyts}, \& {Zamorani}}]{Rodighiero2011}
{Rodighiero}, G., {Daddi}, E., {Baronchelli}, I., {et~al.} 2011, \apjl, 739, L40

\bibitem[{{Rodighiero} {et~al.}(2014){Rodighiero}, {Renzini}, {Daddi}, {Baronchelli}, {Berta}, {Cresci}, {Franceschini}, {Gruppioni}, {Lutz}, {Mancini}, {Santini}, {Zamorani}, {Silverman}, {Kashino}, {Andreani}, {Cimatti}, {S{\'a}nchez}, {Le Floch}, {Magnelli}, {Popesso}, \& {Pozzi}}]{Rodighiero2014}
{Rodighiero}, G., {Renzini}, A., {Daddi}, E., {et~al.} 2014, \mnras, 443, 19

\bibitem[{{Sabater} {et~al.}(2019){Sabater}, {Best}, {Hardcastle}, {Shimwell}, {Tasse}, {Williams}, {Br{\"u}ggen}, {Cochrane}, {Croston}, {de Gasperin}, {Duncan}, {G{\"u}rkan}, {Mechev}, {Morabito}, {Prandoni}, {R{\"o}ttgering}, {Smith}, {Harwood}, {Mingo}, {Mooney}, \& {Saxena}}]{Sabater2019A&A...622A..17S}
{Sabater}, J., {Best}, P.~N., {Hardcastle}, M.~J., {et~al.} 2019, \aap, 622, A17

\bibitem[{{Sabater} {et~al.}(2021){Sabater}, {Best}, {Tasse}, {Hardcastle}, {Shimwell}, {Nisbet}, {Jelic}, {Callingham}, {R{\"o}ttgering}, {Bonato}, {Bondi}, {Ciardi}, {Cochrane}, {Jarvis}, {Kondapally}, {Koopmans}, {O'Sullivan}, {Prandoni}, {Schwarz}, {Smith}, {Wang}, {Williams}, \& {Zaroubi}}]{Sabater2021}
{Sabater}, J., {Best}, P.~N., {Tasse}, C., {et~al.} 2021, \aap, 648, A2

\bibitem[{{Sargent} {et~al.}(2012){Sargent}, {B{\'e}thermin}, {Daddi}, \& {Elbaz}}]{Sargent2012}
{Sargent}, M.~T., {B{\'e}thermin}, M., {Daddi}, E., \& {Elbaz}, D. 2012, \apjl, 747, L31

\bibitem[{{Sargent} {et~al.}(2010){Sargent}, {Schinnerer}, {Murphy}, {Carilli}, {Helou}, {Aussel}, {Le Floc'h}, {Frayer}, {Ilbert}, {Oesch}, {Salvato}, {Smol{\v{c}}i{\'c}}, {Kartaltepe}, \& {Sanders}}]{Sargent2010}
{Sargent}, M.~T., {Schinnerer}, E., {Murphy}, E., {et~al.} 2010, \apjl, 714, L190

\bibitem[{{Schleicher} \& {Beck}(2016)}]{Schleicher2016}
{Schleicher}, D. R.~G. \& {Beck}, R. 2016, \aap, 593, A77

\bibitem[{{Schreiber} {et~al.}(2015){Schreiber}, {Pannella}, {Elbaz}, {B{\'e}thermin}, {Inami}, {Dickinson}, {Magnelli}, {Wang}, {Aussel}, {Daddi}, {Juneau}, {Shu}, {Sargent}, {Buat}, {Faber}, {Ferguson}, {Giavalisco}, {Koekemoer}, {Magdis}, {Morrison}, {Papovich}, {Santini}, \& {Scott}}]{Schreiber2015}
{Schreiber}, C., {Pannella}, M., {Elbaz}, D., {et~al.} 2015, \aap, 575, A74

\bibitem[{{Shimwell} {et~al.}(2022){Shimwell}, {Hardcastle}, {Tasse}, {Best}, {R{\"o}ttgering}, {Williams}, {Botteon}, {Drabent}, {Mechev}, {Shulevski}, {van Weeren}, {Bester}, {Br{\"u}ggen}, {Brunetti}, {Callingham}, {Chy{\.z}y}, {Conway}, {Dijkema}, {Duncan}, {de Gasperin}, {Hale}, {Haverkorn}, {Hugo}, {Jackson}, {Mevius}, {Miley}, {Morabito}, {Morganti}, {Offringa}, {Oonk}, {Rafferty}, {Sabater}, {Smith}, {Schwarz}, {Smirnov}, {O'Sullivan}, {Vedantham}, {White}, {Albert}, {Alegre}, {Asabere}, {Bacon}, {Bonafede}, {Bonnassieux}, {Brienza}, {Bilicki}, {Bonato}, {Calistro Rivera}, {Cassano}, {Cochrane}, {Croston}, {Cuciti}, {Dallacasa}, {Danezi}, {Dettmar}, {Di Gennaro}, {Edler}, {En{\ss}lin}, {Emig}, {Franzen}, {Garc{\'\i}a-Vergara}, {Grange}, {G{\"u}rkan}, {Hajduk}, {Heald}, {Heesen}, {Hoang}, {Hoeft}, {Horellou}, {Iacobelli}, {Jamrozy}, {Jeli{\'c}}, {Kondapally}, {Kukreti}, {Kunert-Bajraszewska}, {Magliocchetti}, {Mahatma}, {Ma{\l}ek}, {Mandal}, {Massaro}, {Meyer-Zhao}, {Mingo}, {Mostert}, {Nair},
  {Nakoneczny}, {Nikiel-Wroczy{\'n}ski}, {Orr{\'u}}, {Pajdosz-{\'S}mierciak}, {Pasini}, {Prandoni}, {van Piggelen}, {Rajpurohit}, {Retana-Montenegro}, {Riseley}, {Rowlinson}, {Saxena}, {Schrijvers}, {Sweijen}, {Siewert}, {Timmerman}, {Vaccari}, {Vink}, {West}, {Wo{\l}owska}, {Zhang}, \& {Zheng}}]{Shimwell2022}
{Shimwell}, T.~W., {Hardcastle}, M.~J., {Tasse}, C., {et~al.} 2022, \aap, 659, A1

\bibitem[{{Shimwell} {et~al.}(2017){Shimwell}, {R{\"o}ttgering}, {Best}, {Williams}, {Dijkema}, {de Gasperin}, {Hardcastle}, {Heald}, {Hoang}, {Horneffer}, {Intema}, {Mahony}, {Mandal}, {Mechev}, {Morabito}, {Oonk}, {Rafferty}, {Retana-Montenegro}, {Sabater}, {Tasse}, {van Weeren}, {Br{\"u}ggen}, {Brunetti}, {Chy{\.z}y}, {Conway}, {Haverkorn}, {Jackson}, {Jarvis}, {McKean}, {Miley}, {Morganti}, {White}, {Wise}, {van Bemmel}, {Beck}, {Brienza}, {Bonafede}, {Calistro Rivera}, {Cassano}, {Clarke}, {Cseh}, {Deller}, {Drabent}, {van Driel}, {Engels}, {Falcke}, {Ferrari}, {Fr{\"o}hlich}, {Garrett}, {Harwood}, {Heesen}, {Hoeft}, {Horellou}, {Israel}, {Kapi{\'n}ska}, {Kunert-Bajraszewska}, {McKay}, {Mohan}, {Orr{\'u}}, {Pizzo}, {Prandoni}, {Schwarz}, {Shulevski}, {Sipior}, {Smith}, {Sridhar}, {Steinmetz}, {Stroe}, {Varenius}, {van der Werf}, {Zensus}, \& {Zwart}}]{Shimwell2017}
{Shimwell}, T.~W., {R{\"o}ttgering}, H.~J.~A., {Best}, P.~N., {et~al.} 2017, \aap, 598, A104

\bibitem[{{Shimwell} {et~al.}(2019){Shimwell}, {Tasse}, {Hardcastle}, {Mechev}, {Williams}, {Best}, {R{\"o}ttgering}, {Callingham}, {Dijkema}, {de Gasperin}, {Hoang}, {Hugo}, {Mirmont}, {Oonk}, {Prandoni}, {Rafferty}, {Sabater}, {Smirnov}, {van Weeren}, {White}, {Atemkeng}, {Bester}, {Bonnassieux}, {Br{\"u}ggen}, {Brunetti}, {Chy{\.z}y}, {Cochrane}, {Conway}, {Croston}, {Danezi}, {Duncan}, {Haverkorn}, {Heald}, {Iacobelli}, {Intema}, {Jackson}, {Jamrozy}, {Jarvis}, {Lakhoo}, {Mevius}, {Miley}, {Morabito}, {Morganti}, {Nisbet}, {Orr{\'u}}, {Perkins}, {Pizzo}, {Schrijvers}, {Smith}, {Vermeulen}, {Wise}, {Alegre}, {Bacon}, {van Bemmel}, {Beswick}, {Bonafede}, {Botteon}, {Bourke}, {Brienza}, {Calistro Rivera}, {Cassano}, {Clarke}, {Conselice}, {Dettmar}, {Drabent}, {Dumba}, {Emig}, {En{\ss}lin}, {Ferrari}, {Garrett}, {G{\'e}nova-Santos}, {Goyal}, {G{\"u}rkan}, {Hale}, {Harwood}, {Heesen}, {Hoeft}, {Horellou}, {Jackson}, {Kokotanekov}, {Kondapally}, {Kunert-Bajraszewska}, {Mahatma}, {Mahony}, {Mandal}, {McKean},
  {Merloni}, {Mingo}, {Miskolczi}, {Mooney}, {Nikiel-Wroczy{\'n}ski}, {O'Sullivan}, {Quinn}, {Reich}, {Roskowi{\'n}ski}, {Rowlinson}, {Savini}, {Saxena}, {Schwarz}, {Shulevski}, {Sridhar}, {Stacey}, {Urquhart}, {van der Wiel}, {Varenius}, {Webster}, \& {Wilber}}]{Shimwell2019}
{Shimwell}, T.~W., {Tasse}, C., {Hardcastle}, M.~J., {et~al.} 2019, \aap, 622, A1

\bibitem[{{Shu} {et~al.}(2022){Shu}, {Yang}, {Liu}, {Wang}, {Wang}, {Han}, {Huang}, {Lim}, {Chang}, {Zheng}, {Zheng}, {Wang}, \& {Kong}}]{Shu2022}
{Shu}, X., {Yang}, L., {Liu}, D., {et~al.} 2022, \apj, 926, 155

\bibitem[{{Shuntov} {et~al.}(2025){Shuntov}, {Ilbert}, {Toft}, {Arango-Toro}, {Akins}, {Casey}, {Franco}, {Harish}, {Kartaltepe}, {Koekemoer}, {McCracken}, {Paquereau}, {Laigle}, {Bethermin}, {Dubois}, {Drakos}, {Faisst}, {Gozaliasl}, {Gillman}, {Hayward}, {Hirschmann}, {Huertas-Company}, {Jespersen}, {Jin}, {Kokorev}, {Lambrides}, {Le Borgne}, {Liu}, {Magdis}, {Massey}, {McPartland}, {Mercier}, {McCleary}, {McKinney}, {Oesch}, {Renzini}, {Rhodes}, {Rich}, {Robertson}, {Sanders}, {Trebitsch}, {Tresse}, {Valentino}, {Vijayan}, {Weaver}, {Weibel}, {Wilkins}, \& {Yang}}]{Shuntov2025}
{Shuntov}, M., {Ilbert}, O., {Toft}, S., {et~al.} 2025, \aap, 695, A20

\bibitem[{{Simpson} {et~al.}(2014){Simpson}, {Swinbank}, {Smail}, {Alexander}, {Brandt}, {Bertoldi}, {de Breuck}, {Chapman}, {Coppin}, {da Cunha}, {Danielson}, {Dannerbauer}, {Greve}, {Hodge}, {Ivison}, {Karim}, {Knudsen}, {Poggianti}, {Schinnerer}, {Thomson}, {Walter}, {Wardlow}, {Wei{\ss}}, \& {van der Werf}}]{Simpson2014}
{Simpson}, J.~M., {Swinbank}, A.~M., {Smail}, I., {et~al.} 2014, \apj, 788, 125

\bibitem[{{Sinha} {et~al.}(2022){Sinha}, {Basu}, {Datta}, \& {Chakraborty}}]{Sinha2022}
{Sinha}, A., {Basu}, A., {Datta}, A., \& {Chakraborty}, A. 2022, \mnras, 514, 4343

\bibitem[{{Skrutskie} {et~al.}(2006){Skrutskie}, {Cutri}, {Stiening}, {Weinberg}, {Schneider}, {Carpenter}, {Beichman}, {Capps}, {Chester}, {Elias}, {Huchra}, {Liebert}, {Lonsdale}, {Monet}, {Price}, {Seitzer}, {Jarrett}, {Kirkpatrick}, {Gizis}, {Howard}, {Evans}, {Fowler}, {Fullmer}, {Hurt}, {Light}, {Kopan}, {Marsh}, {McCallon}, {Tam}, {Van Dyk}, \& {Wheelock}}]{Skrutskie2006}
{Skrutskie}, M.~F., {Cutri}, R.~M., {Stiening}, R., {et~al.} 2006, \aj, 131, 1163

\bibitem[{{Smail} {et~al.}(2021){Smail}, {Dudzevi{\v{c}}i{\={u}}t{\.{e}}}, {Stach}, {Almaini}, {Birkin}, {Chapman}, {Chen}, {Geach}, {Gullberg}, {Hodge}, {Ikarashi}, {Ivison}, {Scott}, {Simpson}, {Swinbank}, {Thomson}, {Walter}, {Wardlow}, \& {van der Werf}}]{Smail2021}
{Smail}, I., {Dudzevi{\v{c}}i{\={u}}t{\.{e}}}, U., {Stach}, S.~M., {et~al.} 2021, \mnras, 502, 3426

\bibitem[{{Smith} {et~al.}(2021){Smith}, {Haskell}, {G{\"u}rkan}, {Best}, {Hardcastle}, {Kondapally}, {Williams}, {Duncan}, {Cochrane}, {McCheyne}, {R{\"o}ttgering}, {Sabater}, {Shimwell}, {Tasse}, {Bonato}, {Bondi}, {Jarvis}, {Leslie}, {Prandoni}, \& {Wang}}]{Smith2021}
{Smith}, D.~J.~B., {Haskell}, P., {G{\"u}rkan}, G., {et~al.} 2021, \aap, 648, A6

\bibitem[{{Smol{\v{c}}i{\'c}} {et~al.}(2017){Smol{\v{c}}i{\'c}}, {Novak}, {Bondi}, {Ciliegi}, {Mooley}, {Schinnerer}, {Zamorani}, {Navarrete}, {Bourke}, {Karim}, {Vardoulaki}, {Leslie}, {Delhaize}, {Carilli}, {Myers}, {Baran}, {Delvecchio}, {Miettinen}, {Banfield}, {Balokovi{\'c}}, {Bertoldi}, {Capak}, {Frail}, {Hallinan}, {Hao}, {Herrera Ruiz}, {Horesh}, {Ilbert}, {Intema}, {Jeli{\'c}}, {Kl{\"o}ckner}, {Krpan}, {Kulkarni}, {McCracken}, {Laigle}, {Middleberg}, {Murphy}, {Sargent}, {Scoville}, \& {Sheth}}]{Smolcic2017}
{Smol{\v{c}}i{\'c}}, V., {Novak}, M., {Bondi}, M., {et~al.} 2017, \aap, 602, A1

\bibitem[{Smolčić {et~al.}(2008)Smolčić, Schinnerer, Scodeggio, Franzetti, Aussel, Bondi, Brusa, Carilli, Capak, Charlot, Ciliegi, Ilbert, Ivezić, Jahnke, McCracken, Obrić, Salvato, Sanders, Scoville, Trump, Tremonti, Tasca, Walcher, \& Zamorani}]{Smolcic2008}
Smolčić, V., Schinnerer, E., Scodeggio, M., {et~al.} 2008, The Astrophysical Journal Supplement Series, 177, 14

\bibitem[{{Speagle} {et~al.}(2014){Speagle}, {Steinhardt}, {Capak}, \& {Silverman}}]{Speagle2014}
{Speagle}, J.~S., {Steinhardt}, C.~L., {Capak}, P.~L., \& {Silverman}, J.~D. 2014, \apjs, 214, 15

\bibitem[{{Sun} {et~al.}(2021){Sun}, {Egami}, {P{\'e}rez-Gonz{\'a}lez}, {Smail}, {Caputi}, {Bauer}, {Rawle}, {Fujimoto}, {Kohno}, {Dudzevi{\v{c}}i{\={u}}t{\.{e}}}, {Atek}, {Bianconi}, {Chapman}, {Combes}, {Jauzac}, {Jolly}, {Koekemoer}, {Magdis}, {Rodighiero}, {Rujopakarn}, {Schaerer}, {Steinhardt}, {Van der Werf}, {Walth}, \& {Weaver}}]{Sun2021}
{Sun}, F., {Egami}, E., {P{\'e}rez-Gonz{\'a}lez}, P.~G., {et~al.} 2021, \apj, 922, 114

\bibitem[{{Tabatabaei} {et~al.}(2016){Tabatabaei}, {Martinsson}, {Knapen}, {Beckman}, {Koribalski}, \& {Elmegreen}}]{Tabatabaei2016}
{Tabatabaei}, F.~S., {Martinsson}, T.~P.~K., {Knapen}, J.~H., {et~al.} 2016, \apjl, 818, L10

\bibitem[{{Talia} {et~al.}(2021){Talia}, {Cimatti}, {Giulietti}, {Zamorani}, {Bethermin}, {Faisst}, {Le F{\`e}vre}, \& {Smol{\c{c}}i{\'c}}}]{Talia2021}
{Talia}, M., {Cimatti}, A., {Giulietti}, M., {et~al.} 2021, \apj, 909, 23

\bibitem[{{Tasse} {et~al.}(2021){Tasse}, {Shimwell}, {Hardcastle}, {O'Sullivan}, {van Weeren}, {Best}, {Bester}, {Hugo}, {Smirnov}, {Sabater}, {Calistro-Rivera}, {de Gasperin}, {Morabito}, {R{\"o}ttgering}, {Williams}, {Bonato}, {Bondi}, {Botteon}, {Br{\"u}ggen}, {Brunetti}, {Chy{\.z}y}, {Garrett}, {G{\"u}rkan}, {Jarvis}, {Kondapally}, {Mandal}, {Prandoni}, {Repetti}, {Retana-Montenegro}, {Schwarz}, {Shulevski}, \& {Wiaux}}]{Tasse2021}
{Tasse}, C., {Shimwell}, T., {Hardcastle}, M.~J., {et~al.} 2021, \aap, 648, A1

\bibitem[{{Taylor} {et~al.}(2023){Taylor}, {Barger}, {Cowie}, {Hasinger}, {Hu}, \& {Songaila}}]{Taylor2023}
{Taylor}, A.~J., {Barger}, A.~J., {Cowie}, L.~L., {et~al.} 2023, \apjs, 266, 24

\bibitem[{{Thorne} {et~al.}(2021){Thorne}, {Robotham}, {Davies}, {Bellstedt}, {Driver}, {Bravo}, {Bremer}, {Holwerda}, {Hopkins}, {Lagos}, {Phillipps}, {Siudek}, {Taylor}, \& {Wright}}]{Thorne2021}
{Thorne}, J.~E., {Robotham}, A. S.~G., {Davies}, L. J.~M., {et~al.} 2021, \mnras, 505, 540

\bibitem[{{Vaccari}(2015)}]{Vaccari2015salt.confE..17V}
{Vaccari}, M. 2015, in SALT Science Conference 2015 (SSC2015), ed. D.~{Buckley} \& A.~{Schroeder}, 17

\bibitem[{{van der Vlugt} {et~al.}(2021){van der Vlugt}, {Algera}, {Hodge}, {Novak}, {Radcliffe}, {Riechers}, {R{\"o}ttgering}, {Smol{\v{c}}i{\'c}}, \& {Walter}}]{vanderVlugt2021ApJ...907....5V}
{van der Vlugt}, D., {Algera}, H.~S.~B., {Hodge}, J.~A., {et~al.} 2021, \apj, 907, 5

\bibitem[{{van der Vlugt} {et~al.}(2022){van der Vlugt}, {Hodge}, {Algera}, {Smail}, {Leslie}, {Radcliffe}, {Riechers}, \& {R{\"o}ttgering}}]{vanderVlugt2022ApJ...941...10V}
{van der Vlugt}, D., {Hodge}, J.~A., {Algera}, H.~S.~B., {et~al.} 2022, \apj, 941, 10

\bibitem[{{van der Vlugt} {et~al.}(2023){van der Vlugt}, {Hodge}, {Jin}, {Algera}, {Leslie}, {Riechers}, {R{\"o}ttgering}, {Smol{\v{c}}i{\'c}}, \& {Walter}}]{vanderVlugt2023ApJ...951..131V}
{van der Vlugt}, D., {Hodge}, J.~A., {Jin}, S., {et~al.} 2023, \apj, 951, 131

\bibitem[{{van Haarlem, M. P.} {et~al.}(2013){van Haarlem, M. P.}, {Wise, M. W.}, {Gunst, A. W.}, {Heald, G.}, {McKean, J. P.}, {Hessels, J. W. T.}, {de Bruyn, A. G.}, {Nijboer, R.}, {Swinbank, J.}, {Fallows, R.}, {Brentjens, M.}, {Nelles, A.}, {Beck, R.}, {Falcke, H.}, {Fender, R.}, {H\"orandel, J.}, {Koopmans, L. V. E.}, {Mann, G.}, {Miley, G.}, {R\"ottgering, H.}, {Stappers, B. W.}, {Wijers, R. A. M. J.}, {Zaroubi, S.}, {van den Akker, M.}, {Alexov, A.}, {Anderson, J.}, {Anderson, K.}, {van Ardenne, A.}, {Arts, M.}, {Asgekar, A.}, {Avruch, I. M.}, {Batejat, F.}, {B\"ahren, L.}, {Bell, M. E.}, {Bell, M. R.}, {van Bemmel, I.}, {Bennema, P.}, {Bentum, M. J.}, {Bernardi, G.}, {Best, P.}, {B\^{\i}rzan, L.}, {Bonafede, A.}, {Boonstra, A.-J.}, {Braun, R.}, {Bregman, J.}, {Breitling, F.}, {van de Brink, R. H.}, {Broderick, J.}, {Broekema, P. C.}, {Brouw, W. N.}, {Br\"uggen, M.}, {Butcher, H. R.}, {van Cappellen, W.}, {Ciardi, B.}, {Coenen, T.}, {Conway, J.}, {Coolen, A.}, {Corstanje, A.}, {Damstra, S.}, {Davies,
  O.}, {Deller, A. T.}, {Dettmar, R.-J.}, {van Diepen, G.}, {Dijkstra, K.}, {Donker, P.}, {Doorduin, A.}, {Dromer, J.}, {Drost, M.}, {van Duin, A.}, {Eisl\"offel, J.}, {van Enst, J.}, {Ferrari, C.}, {Frieswijk, W.}, {Gankema, H.}, {Garrett, M. A.}, {de Gasperin, F.}, {Gerbers, M.}, {de Geus, E.}, {Grie\ss{}meier, J.-M.}, {Grit, T.}, {Gruppen, P.}, {Hamaker, J. P.}, {Hassall, T.}, {Hoeft, M.}, {Holties, H. A.}, {Horneffer, A.}, {van der Horst, A.}, {van Houwelingen, A.}, {Huijgen, A.}, {Iacobelli, M.}, {Intema, H.}, {Jackson, N.}, {Jelic, V.}, {de Jong, A.}, {Juette, E.}, {Kant, D.}, {Karastergiou, A.}, {Koers, A.}, {Kollen, H.}, {Kondratiev, V. I.}, {Kooistra, E.}, {Koopman, Y.}, {Koster, A.}, {Kuniyoshi, M.}, {Kramer, M.}, {Kuper, G.}, {Lambropoulos, P.}, {Law, C.}, {van Leeuwen, J.}, {Lemaitre, J.}, {Loose, M.}, {Maat, P.}, {Macario, G.}, {Markoff, S.}, {Masters, J.}, {McFadden, R. A.}, {McKay-Bukowski, D.}, {Meijering, H.}, {Meulman, H.}, {Mevius, M.}, {Middelberg, E.}, {Millenaar, R.}, {Miller-Jones, J.
  C. A.}, {Mohan, R. N.}, {Mol, J. D.}, {Morawietz, J.}, {Morganti, R.}, {Mulcahy, D. D.}, {Mulder, E.}, {Munk, H.}, {Nieuwenhuis, L.}, {van Nieuwpoort, R.}, {Noordam, J. E.}, {Norden, M.}, {Noutsos, A.}, {Offringa, A. R.}, {Olofsson, H.}, {Omar, A.}, {Orr\'u, E.}, {Overeem, R.}, {Paas, H.}, {Pandey-Pommier, M.}, {Pandey, V. N.}, {Pizzo, R.}, {Polatidis, A.}, {Rafferty, D.}, {Rawlings, S.}, {Reich, W.}, {de Reijer, J.-P.}, {Reitsma, J.}, {Renting, G. A.}, {Riemers, P.}, {Rol, E.}, {Romein, J. W.}, {Roosjen, J.}, {Ruiter, M.}, {Scaife, A.}, {van der Schaaf, K.}, {Scheers, B.}, {Schellart, P.}, {Schoenmakers, A.}, {Schoonderbeek, G.}, {Serylak, M.}, {Shulevski, A.}, {Sluman, J.}, {Smirnov, O.}, {Sobey, C.}, {Spreeuw, H.}, {Steinmetz, M.}, {Sterks, C. G. M.}, {Stiepel, H.-J.}, {Stuurwold, K.}, {Tagger, M.}, {Tang, Y.}, {Tasse, C.}, {Thomas, I.}, {Thoudam, S.}, {Toribio, M. C.}, {van der Tol, B.}, {Usov, O.}, {van Veelen, M.}, {van der Veen, A.-J.}, {ter Veen, S.}, {Verbiest, J. P. W.}, {Vermeulen, R.}, {Vermaas,
  N.}, {Vocks, C.}, {Vogt, C.}, {de Vos, M.}, {van der Wal, E.}, {van Weeren, R.}, {Weggemans, H.}, {Weltevrede, P.}, {White, S.}, {Wijnholds, S. J.}, {Wilhelmsson, T.}, {Wucknitz, O.}, {Yatawatta, S.}, {Zarka, P.}, {Zensus, A.}, \& {van Zwieten, J.}}]{vanHaarlem2013}
{van Haarlem, M. P.}, {Wise, M. W.}, {Gunst, A. W.}, {et~al.} 2013, A\&A, 556, A2

\bibitem[{{Vogelsberger} {et~al.}(2020){Vogelsberger}, {Marinacci}, {Torrey}, \& {Puchwein}}]{Vogelsberger2020NatRP...2...42V}
{Vogelsberger}, M., {Marinacci}, F., {Torrey}, P., \& {Puchwein}, E. 2020, Nature Reviews Physics, 2, 42

\bibitem[{{Wang} {et~al.}(2019){Wang}, {Gao}, {Duncan}, {Williams}, {Rowan-Robinson}, {Sabater}, {Shimwell}, {Bonato}, {Calistro-Rivera}, {Chy{\.z}y}, {Farrah}, {G{\"u}rkan}, {Hardcastle}, {McCheyne}, {Prandoni}, {Read}, {R{\"o}ttgering}, \& {Smith}}]{Wang2019}
{Wang}, L., {Gao}, F., {Duncan}, K.~J., {et~al.} 2019, \aap, 631, A109

\bibitem[{{Weaver} {et~al.}(2023){Weaver}, {Davidzon}, {Toft}, {Ilbert}, {McCracken}, {Gould}, {Jespersen}, {Steinhardt}, {Lagos}, {Capak}, {Casey}, {Chartab}, {Faisst}, {Hayward}, {Kartaltepe}, {Kauffmann}, {Koekemoer}, {Kokorev}, {Laigle}, {Liu}, {Long}, {Magdis}, {McPartland}, {Milvang-Jensen}, {Mobasher}, {Moneti}, {Peng}, {Sanders}, {Shuntov}, {Sneppen}, {Valentino}, {Zalesky}, \& {Zamorani}}]{Weaver2023}
{Weaver}, J.~R., {Davidzon}, I., {Toft}, S., {et~al.} 2023, \aap, 677, A184

\bibitem[{{Weaver} {et~al.}(2022){Weaver}, {Kauffmann}, {Ilbert}, {McCracken}, {Moneti}, {Toft}, {Brammer}, {Shuntov}, {Davidzon}, {Hsieh}, {Laigle}, {Anastasiou}, {Jespersen}, {Vinther}, {Capak}, {Casey}, {McPartland}, {Milvang-Jensen}, {Mobasher}, {Sanders}, {Zalesky}, {Arnouts}, {Aussel}, {Dunlop}, {Faisst}, {Franx}, {Furtak}, {Fynbo}, {Gould}, {Greve}, {Gwyn}, {Kartaltepe}, {Kashino}, {Koekemoer}, {Kokorev}, {Le F{\`e}vre}, {Lilly}, {Masters}, {Magdis}, {Mehta}, {Peng}, {Riechers}, {Salvato}, {Sawicki}, {Scarlata}, {Scoville}, {Shirley}, {Silverman}, {Sneppen}, {Smolc̆i{\'c}}, {Steinhardt}, {Stern}, {Tanaka}, {Taniguchi}, {Teplitz}, {Vaccari}, {Wang}, \& {Zamorani}}]{Weaver2022}
{Weaver}, J.~R., {Kauffmann}, O.~B., {Ilbert}, O., {et~al.} 2022, \apjs, 258, 11

\bibitem[{{Whitaker} {et~al.}(2014){Whitaker}, {Franx}, {Leja}, {van Dokkum}, {Henry}, {Skelton}, {Fumagalli}, {Momcheva}, {Brammer}, {Labb{\'e}}, {Nelson}, \& {Rigby}}]{Whitaker2014}
{Whitaker}, K.~E., {Franx}, M., {Leja}, J., {et~al.} 2014, \apj, 795, 104

\bibitem[{{Whitaker} {et~al.}(2017){Whitaker}, {Pope}, {Cybulski}, {Casey}, {Popping}, \& {Yun}}]{Whitaker2017ApJ...850..208W}
{Whitaker}, K.~E., {Pope}, A., {Cybulski}, R., {et~al.} 2017, \apj, 850, 208

\bibitem[{{Whittam} {et~al.}(2022){Whittam}, {Jarvis}, {Hale}, {Prescott}, {Morabito}, {Heywood}, {Adams}, {Afonso}, {An}, {Ao}, {Bowler}, {Collier}, {Deane}, {Delhaize}, {Frank}, {Glowacki}, {Hatfield}, {Maddox}, {Marchetti}, {Matthews}, {Prandoni}, {Randriamampandry}, {Randriamanakoto}, {Smith}, {Taylor}, {Thomas}, \& {Vaccari}}]{Whittam2022}
{Whittam}, I.~H., {Jarvis}, M.~J., {Hale}, C.~L., {et~al.} 2022, \mnras, 516, 245

\bibitem[{{Williams} {et~al.}(2024){Williams}, {Alberts}, {Ji}, {Hainline}, {Lyu}, {Rieke}, {Endsley}, {Suess}, {Sun}, {Johnson}, {Florian}, {Shivaei}, {Rujopakarn}, {Baker}, {Bhatawdekar}, {Boyett}, {Bunker}, {Cameron}, {Carniani}, {Charlot}, {Curtis-Lake}, {DeCoursey}, {de Graaff}, {Egami}, {Eisenstein}, {Gibson}, {Hausen}, {Helton}, {Maiolino}, {Maseda}, {Nelson}, {P{\'e}rez-Gonz{\'a}lez}, {Rieke}, {Robertson}, {Saxena}, {Tacchella}, {Willmer}, \& {Willott}}]{Williams2024}
{Williams}, C.~C., {Alberts}, S., {Ji}, Z., {et~al.} 2024, \apj, 968, 34

\bibitem[{{Yun} {et~al.}(2001){Yun}, {Reddy}, \& {Condon}}]{Yun2001}
{Yun}, M.~S., {Reddy}, N.~A., \& {Condon}, J.~J. 2001, \apj, 554, 803

\end{thebibliography}

\begin{appendix}
\onecolumn

\section{Fitting the SMFs}\label{app:a_1}

\begin{figure*}[h!]
    \centering
    \includegraphics[width=0.5\textwidth]{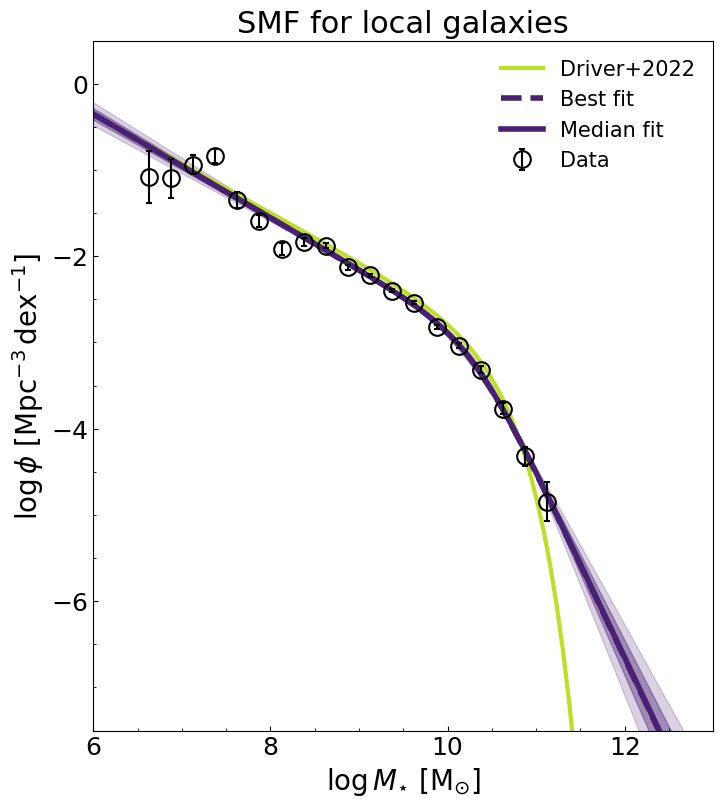}
    \caption{Results of the fit of a Double Power Law (dark blue curve) to the observed SMFs for local ($z<0.08$) SFGs from \cite{Driver2022}. Solid lines refer to the median of the posterior distribution, while dotted lines represent the maximum likelihood model. Shaded areas represent the 1- and 2- $\sigma$ uncertainty interval obtained from the 2nd, 16th, 84th and 97th percentiles of the posterior distribution.
    Our fits are compared to the results of \cite{Driver2022} (green curve) inferred from a single Schechter function. Points are the observed SMFs for local SFGs of \cite{Driver2022}.}
    \label{fig:fit_SMF_Driver_vs_dpl}
\end{figure*}

\begin{figure*}
    \centering
    \includegraphics[width=\textwidth]{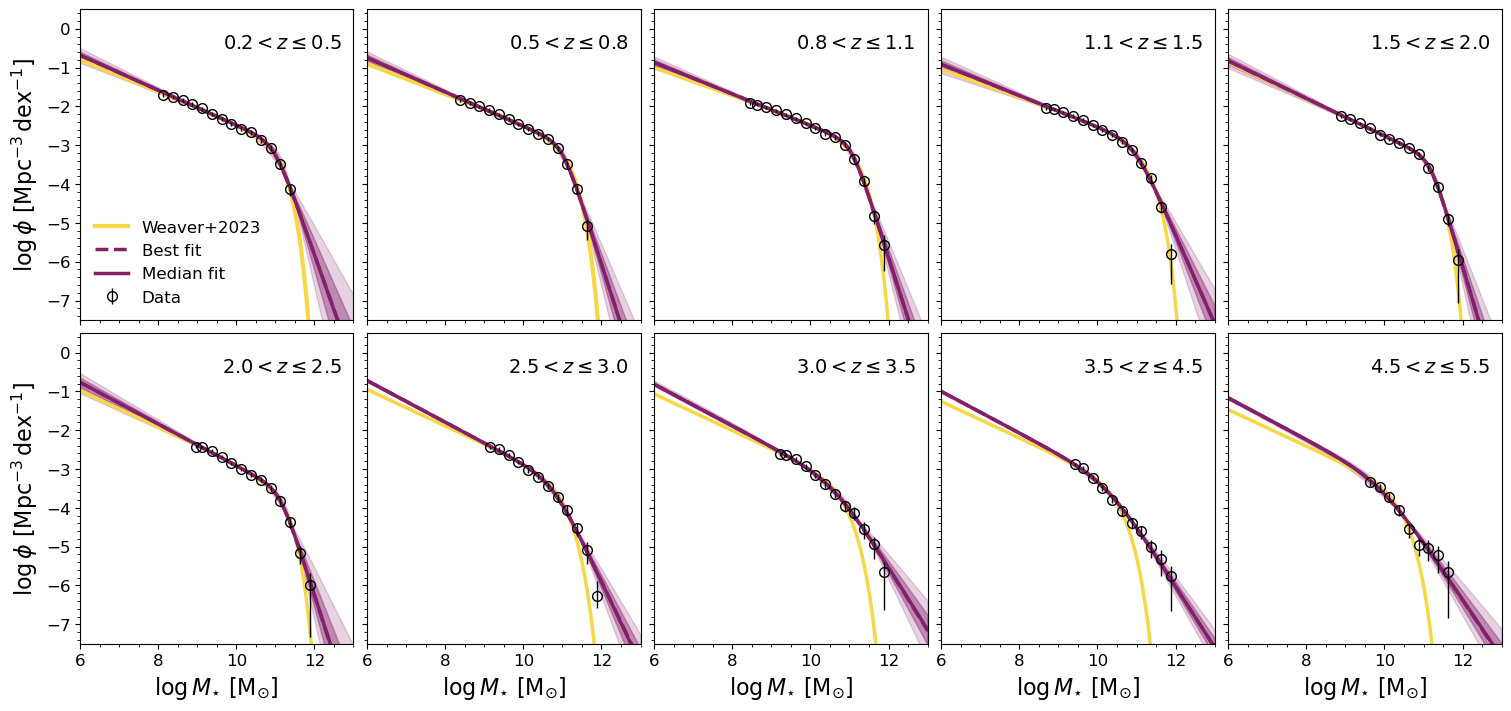}
    \caption{Same as Fig. \ref{fig:fit_SMF_Driver_vs_dpl} but for the observed SMFs for SFGs of \citetalias{Weaver2023} in different redshift bins. 
    Our fits (purple lines) are compared to the results of \citetalias{Weaver2023} (gold curves) for the double ($z<3$) and single ($z>3$) Schechter model. Points are the observed SMFs for SFGs of \citetalias{Weaver2023}.}
    \label{fig:fit_SMF_Weaver_vs_dpl}
\end{figure*}

Here we detail the fitting procedure of the observed SMFs of \citetalias{Weaver2023} and \cite{Driver2022} at different redshift bins.
Our MCMC maximises the likelihood: 
\vspace{-0.25cm}
\begin{table}[h!]
    \centering
    \renewcommand{\arraystretch}{0.9} 
    \setlength{\abovecaptionskip}{0pt} 
    \setlength{\belowcaptionskip}{0pt}
    \setlength{\extrarowheight}{-5pt} 
    \begin{tabular}{c}
        $\displaystyle
        \mathcal{L}(\theta) = -\frac{1}{2} \sum \left( \frac{y - \mathcal{M}(\theta)}{\sigma_{\text{y}}} \right)^2,
        $
    \end{tabular}
\end{table}
\vspace{-0.25cm}

\noindent
where $\mathcal{M}(\theta)$ are the expectations from the empirical model compared with the observed data $y$ and their uncertainties $\sigma_y$. We adopt flat priors ($\pi (\theta)$) for each parameter, initialised to the average value inside the reasonably large intervals: $-1 \leq \alpha \leq 0$, $0 \leq \beta \leq 4$ and $9 \leq \log M_0 (M_{\odot})\leq 12.5$ to allow the chain to span a sufficient range of values. We find the posterior distribution $\mathcal{P}(\theta) \propto \mathcal{L}\pi(\theta)$ to be well-sampled by adopting 500 walkers and 5000 iterations. To allow the chain to reach statistical equilibrium, we discard the first 500 iterations of the MCMC, corresponding to 10\% of the total number of iterations. 
To allow the chain to converge, we fixed the normalisation of the second power-law to the value $\log \Phi_2=3.5$ for all the redshift bins. Because of the lack of data points in the low-mass regime at $z>2.5$ the parameter $\alpha$ is fixed at the maximum likelihood value of the preceding redshift bin ($2.0<z\leq 2.5$). Analogously, $\log \Phi_1$ is fixed for $z>3.5$.   
Fig. \ref{fig:fit_SMF_Driver_vs_dpl} and Fig. \ref{fig:fit_SMF_Weaver_vs_dpl} show the resulting fitting solutions corresponding to the median posterior distribution and maximum likelihood parameters for the local and high-z SMFs, respectively. Our solutions are compared with the results obtained by \cite{Driver2022} (for the local SMF) and \citetalias{Weaver2023} by adopting a Single and Double Schechter profile, respectively. In Tab. \ref{tab:best_fit_parameters}, we report the values of the best-fit parameters for the Double Power Law profile obtained from our analysis.
\noindent
The trend of the observed data is well reproduced by the Double Power Law function for all the redshift bins. In particular, it traces the high-masses points for $z>3$, where the Double Schechter profile fails.
Additionally, we fit the evolution of the parameters obtained from the fit at fixed redshift bins. This is done to retrieve the time and shape evolution of the SMF in the redshift range $0<z<5$. We note that the redshift evolution of our resulting parameters cannot be well reproduced with a simple polynomial function. For this reason, we adopt a cubic spline fit with the number of knots equal to the number of redshift bins.

\begin{table*}[h!]
\caption{List of the parameters obtained by fitting the double power-law profile to \citetalias{Weaver2023} SMFs data. Solutions derived from the median posterior distribution are shown with 1$\sigma$ uncertainty and the maximum likelihood values are reported in square brackets. The second power-law normalisation is fixed ($\log \Phi_2=3.5$) for all redshift bins.}
\label{tab:best_fit_parameters}
\centering
\renewcommand{\arraystretch}{1.3}
\scalebox{1.0}{\begin{tabular}{lcccc} 
\hline
z-bin                 & $\alpha$ & $\log \Phi_1$ & $\log M$ & $\beta$  \\ 
                      &  & $[\rm Mpc^{-3} \, dex^{-1}]$ & $[M_{\odot}]$ &  \\ 
\hline
$0.0\leq z \leq 0.08$ &  $-0.39^{+0.02}_{-0.02}$ [$-0.39$] &  $3.09_{-0.05}^{+0.04}$ [$3.09$] & $10.55_{-0.02}^{+0.02}$ [$10.55$] & $1.2_{-0.2}^{+0.2}$ [$1.2$]       \\
$0.2 < z \leq 0.5$    &  $-0.55^{+0.03}_{-0.03}$ [$-0.55$] &   $3.02_{-0.07}^{+0.06}$ [$3.01$] &  $11.19_{-0.04}^{+0.04}$ [$11.18$] & $1.84_{-0.57}^{+0.69}$ [$1.8$]      \\
$0.5 < z \leq 0.8$    &  $-0.56^{+0.02}_{-0.02}$ [$-0.56$] &   $3.03_{-0.06}^{+0.05}$ [$3.03$] &  $11.18_{-0.03}^{+0.03}$ [$11.18$] & $2.1_{-0.4}^{+0.5}$ [$2.1$]      \\
$0.8 < z \leq 1.1$    &  $-0.59^{+0.02}_{-0.02}$ [$-0.59$]  &   $2.99_{-0.05}^{+0.05}$ [$2.99$] & $11.24_{-0.03}^{+0.03}$ [$11.24$] & $2.2_{-0.4}^{+0.4}$ [$2.2$]      \\
$1.1 < z \leq 1.5$    &  $-0.59^{+0.03}_{-0.03}$ [$-0.59$]  &  $3.02_{-0.07}^{+0.07}$ [$3.01$] & $11.21_{-0.03}^{+0.03}$ [$11.21$] &  $1.3_{-0.3}^{+0.3}$ [$1.3$]       \\
$1.5 < z \leq 2.0$    &  $-0.52^{+0.02}_{-0.02}$ [$-0.52$]   &   $3.35_{-0.05}^{+0.05}$ [$3.35$] & $11.23_{-0.04}^{+0.04}$ [$11.22$] &  $2.40_{-0.4}^{+0.5}$ [$2.4$]      \\
$2.0 < z \leq 2.5$    &  $-0.46^{+0.03}_{-0.04}$ [$-0.46$]  &  $3.52_{-0.08}^{+0.08}$ [$3.52$] & $11.12_{-0.06}^{+0.06}$ [$11.12$] &   $2.1_{-0.6}^{+0.7}$ [$2.1$]     \\
$2.5 < z \leq 3.0$    &  $-0.46 $                        &  $3.33_{-0.05}^{+0.05}$ [$3.33$] &    $10.87_{-0.05}^{+0.05}$ [$10.86$] &  $1.1_{-0.2}^{+0.3}$ [$1.1$]        \\
$3.0 < z \leq 3.5$    &   $-0.46$                        &  $3.36_{-0.07}^{+0.07}$ [$3.36$] &    $10.75_{-0.07}^{+0.07}$ [$10.74$] &  $0.6_{-0.2}^{+0.2}$ [$0.6$]      \\
$3.5 < z \leq 4.5$    &   $-0.46$                        &    $3.36$ &    $10.40_{-0.03}^{+0.03}$ [$10.39$] &   $0.6_{-0.1}^{+0.1}$ [$0.6$]    \\
$4.5 < z \leq 5.5$    &   $-0.46$                        &    $3.36$ &     $10.08_{-0.05}^{+0.04}$ [$10.08$] &   $0.5_{-0.1}^{+0.1}$ [$0.5$]    \\
\hline
\end{tabular}
}
\end{table*}

\end{appendix}

\end{document}